# Template-based eukaryotic genome editing directed by *Svi*Cas3

## Authors:

Wang-Yu Tong*, Yong Li, Shou-Dong Ye, An-Jing Wang, Yan-Yan Tang, Mei-Li Li, Zhong-Fan Yu, Ting-Ting Xia, Qing-Yang Liu and Si-Qi Zhu

## Affiliations:

*Integrated Biotechnology Laboratory, School of Life Sciences, Anhui University, 111 Jiulong Road, Hefei 230601, China*

** Corresponding author:* tongwy@ahu.edu.cn

*Tel.: +86-551-63861282*

*Fax: +86-551-63861282*

## Abstract

RNA-guided gene editing based on the CRISPR-Cas system is currently the most effective genome editing technique. Here, we report that the *Svi*Cas3 from the subtype I-B-*Svi* Cas system in *Streptomyces virginiae* IBL14 is an RNA-guided and DNA-guided DNA endonuclease suitable for the HDR-directed gene and/or base editing of eukaryotic cell genomes. The genome editing efficiency of *Svi*Cas3 guided by DNA is no less than that of *Svi*Cas3 guided by RNA. In particular, t-DNA, as a template and a guide, does not require a proto-spacer-adjacent motif, demonstrating that CRISPR, as the basis for crRNA design, is not required for the *Svi*Cas3-mediated gene and base editing. This discovery will broaden our understanding of enzyme diversity in CRISPR-Cas systems, will provide important tools for the creation and modification of living things and the treatment of human genetic diseases, and will usher in a new era of DNA-guided gene editing and base editing.

## Introduction

Among the many epoch-making achievements in the history of life sciences, the answers to these five fundamental questions constitute the cornerstone of life sciences: (i) what is the nature of heredity? the DNA [1,2], (ii) how is DNA inherited? the DNA double helix and its semiconservative replication [3,4], (iii) how do genetics and physiology relate?- the genetic central dogma and the genetic codon [5,6], (iv) how can genetics serve humans? the genetic engineering of extrachromosomal DNA based on restriction endonuclease (s) and polymerase chain reaction (PCR) [7,8], and (v) How do we create and modify living things? gene editing and base editing based on the mechanism of DNA-orientation, including the protein-guided, RNA-guided and DNA-guided gene editing and base editing. [9-13]. Template-based genome editing is a type of genetic engineering in which a specific site in the target sequence of a cell's genomic DNA is cleaved or degraded by a specific endonuclease or its complex and the target sequence is replaced by a template-DNA donor (t-DNA) through the target cell's own HDR system. From the perspective of genome editing, in the presence of t-DNA, the key step to trigger the cell's own HDR machinery is to create an accurate double-strand DNA (dsDNA) break or single-strand DNA (ssDNA) in the cells' genomic DNA by an engineered nuclease [14].

Cas9 proteins from type II CRISPR-Cas (clustered regularly interspaced short palindromic repeats and CRISPR-associated system) systems are widely believed to perform CRISPR-RNA/crRNA-guided, proto-spacer-adjacent motif (PAM)-dependent cleavage of the target DNA strand (TS) and non-target DNA strand (NTS) using an HNH-like nuclease domain and a RuvC-like nuclease domain, respectively [15,16]. Cas3 superfamily proteins from type I CRISPR-Cas systems perform crRNA-guided, PAM-dependent cleavage sequentially on both the NTS and TS using the N-terminal histidine-aspartate (HD) domain with the help of the C-terminal ATP-dependent superfamily 2 helicase (Hel) domain in the presence of ATP [17-21]. Moreover, Cas3 appears to be neither a component for crRNA processing nor a stable factor for Cascade-crRNA complexes [19,20,22]. In previous genome editing studies based on the subtype I-B-*Svi* CRISPR-Cas system, we have demonstrated that *Svi*Cas3 alone can conduct prokaryotic genome editing, i.e., without the involvement of a Cascade (see the sister article: Prokaryotic genome editing based on the subtype I-B-*Svi* CRISPR-Cas system). Particularly in the experiments to optimize the gene editing process of *Saccharomyces cerevisiae* LYC4, we were fortunate to find that guide-DNA was not necessary. That is, crRNA design and PAM are not needed. This discovery suggests that the *Svi*Cas3 from the subtype I-*Svi* Cas system in *Streptomyces virginiae* IBL14 may be not only an RNA-guided (recognizing R-loop), but also a DNA-guided (recognizing D-loop) endonuclease, and can be developed as base and/or gene editing tools for eukaryotic cell genomes.

The aim of this paper is to present a new DNA-editing platform, namely a set of HDR-directed gene and base editing methods based on the single *Svi*Cas3 enzyme, which can be guiuded by RNA and DNA. We developed different gene editing tools for different host cells based on two strategies: (i) the gene *cas* and t-DNA plus g-DNA in two different plasmids (Cas expression plasmid and gene editing plasmid, named plasmid-*cas gene abbreviation* and plasmid-t/g-*target gene abbreviation*), respectively and (ii) the gene *cas* and t-DNA in one vector (all-in-one plasmids, named plasmid-*cas*3-t-*target gene abbreviation*). Fortunately, all *Svi*Cas3-based, HDR-directed gene and base editing tools guided by RNA and/or DNA can be successfully used for eukaryotic genomic DNA editing.

# Results

## RNA-guided gene editing in *S. cerevisiae*

Given that the gene editing tools developed based on the subtype I-B-*Svi* Cas system are effective in the genome editing of prokaryotic species (see the sister article: Prokaryotic genome editing based on the subtype I-B-*Svi* CRISPR-Cas system), we inferred that it is possible for the I-B-*Svi* CRISPR-Cas system to be developed into gene editing tools for use in the genome editing of eukaryotic cells because the function of *Svi*Cas3 is only to perform site-specific cleavage of endogenous DNA in the genome editing of organisms mediated by the *Svi*Cas system. Accordingly, we selected the gene *crtE*, consisting of 1131 nt on chromosome IV, in an engineered *Saccharomyces cerevisiae* LYC4 (which carries four lycopene biosynthesis genes: *crtE* encoding geranylgeranyl pyrophosphate synthase / GGPS, *crtB* encoding phytoene synthase (PSY), *crtI* encoding phytoene desaturase (PDS) and *zds* encoding zeta-carotene desaturase (ZDS)) as a target because of the advantages of its red-white selection and status as a genetic marker of auxotroph [23,24], and we designed and constructed the gene editing plasmid pYES2-NTA-t/g-Δ*crtE* (Figures 1A and 1C) and two Cas expression plasmids (pRS415-*cas*7-5-3r and pRS415-*cas*3r, respectively harbouring raw gene fragments *cas*7-5-3 and *cas*3 from the strain *S. virginiae* IBL14 genome). The arbitrarily designed fragment sizes of t-DNA (UHA plus DHA) and deletion in the gene editing of *crtE* were 807 nt (353+454 nt) and 252 nt, respectively and the selected PAM was ttc (Figure 1C, Figures S1A and S1C, Table S2). No correctly gene-edited mutants were obtained owing to the differences in base preference and cistron between prokaryotic and eukaryotic species. Subsequently, we designed and constructed two Cas expression plasmids pRS415-*cas*7-5-3 (carrying the three codon optimized genes, *Svicas*7, *Svicas*5 and *Svicas*3 in the form of monocistronic transcription) and pRS415-*cas*3 (harbouring only the codon optimized gene *Svicas*3), with a nuclear localization signal of SV40 large T antigen at the C-terminus of each *Svi*Cas (Figures S1B and S1C, Tables S1 and S2). After delivering the two sets of gene editing tools (pRS415-*cas*7-5-3 plus pYES2-NTA-t/g-Δ*crtE* and pRS415-*cas*3 plus pYES2-NTA-t/g-Δ*crtE*) into the competent cells of *S. cerevisiae* LYC4, we obtained white gene-edited mutants (Figure 1A) and validated them by PCR, DNA electrophoresis and sequencing analysis (Figures 1B and 1C).

**A**

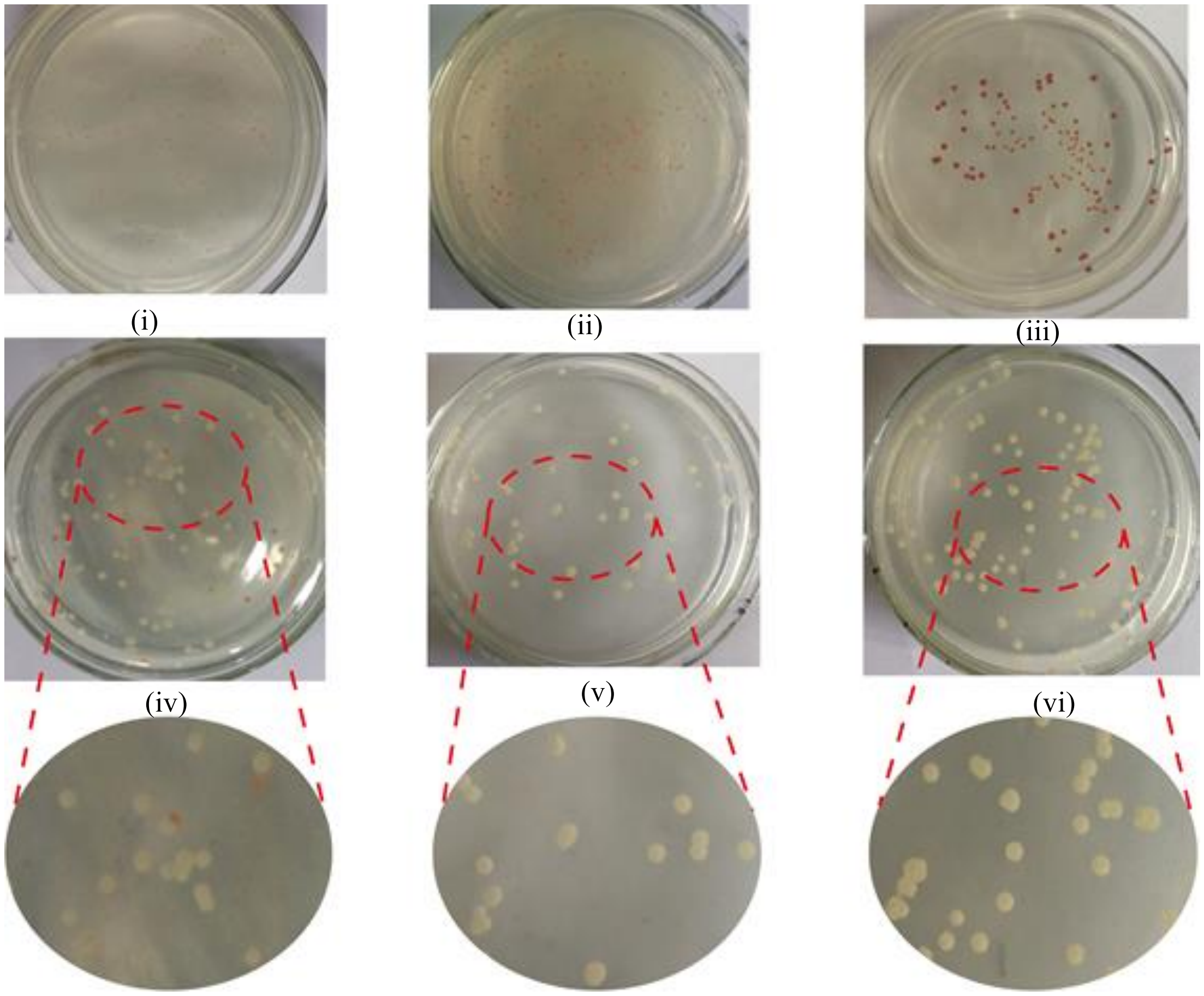

i-iii: *S.cerevisiae* LYC4, or transformants by pRS415-*cas7*-5-3 or pRS415-*cas*3;
iv: transformants by pYES2-t/g-Δ*crtE;*
iv-vi: transformants by pRS415-*cas*7-5-3+pYES2-t/g-Δ*crtE* or pRS415-*cas*7-5-3+pYES2-t-Δ*crtE*
or transformants by pRS415-*cas*3+pYES2-t/g-Δ*crtE* or pRS415-*cas*3+pYES2-t-Δ*crtE*

**B**

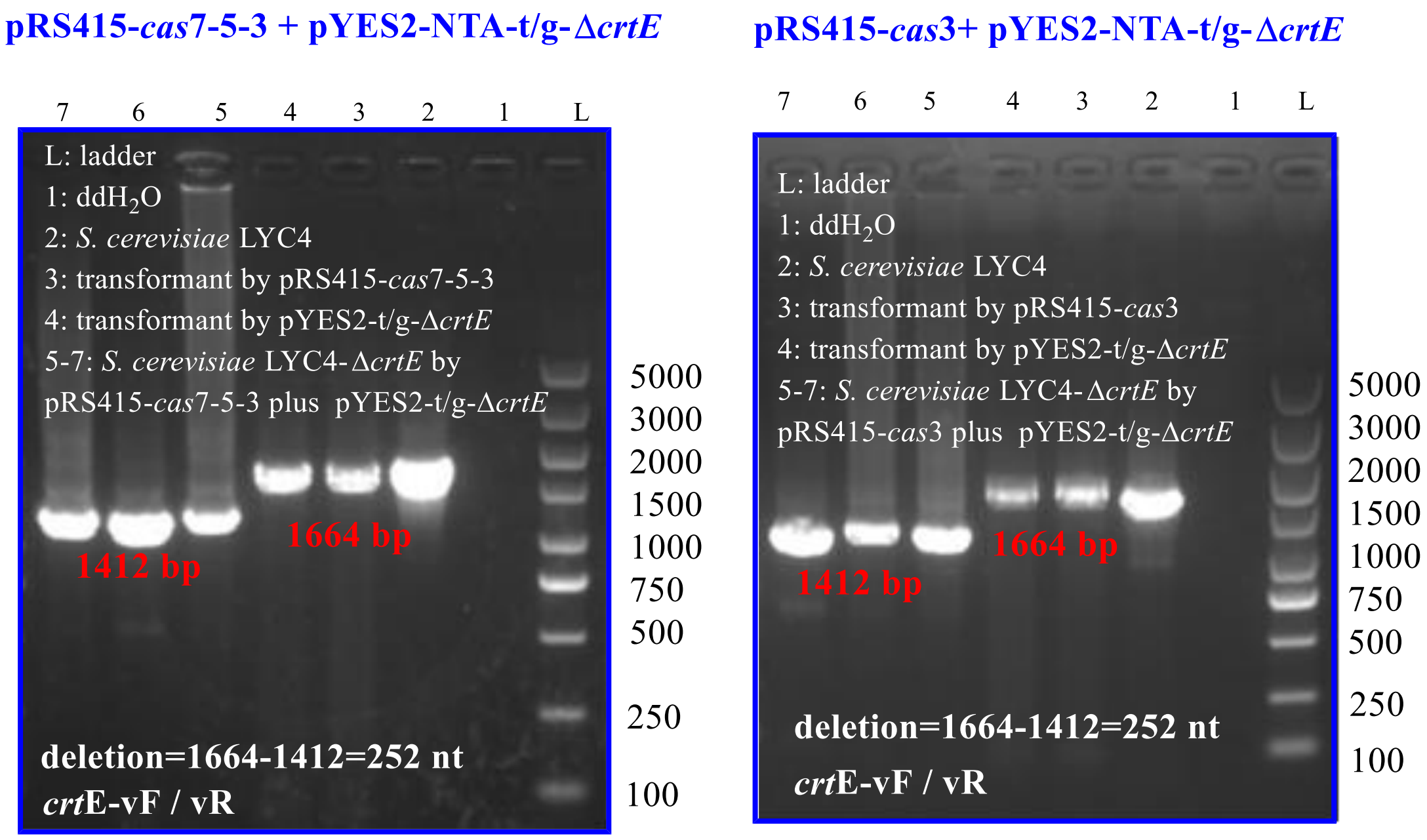
pRS415-cas7-5-3 + pYES2-NTA-t/g-ΔcrtE
7 6 5 4 3 2 1 L
L: ladder
1: ddH2O
2: S. cerevisiae LYC4
3: transformant by pRS415-cas7-5-3
4: transformant by pYES2-t/g-ΔcrtE
5-7: S. cerevisiae LYC4-ΔcrtE by pRS415-cas7-5-3 plus pYES2-t/g-ΔcrtE
1412 bp
1664 bp
5000
3000
2000
1500
1000
750
500
250
100
deletion=1664-1412=252 nt
crtE-vF / vR
pRS415-cas3+ pYES2-NTA-t/g-ΔcrtE
7 6 5 4 3 2 1 L
L: ladder
1: ddH2O
2: S. cerevisiae LYC4
3: transformant by pRS415-cas3
4: transformant by pYES2-t/g-ΔcrtE
5-7: S. cerevisiae LYC4-ΔcrtE by pRS415-cas3 plus pYES2-t/g-ΔcrtE
1412 bp
1664 bp
5000
3000
2000
1500
1000
750
500
250
100
deletion=1664-1412=252 nt
crtE-vF / vR

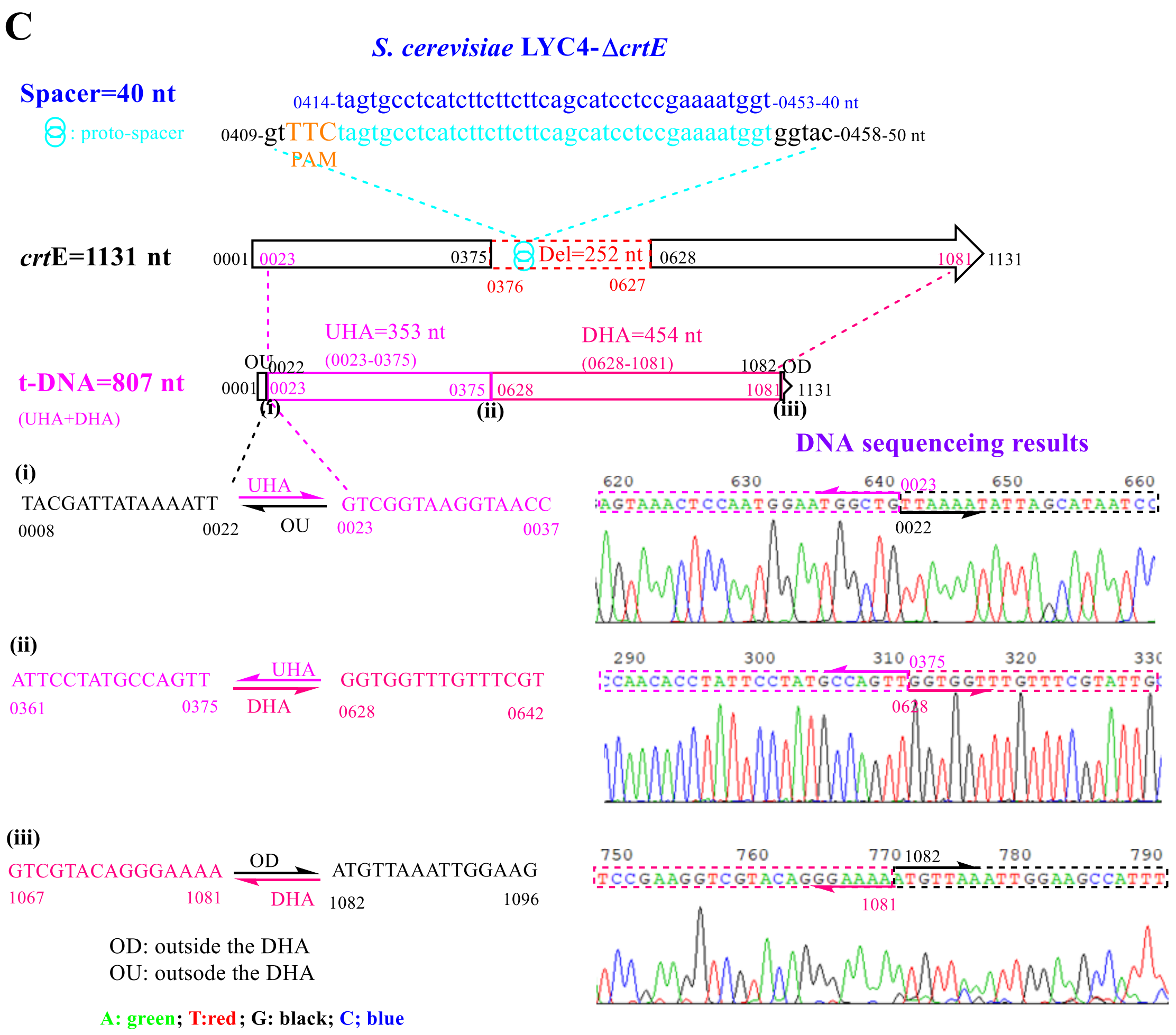

**Figure 1. Transformant features on plates and mutant verification in RNA-guided genome editing of *S. cerevisiae* LYC4. (A)** Typical features of both host *S.cerevisiae* LYC4 and its transformants on corresponding plates. **(B)** DNA gel electrophoresis of the PCR products of the target sequences in the RNA-guided genome editing of *S. cerevisiae* LYC4 (transformed by pRS415-*cas*7-5-3 plus pYES2-NTA-t/g-Δ*crtE* and pRS415-*cas*3 plus pYES2-NTA-t/g-Δ*crtE*, respectively). **(C)** The selected proto-spacer, the engineered t-DNA, and the DNA sequencing results of the PCR products of the edited sequences in the gene-edited mutant *S. cerevisiae* LYC4-Δ*crtE* genome (the junction sequences of approximately 40 nt between the outside of UHA and UHA, UHA and DHA, and DHA and the outside of DHA are highlighted in Figure 1C).

In the genome of the gene-edited mutant *S. cerevisiae* LYC4-Δ*crtE*, the band size of the PCR product of the target sequences in the DNA gel electrophoretogram (Figure 1B) is approximately 1412 nt, consistent with the size of the predesigned DNA fragment, and the result of targeted sequencing was completely consistent with the design sequence (Figure 1C, Table S2 ). This result again demonstrated that the *Svi*Cas3 protein is necessary and that both *Svi*Cas7 and *Svi*Cas5 are auxiliary in non-self-targeting genome editing and that the crRNA fragment composed of R-S-R could also guide the *Svi*Cas3 enzyme or the *Svi*Cascade to perform site-specific cleavage of a target gene and further trigger genome editing in eukaryotic cells. Additionally, in DNA gel electrophoresis (Lanes 5-7 in Figure 1B), we can see that the 1412 bp-band of the PCR products from the correctly gene-edited mutant *S. cerevisiae* LYC4-Δ*crtE* genome is much more prominent than the 1664 bp-band of raw target sequences (not successfully gene-edited transformants), suggesting that the homologous recombination efficiency / HRE in the genome editing of *S. cerevisiae* LYC4 is quite high (Table S3). It is worth noting that the transformants of *S. cerevisiae* LYC4 transformed by the single plasmid pYES2-NTA-t/g-Δ*crtE* are either white or red (Figure 1). Clearly, the white colonies are not the *Svi*Cas3-mediated, Δ*crtE*-edited mutants, but may be the result of transcriptional regulation of complementary sequences between t/g-Δ*crtE* and the gene target site.

**RNA-guided gene editing in mammalian cells**

To further verify the effectiveness of the single *Svi*Cas3 in genome editing of mammalian cells, two commonly used cell lines (HEK293T: human embryonic kidney 293 cells expressing a temperature-sensitive mutant of the SV40 large T antigen; NIH-3T3: mouse embryonic fibroblast cell line as the standard fibroblast cell line) and three genes of interest [(i) *DROSHA*-encoding an RNase III enzyme, a core nuclease that initiates miRNA processing in the nucleus, (ii) *CAMKMT*-encoding a calmodulin-lysine N-methyltransferase that catalyses the trimethylation of Lys-116 in calmodulin and (iii) *Lepr*-encoding a high affinity receptor that mediates the regulation of the *Leptin* gene] (https://www.ncbi.nlm.nih.gov) [25] were selected as the target cells and the target genes. In the genome editing of the three genes *DROSHA* (HEK293T, Homo sapiens-chr V), *CAMKMT* (HEK293T, Homo sapiens-chr II) and *Lepr* (NIH-3T3, Mus musculus-chr IV), we first selected the expression vector AIO-mCherry (http://www.addgene.org/) (mCherry can be expressed in cells after the vector is delivered into HEK293T and NIH-3T3 cells) as the original vector of gene editing tools, and then designed and constructed three gene editing vectors: (i) AIO-mCherry-*cas*3-t/g-Δ*DROSHA*::*egfp*, (ii) AIO-mCherry-*cas*3-t/g-Δ*CAMKMT*::*egfp* and (iii) AIO-mCherry-*cas*3-t/g-Δ*Lepr*::*egfp* (Figure S2, Tables S1 and S2), following the procedures described in the section "Construction of gene editing tools for genome editing in mammalian cells". The arbitrarily designed fragment sizes of both t-DNA (UHA plus *egfp* plus DHA) and deletion were: 299+726+365=1390 nt and 264 nt (Δ*DROSHA*::*egfp*), 351+726+429=1506 nt and 306 nt (Δ*CAMKMT*::*egfp*) and 351+726+153=1230 nt and 84 nt (Δ*Lepr*::*egfp*), respectively. The selected PAMs respectively was ttc (Δ*DROSHA*::*egfp*), tcc (Δ*CAMKMT*::*egfp*) and tac (Δ*Lepr*::*egfp*) (Figure 2C and Figure S2, Table S2). After transfecting each of the three gene editing vectors (AIO-mCherry-*cas*3-t/g-Δ*DROSHA*::*egfp*, AIO-mCherry-*cas*3-t/g-Δ*CAMKMT*::*egfp*, and AIO-mCherry-*cas*3-t/g-Δ*Lepr*::*egfp*.) into corresponding cells, we obtained three potential gene-edited variants with green fluorescence (Figure 2A) (Note: Each UHA of the three gene editing vectors has no promoter, so EGFP can be expressed only by inserting the *egfp* gene into the framework of the three target genes, which allows green fluorescence to be detected.): HEK293T-Δ*DROSHA::egfp*, HEK293T-Δ*CAMKMT*::*egfp* and NIH-3T3-Δ*Lepr*::*egfp*, and tested them using a basic PCR (primers: *DROSHA*-vF / vR and *DROSHA*-*egfp*-vF / *DROSHA*-vR, *CAMKMT*-vF / vR and *CAMKMT*-*egfp*-vF/*CAMKM*T-vR, *Lepr*-vF / vR and *Lepr*-vF / *Lepr*-*egfp*-vR) and DNA sequencing analysis (Figures 2B and 2C, Table S2).

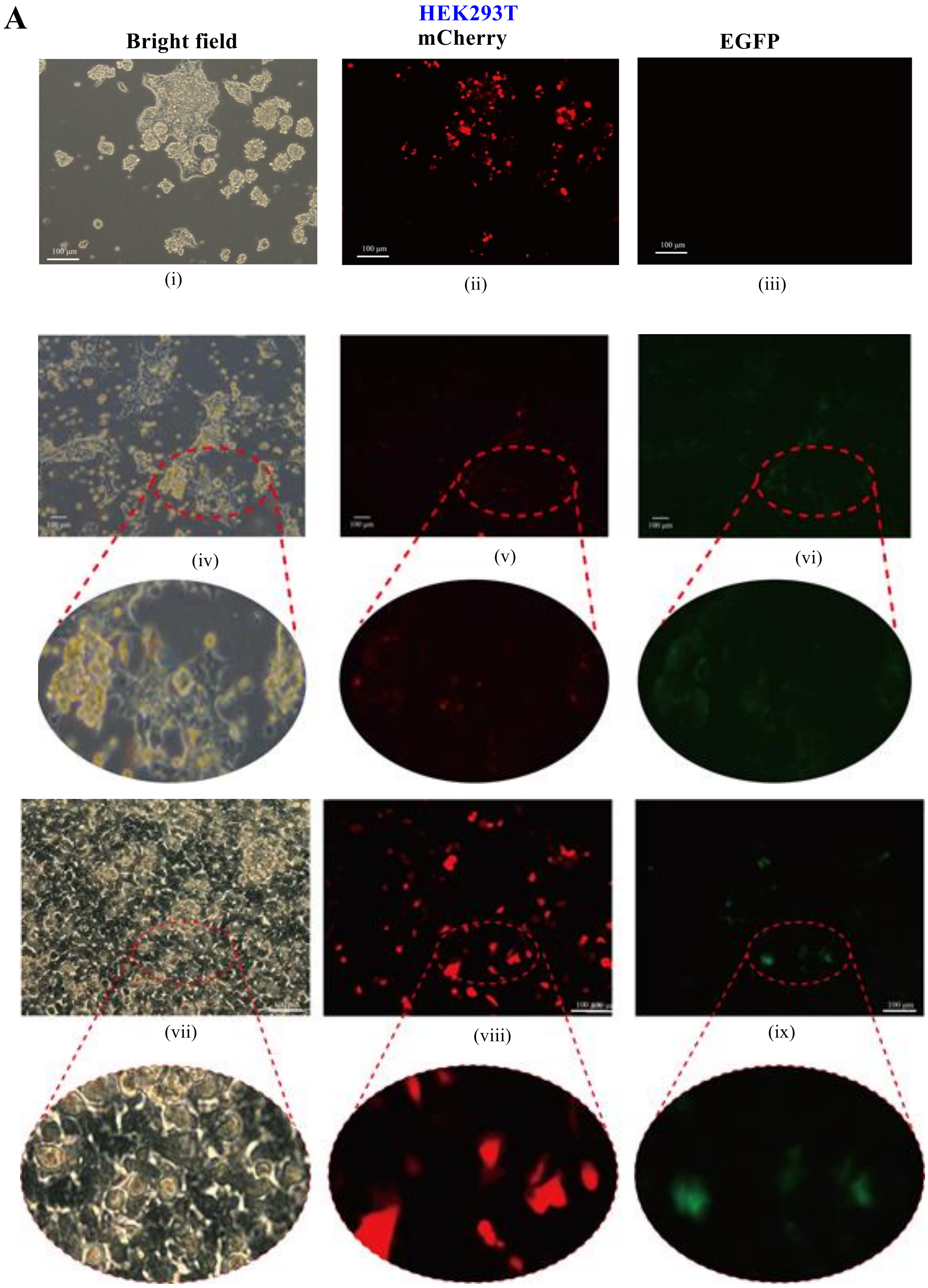

(i-iii: transformants by AIO-mCherry-*cas*3, AIO-mCherry-t/g-Δ*DROSHA*::*egfp*, AIO-mCherry-t/g-Δ*CAMKMT*::*egfp,* AIO-mCherry-t-Δ*DROSHA*::*egfp*, AIO-mCherry-t-Δ*CAMKMT*::*egfp* or AIO-mCherry-$t_{g1}/t_{b2}$-Δ*CAMKMT*::*egfp*;
iv-ix: HEK293T-Δ*DROSHA*::*egfp* by AIO-mCherry-*cas*3-t/g-Δ*DROSHA*::*egfp*, AIO-mCherry-*cas*3-t/g-Δ*CAMKMT*::*egfp*, AIO-mCherry-*cas*3-t-Δ*DROSHA*::*egfp*, AIO-mCherry-*cas*3-t-Δ*CAMKMT*::*egfp* or AIO-mCherry-*cas*3-$t_{g1}/t_{b2}$-Δ*CAMKMT*::*egfp*)

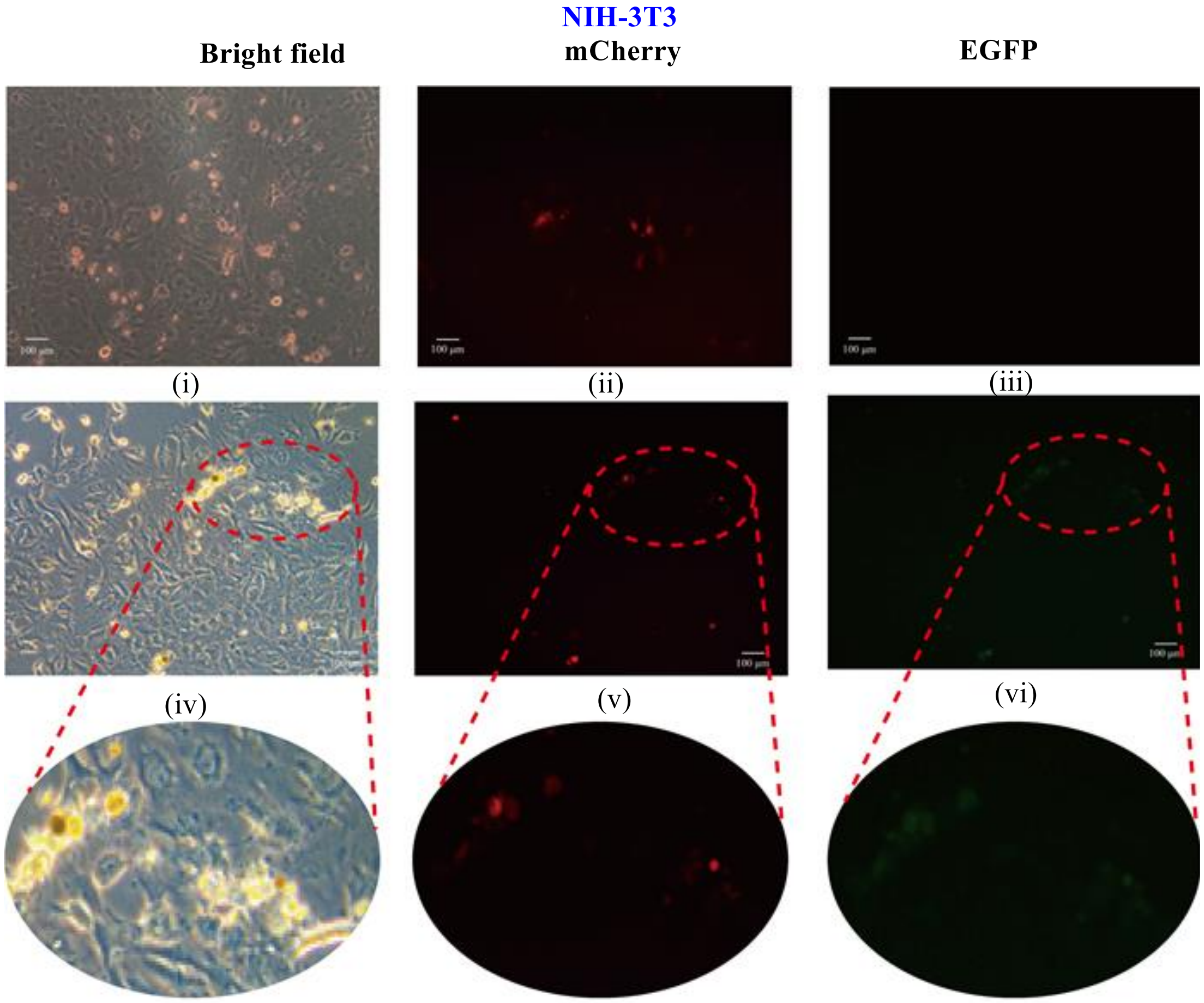

(i-iii: transformants by AIO-mCherry-*cas*3, AIO-mCherry-t/g-Δ*Lepr*::*egfp* or AIO-mCherry-t-Δ*Lepr*::*egfp*; iv-ix: NIH3T3-Δ*Lepr*::*egfp* by AIO-mCherry-*cas*3-t/g-Δ*Lepr*::*egfp* or AIO-mCherry-*cas*3-t-Δ*Lepr*::*egfp*)

**B**

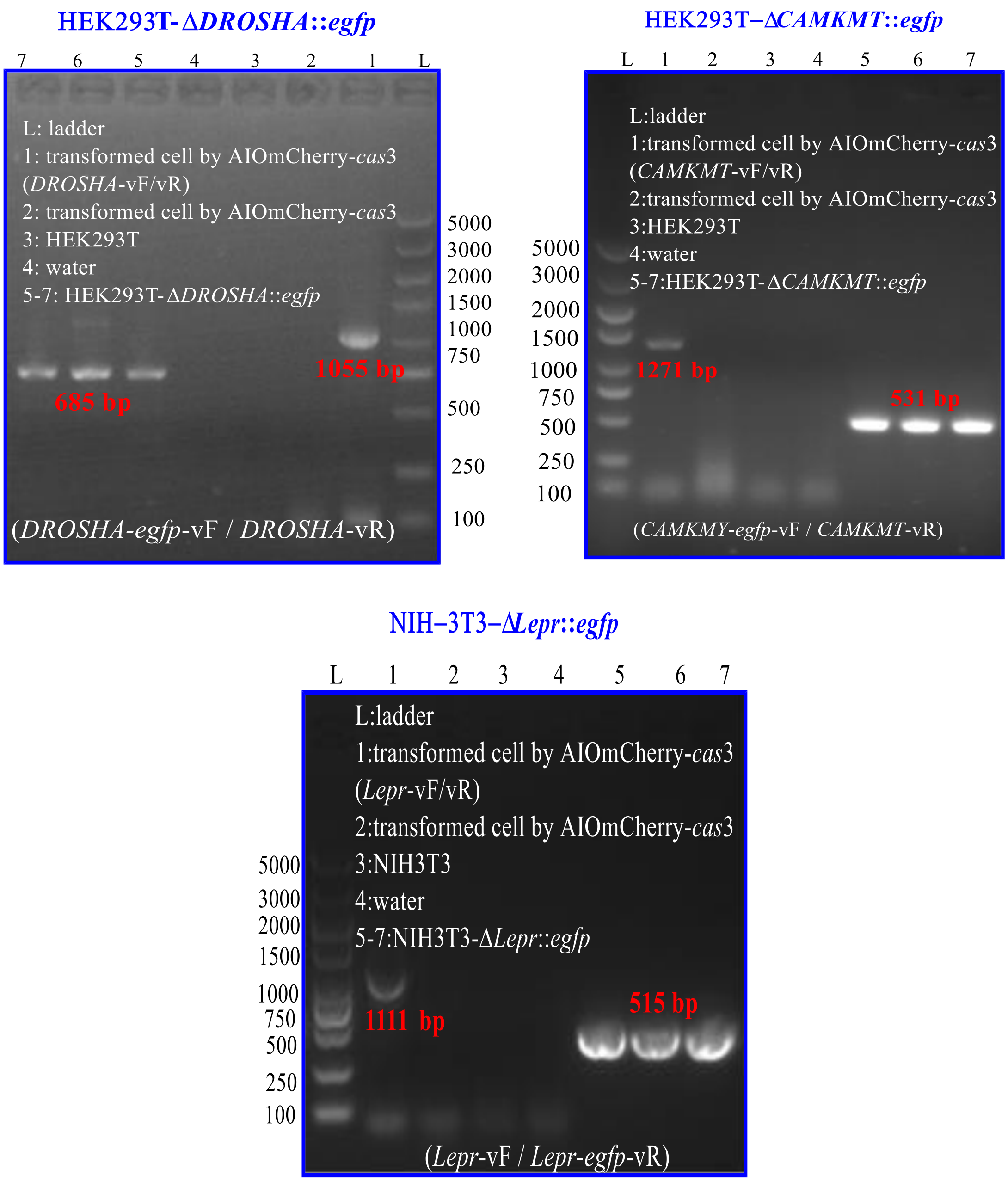
HEK293T-ΔDROSHA::egfp
7 6 5 4 3 2 1 L
L: ladder
1: transformed cell by AIOmCherry-cas3
(DROSHA-vF/vR)
2: transformed cell by AIOmCherry-cas3
3: HEK293T
4: water
5-7: HEK293T-ΔDROSHA::egfp
1055 bp
685 bp
5000
3000
2000
1500
1000
750
500
250
100
(DROSHA-egfp-vF / DROSHA-vR)
HEK293T–ΔCAMKMT::egfp
L 1 2 3 4 5 6 7
L:ladder
1:transformed cell by AIOmCherry-cas3
(CAMKMT-vF/vR)
2:transformed cell by AIOmCherry-cas3
3:HEK293T
4:water
5-7:HEK293T-ΔCAMKMT::egfp
5000
3000
2000
1500
1000
750
500
250
100
1271 bp
531 bp
(CAMKMY-egfp-vF / CAMKMT-vR)
NIH–3T3–ΔLepr::egfp
L 1 2 3 4 5 6 7
L:ladder
1:transformed cell by AIOmCherry-cas3
(Lepr-vF/vR)
2:transformed cell by AIOmCherry-cas3
3:NIH3T3
4:water
5-7:NIH3T3-ΔLepr::egfp
5000
3000
2000
1500
1000
750
500
250
100
1111 bp
515 bp
(Lepr-vF / Lepr-egfp-vR)

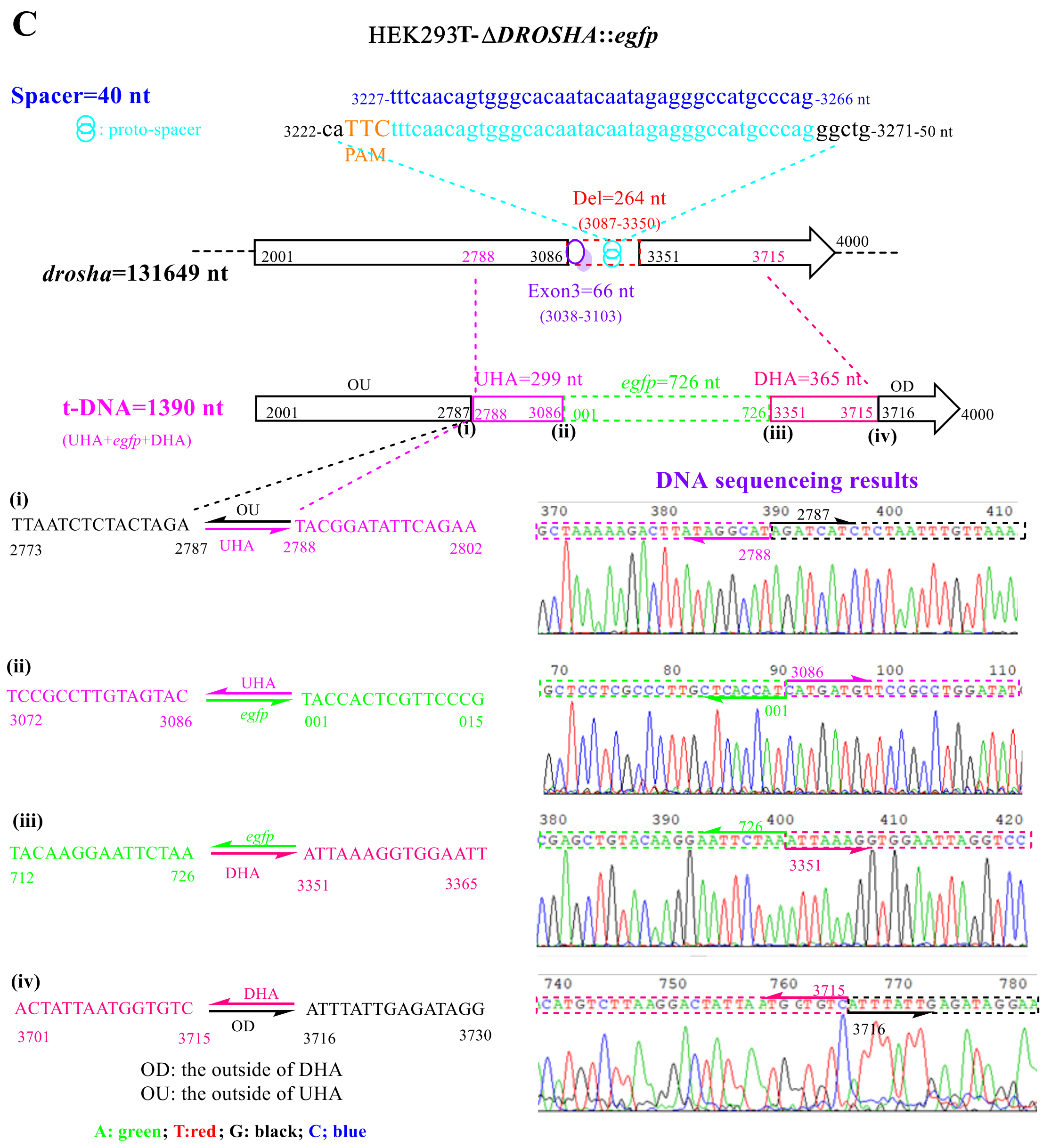
C
HEK293T-Δ*DROSHA*::*egfp*
Spacer=40 nt
: proto-spacer
3227-tttcaacagtgggcacaatacaatagagggccatgcccag-3266 nt
3222-caTTCtttcaacagtgggcacaatacaatagagggccatgcccagggctg-3271-50 nt
PAM
Del=264 nt
(3087-3350)
*drosha*=131649 nt
2001
2788
3086
3351
3715
4000
Exon3=66 nt
(3038-3103)
t-DNA=1390 nt
(UHA+*egfp*+DHA)
OU
UHA=299 nt
*egfp*=726 nt
DHA=365 nt
OD
2001
2787
2788
3086
001
726
3351
3715
3716
4000
(i)
(ii)
(iii)
(iv)
DNA sequenceing results
(i)
TTAATCTCTACTAGA
OU
TACGGATATTCAGAA
2773
2787
UHA
2788
2802
(ii)
TCCGCCTTGTAGTAC
UHA
TACCACTCGTTCCCG
3072
3086
*egfp*
001
015
(iii)
TACAAGGAATTCTAA
*egfp*
ATTAAAGGTGGAATT
712
726
DHA
3351
3365
(iv)
ACTATTAATGGTGTC
DHA
ATTTATTGAGATAGG
3701
3715
OD
3716
3730
OD: the outside of DHA
OU: the outside of UHA
A: green; T:red; G: black; C; blue

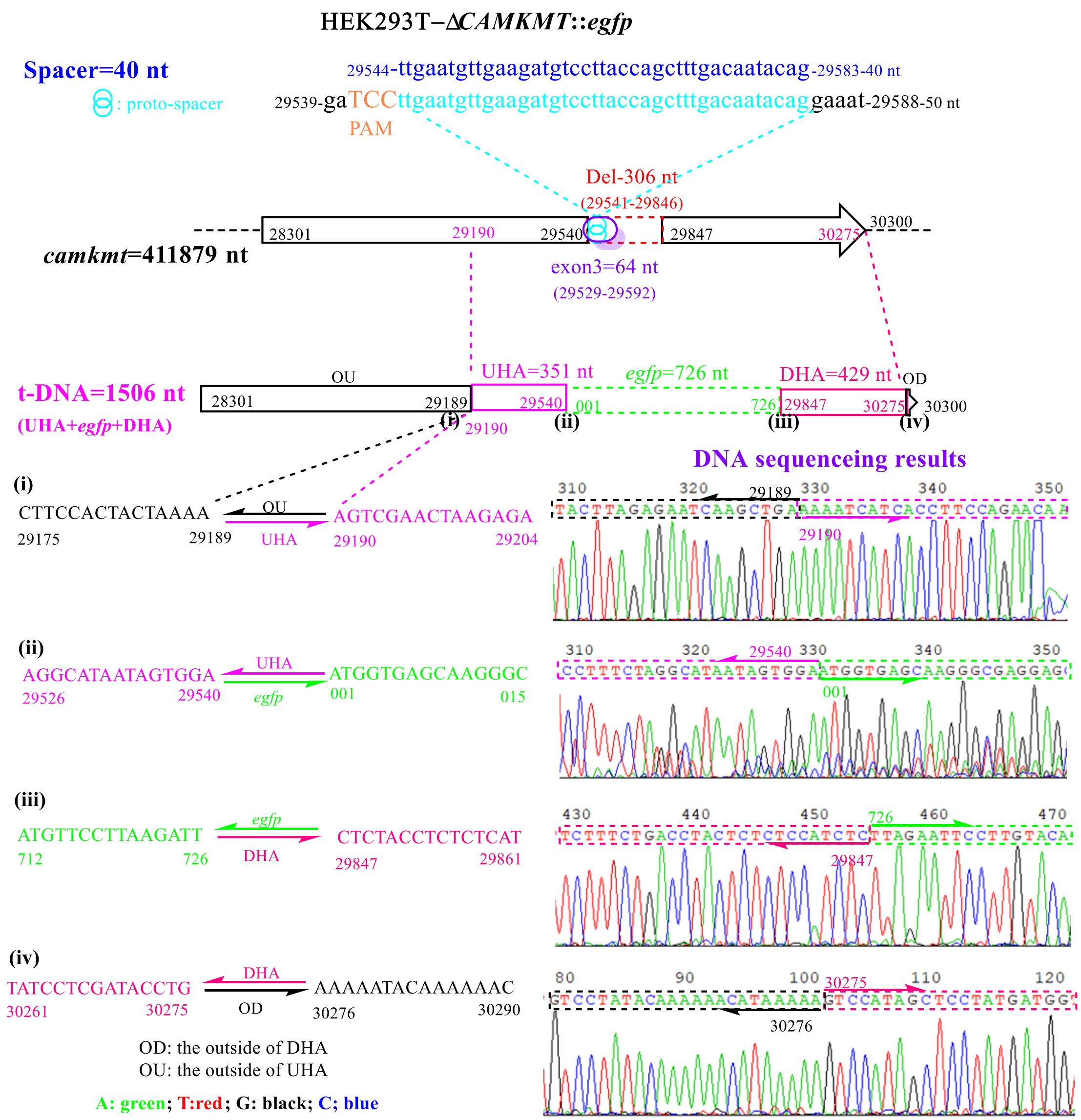
HEK293T−ΔCAMKMT::egfp
Spacer=40 nt
: proto-spacer
29544-ttgaatgttgaagatgtccttaccagctttgacaatacag-29583-40 nt
29539-gaTCCttgaatgttgaagatgtccttaccagctttgacaatacaggaaat-29588-50 nt
PAM
Del-306 nt
(29541-29846)
camkmt=411879 nt
28301
29190
29540
29847
30275
30300
exon3=64 nt
(29529-29592)
t-DNA=1506 nt
(UHA+egfp+DHA)
OU
UHA=351 nt
egfp=726 nt
DHA=429 nt
OD
DNA sequenceing results
(i)
CTTCCACTACTAAAA
AGTCGAACTAAGAGA
29175
29189
29190
29204
(ii)
AGGCATAATAGTGGA
ATGGTGAGCAAGGGC
29526
29540
001
015
(iii)
ATGTTCCTTAAGATT
CTCTACCTCTCTCAT
712
726
29847
29861
(iv)
TATCCTCGATACCTG
AAAAATACAAAAAAC
30261
30275
30276
30290
OD: the outside of DHA
OU: the outside of UHA
A: green; T:red; G: black; C; blue

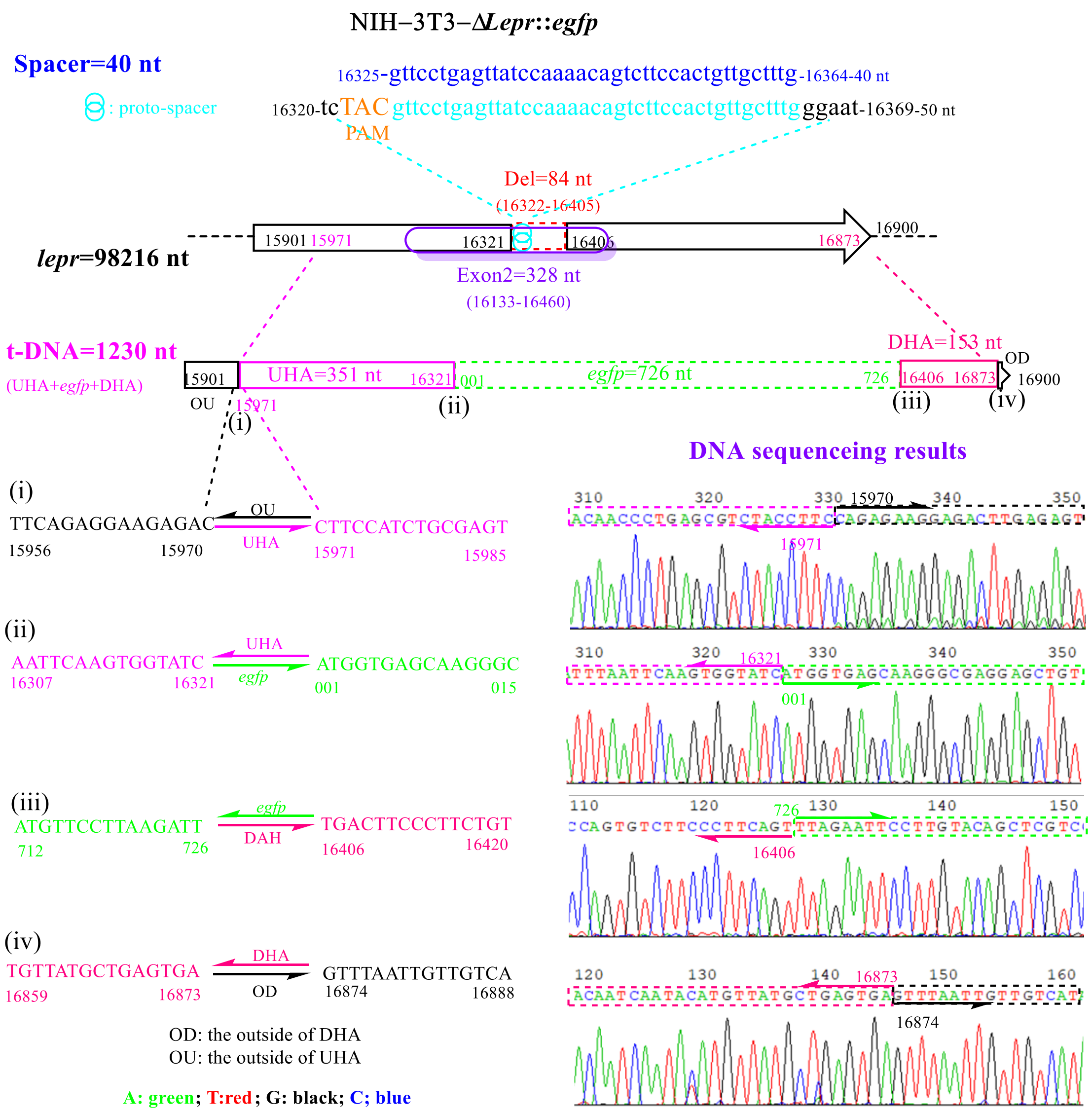

**Figure 2. Micrographs of the gene editing process and mutant verification in the RNA-guided genome editing of HEK293T and NIH-3T3 cells. (A)** Typical micrographs of the gene editing process in the RNA-guided or DNA-guided genome editing of HEK293T and NIH-3T3 cells (Bright field: total cell numbers of HEK293T/NIH-3T3=cell numbers of HEK293T/NIH-3T3 with and without AIO-mCherry-*cas*3-t/g-Δ*gene*::*egfp*, AIO-mCherry-*cas*3-t-Δ*gene*::*egfp* or AIO-mCherry-*cas*3-$t_{g1}$/$t_{b2}$-Δ*CAMKMT*::*egfp*; mCherry: cell numbers of HEK293T/NIH-3T3 harbouring AIO-mCherry-*cas*3-t/g-Δ*gene*::*egfp*, AIO-mCherry-*cas*3-t-Δ*gene*::*egfp* or AIO-mCherry-*cas*3-$t_{g1}$/$t_{b2}$-

ΔCAMKMT::egfp; EGFP: cell numbers of correctly gene-edited or base-edited cells of HEK293T / NIH-3T3) (mCherry and EGFP were detected by Leica Microsystems CMS GmbH-DMi8, Germany). **(B)** DNA gel electrophoresis of the PCR products of target sequences in the RNA-guided genome editing of mammalian cells. **(C)** The selected proto-spacers, the engineered t-DNAs, and the DNA sequencing result of the PCR products of edited sequences in the genomic DNA of the gene-edited variants HEK293T-Δ*DROSHA::egfp*, HEK293T-Δ*CAMKMT::egfp* and NIH-3T3-Δ*Lepr::egfp* (The junction sequences between the outside of UHA and UHA, UHA and *egfp*, *egfp* and DHA, and DHA and the outside of DHA are highlighted in Figure 2C).

To quickly find correctly genome-edited mutants and reduce the impact of large mammalian genomes on PCR results, verification primers with exogenous insertion sequences (e.g., gene *egfp*), *target gene abbreviation-egfp*-vF / vR, were designed in genome editing of mammanlian cells. PCR products only occur in successfully gene-edited cells of HEK293T-Δ*DROSHA::egfp*, HEK293T-Δ*CAMKMT::egfp* and NIH-3T3-Δ*Lepr::egfp*, because the primers *DROSHA-egfp*-vF, *CAMKMT-egfp*-vF and *Lepr-egfp*-vR are portion of the gene *egfp* in t-Δ*DROSHA::egfp*, t-*CAMKMT::egfp* and t-*Lepr::egfp*, and the primers *DROSHA*-vR, *CAMKMT*-vR and *Lepr*-vF are part of the genes *DROSHA*, *CAMKMT* and *Lepr* but are not contained in t-Δ*DROSHA::egfp*, t-*CAMKMT::egfp* and t-*Lepr::egfp*, respectively.

We first observed the potential genome-edited mutants with green fluorescence (Figure 2A), then obesrved the PCR products on a DNA gel electrophoretogram using the verification primers (Figure 2B), and finaly verified the PCR products through DNA sequencing (Figure 2C). All the analytical results were consistent with rational expectations and designed t-DNA (Figure 2 and Table S2), demonstrating again that, similar to *Sp*Cas9, *Svi*Cas3 alone in vivo can perform site-specific cleavage at a target site and further result in template-based genome editing accomplished by the host cell's own HDR system in prokaryotic and eukaryotic cells. The results also support the view that if there is no biocompatibility problem between a restriction endonuclease and host cells, a restriction endonuclease effective for DNA in prokaryotic genomes should also be effective for DNA in eukaryotic genomes, and vice versa. In fact, this view could be inferred from the fact that in vitro, any restriction deoxyribonuclease can cleave DNA molecules of any origin, whether natural or chemically synthesized, at the corresponding restriction enzyme site. Notably, some gene-edited cells showing EGFP did not show mCherry due to loss of the gene editing vector, and the number of dead cells among the gene-edited cells varied from gene to gene and from experimental batch to experiment batch (for example, in the gene editing of *lepr*, the proportion of dead cells was particularly high) (Figure 2A).

**DNA-guided gene editing in *S. cerevisiae***

After confirming that the subtype I-B-*Svi* CRISPR-Cas system in *S. virginiae* IBL14 could be developed into genome editing tools for application in prokaryotic and eukaryotic species, we selected several plasmids, (i) pYES2-NTA-t/g-Δ*crtE*, (ii) pRS415-*cas*7-5-3, (iii) pRS415-*cas*3, (iv) pRS415-*cas*7-5-3 plus pYES2-NTA-t/g-Δ*crtE* and (v) pRS415-*cas*3 plus pYES2-NTA-t-Δ*crtE*, and delivered them into the host *S. cerevisiae* LYC4 cells. All the experimental designs and operational procedures are described in the section "Materials and Methods" except that the plasmid pYES2-NTA-t-Δ*crtE* lacks g-DNA, compared with the plasmid pYES2-NTA-t/g-Δ*crtE* (Figure 3 and Figure S3, Tables S1 and S2). Surprisingly, in the gene-edited mutants by pRS415-*cas*3 plus pYES2-NTA-t-Δ*crtE*, many white colonies occurred on the SD-leu/ura plates (similar to Figure 1A). Furthermore, the band size (Figure 3A) observed after DNA gel electrophoresis of the PCR product of the target gene *crtE* sequence in the white gene-edited mutants obtained with pRS415-*cas*3 plus pYES2-NTA-t-Δ*crtE* was the same (1412 nt) as that in the white gene-edited mutants obtained with pRS415-*cas*7-5-3 plus pYES2-NTA-t/g-Δ*crtE* and by pRS415-*cas*3 plus pYES2-NTA-t/g-Δ*crtE* (Figures 1B and 3A). Subsequently, we repeatedly performed this experiment and further tested the results through DNA electrophoresis and DNA sequencing. Importantly, the band size in the DNA gel electrophoresis (Figure 3B; primer: *crtE*-UF/DR) and the results of DNA sequencing were consistent with the expected DNA fragment size (825 nt) and the t-DNA sequence (Figure 3C, Table S2), indicating that the *Svi*Cas3 enzyme can also be programmed for DNA-guided genome editing.

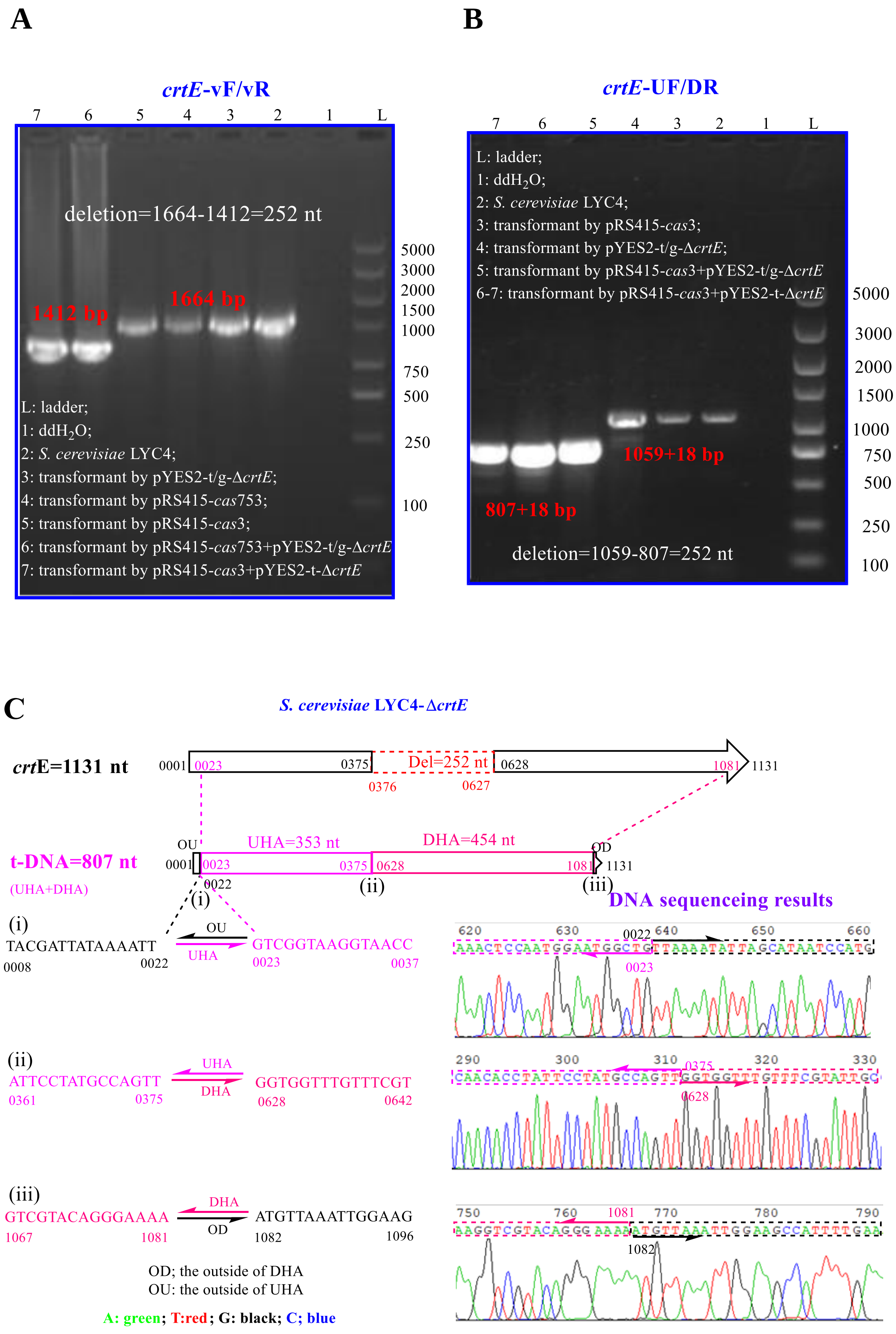
A
crtE-vF/vR
7 6 5 4 3 2 1 L
deletion=1664-1412=252 nt
1412 bp
1664 bp
5000
3000
2000
1500
1000
750
500
250
100
L: ladder;
1: ddH2O;
2: S. cerevisiae LYC4;
3: transformant by pYES2-t/g-ΔcrtE;
4: transformant by pRS415-cas753;
5: transformant by pRS415-cas3;
6: transformant by pRS415-cas753+pYES2-t/g-ΔcrtE
7: transformant by pRS415-cas3+pYES2-t-ΔcrtE
B
crtE-UF/DR
7 6 5 4 3 2 1 L
L: ladder;
1: ddH2O;
2: S. cerevisiae LYC4;
3: transformant by pRS415-cas3;
4: transformant by pYES2-t/g-ΔcrtE;
5: transformant by pRS415-cas3+pYES2-t/g-ΔcrtE
6-7: transformant by pRS415-cas3+pYES2-t-ΔcrtE
5000
3000
2000
1500
1000
750
500
250
100
1059+18 bp
807+18 bp
deletion=1059-807=252 nt
C
S. cerevisiae LYC4-ΔcrtE
crtE=1131 nt
0001 0023 0375 Del=252 nt 0628 1081 1131
0376 0627
t-DNA=807 nt
(UHA+DHA)
OU UHA=353 nt DHA=454 nt OD
0001 0023 0375 0628 1081 1131
0022
(i) (ii) (iii)
DNA sequenceing results
(i)
TACGATTATAAAATT
0008 0022
OU
UHA
GTCGGTAAGGTAACC
0023 0037
(ii)
ATTCCTATGCCAGTT
0361 0375
UHA
DHA
GGTGGTTTGTTTCGT
0628 0642
(iii)
GTCGTACAGGGAAAA
1067 1081
DHA
OD
ATGTTAAATTGGAAG
1082 1096
OD; the outside of DHA
OU: the outside of UHA
A: green; T:red; G: black; C; blue
620 630 0022 640 650 660
AAACTCCAATGGAATGGCTGTTAAAATATTAGCATAATCCATG
0023
290 300 310 0375 320 330
CAACACCTATTCCTATGCCAGTTGGTGGTTTGTTTCGTATTGC
0628
750 760 1081 770 780 790
AAGGTCGTACAGGGAAAAATGTTAAATTGGAAGCCATTTTGAA
1082

**Figure 3. Mutant verification in DNA-guided genome editing of *S. cerevisiae* LYC4. (A and B)** DNA gel electrophoresis of the PCR products of target sequences in the DNA-guided gene editing of *S. cerevisiae* LYC4 genomic DNA through PCR primers *crtE*-vF/vR (A) and *crtE*-UF / DR (B), respectively. **(C)** The designed t-DNA, as well as the DNA sequencing results of the PCR products of the edited sequences in the gene-edited mutant *S. cerevisiae* LYC4-Δ*crtE* genome. The junction sequences between the outside of UHA and UHA, UHA and DHA, and DHA and the outside of DHA are highlighted in Figure 3C.

We have previously shown that the PAMs ("tac" was used in *Lepr*, "tcc" in *svipam*1 and *CAMKMT*, "tgc" in *sviomt*07, and "ttc" in the other target genes) and the repeat sequences (crRNA: $R_{7\ bp}$-$S_{40\ bp}$-$R_{23\ bp}$ in *CAMKMT* and *Lepr* and $R_{30\ bp}$-$S_{40\ bp}$-$R_{30\ bp}$ in the other genes; Table S2) in RNA-guided genome editing of prokaryotic and eukaryotic cells mediated by the *Svi*Cas3 protein are undemanding. Why is that? The compelling fact that the *Svi*Cas3 protein can perform DNA-guided genome editing in the absence of g-DNA suggests that the *Svi*Cas3 enzyme is also a DNA-guided endonuclease. The formation of a full R-loop is known to recuit RNA-guided Cas3 to bind to the NTS bulge, triggering further genome editing and reducing off-target effects [4,19,20,26]. This conclusion indirectly supports that the DNA-guided *Svi*Cas3 enzyme can efficiently generate site-specific single- or double-stranded DNA breaks at full D-loop sites caused by the loading of t-DNA, thus triggering further template-based genome editing.

**DNA-guided gene editing in mammalian cells**

In the study of genome editing of *Saccharomyces cerevisiae* LYC4 based on the subtype I-B-*Svi* CRISPR-Cas system, we unexpectedly found that g-DNA was unnecessary. That is, crRNA and PAM are also not needed. To further validate this discovery and facilitate comparison, the two cell lines (HEK293T and NIH-3T3) and the three target genes (*DROSHA*, *CAMKMT* and *Lepr*) that were investigated in the RNA-guided genome editing were selected as target cells and target genes in the DNA-guided genome editing of mammalian cells. In the DNA-guided genome editing of the three genes (*DROSHA*: HEK293T/chrV-Homo sapiens, *CAMKMT*: HEK293T/chrII-Homo sapiens and *Lepr*: NIH-3T3/chrIV-Mus musculus), we first designed and constructed three gene editing vectors: (i) AIO-mCherry-*cas*3-t-Δ*DROSHA*::*egfp*, (ii) AIO-mCherry-*cas*3-t-Δ*CAMKMT*::*egfp* and (iii) AIO-mCherry-*cas*3-t-Δ*Lepr*::*egfp* (Figure S4, Tables S1 and S2), according to the same procedures as described in the section "Materials and Methods" but with t-DNA instead of template/guide-DNA (t/g-DNA). The designed fragment sizes of t-DNA (UHA+*egfp*+DHA) and the deletion were also the same as those used in the RNA-guided genome editing of mammalian cells (Figures 2C and 4B, Table S2). After transfecting the three vectors AIO-mCherry-*cas*3-t-Δ*DROSHA::egfp* (HEK293T), AIO-mCherry-*cas*3-t-Δ*CAMKMT::egfp* (HEK293T) and AIO-mCherry-*cas*3-t-Δ*Lepr*::*egfp* (NIH-3T3) into corresponding cells, we obtained potential gene-edited cells with green fluorescence (similar to Figure 2A), namely, HEK293T-Δ*DROSHA::egfp*, HEK293T-Δ*CAMKMT*::*egfp* and NIH-3T3-Δ*Lepr*::*egfp*, and validated them through DNA gel electrophoresis and DNA sequencing analysis of the PCR products of edited sequences, following the same procedures as the RNA-guided genome editing in mammalian cells described above.

A

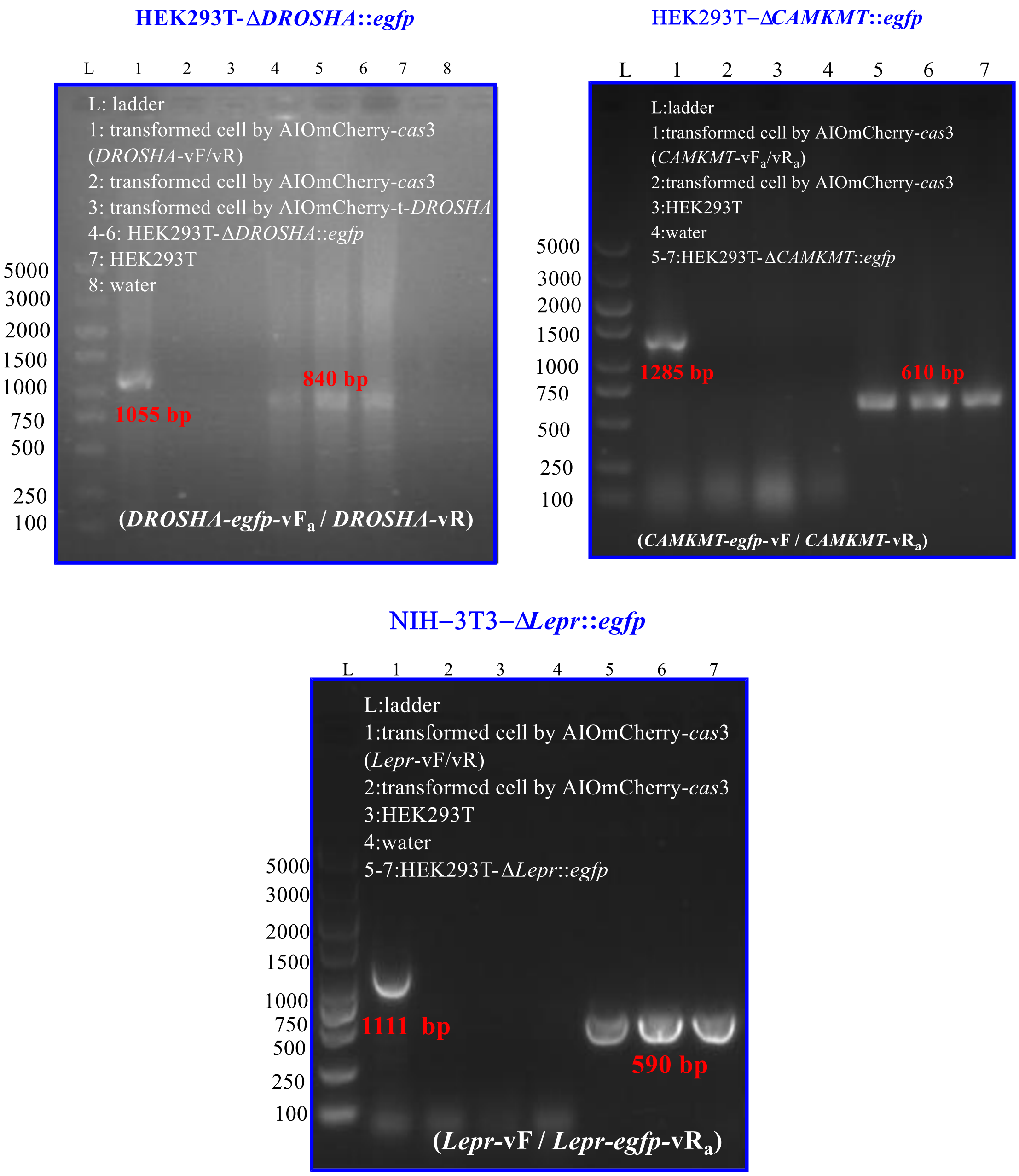

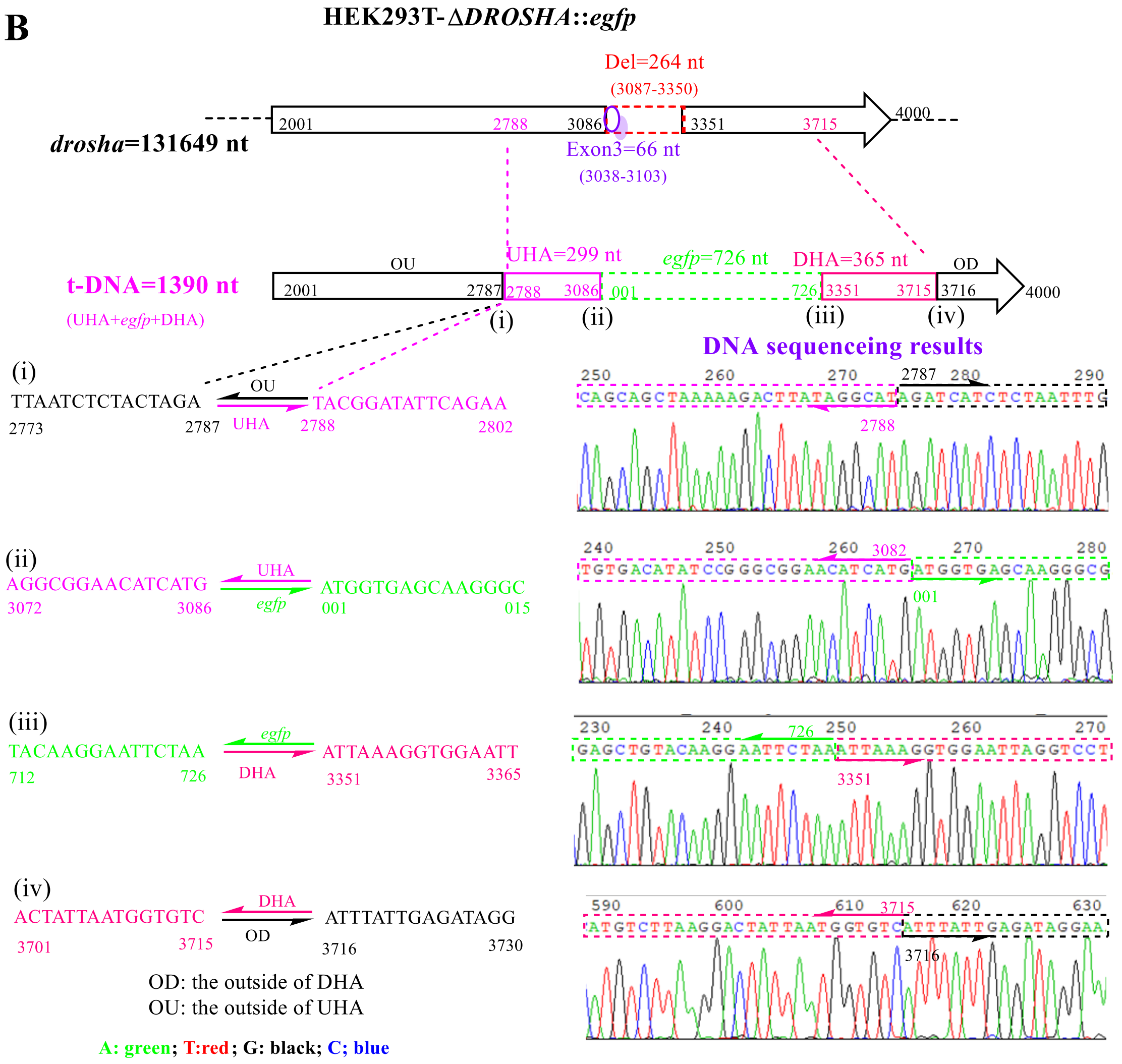

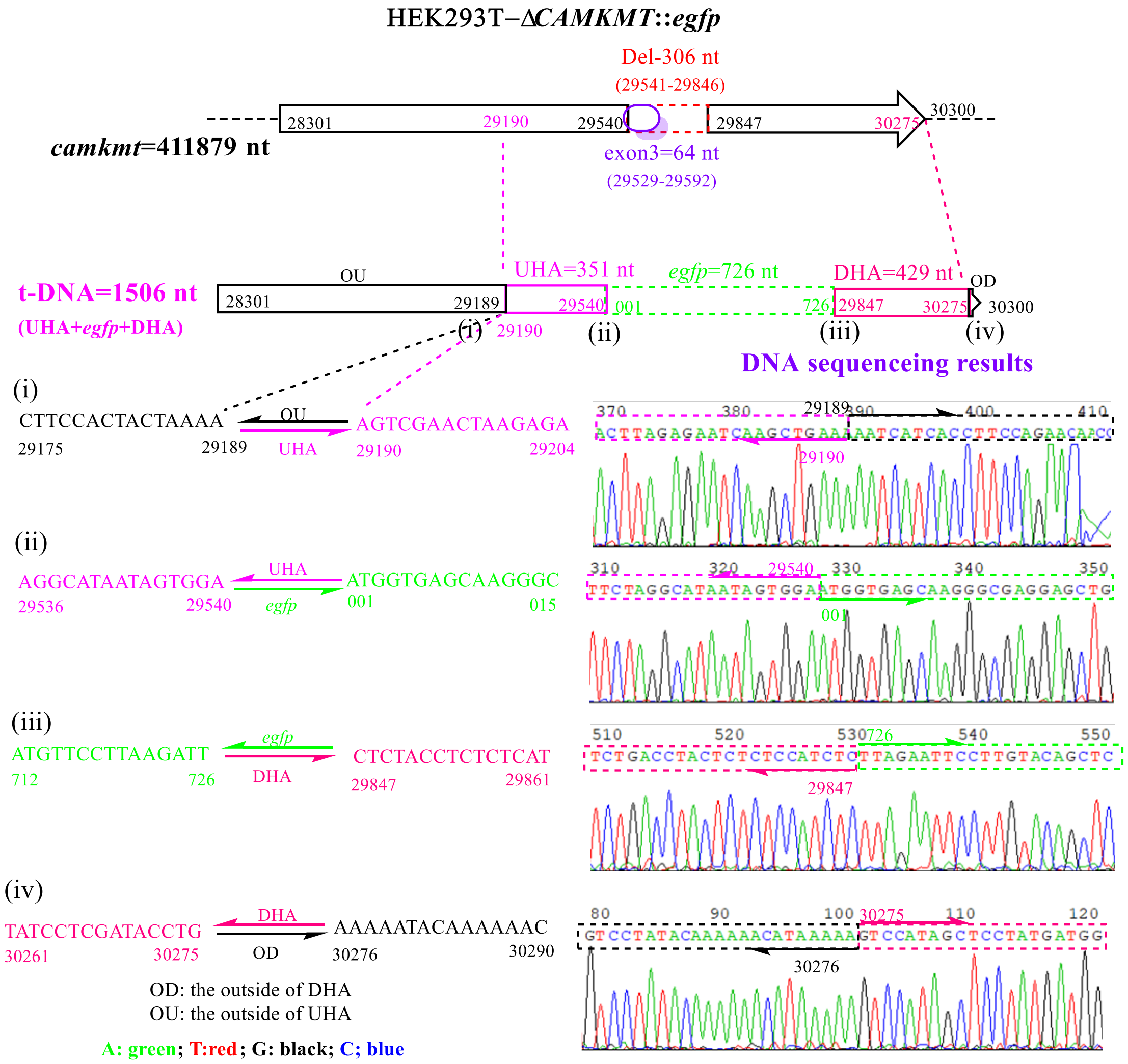

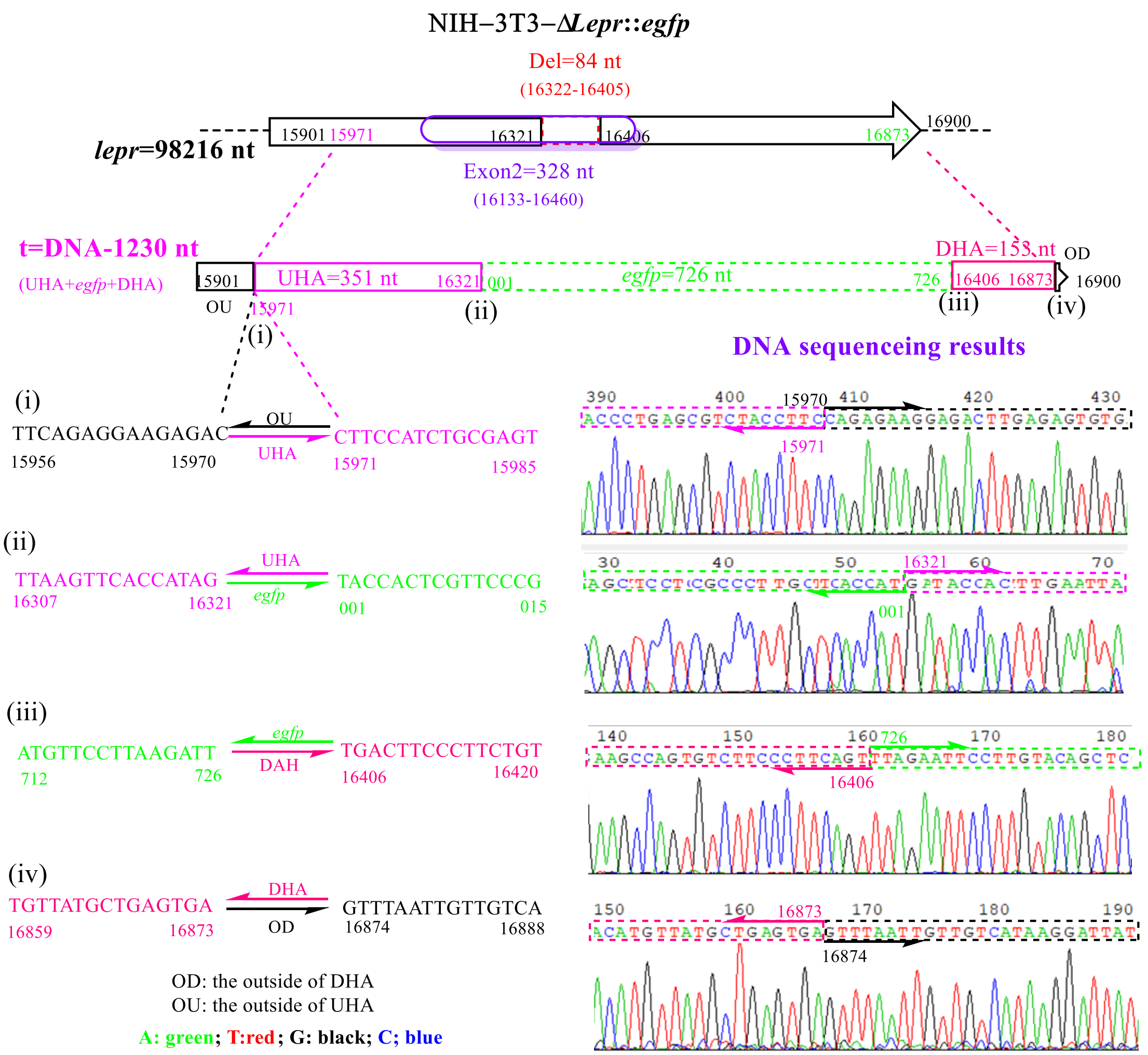

**Figure 4. Mutant verification in the DNA-guided genome editing of HEK293T and NIH-3T3 cells. (A)** DNA gel electrophoresis of the PCR products of target sequences in the DNA-guided genome editing of mammalian cells. **(B)** The designed t-DNAs, and the DNA sequencing result of the PCR products of edited sequences in the cell genome of the gene-edited variants HEK293T-Δ*DROSHA::egfp*, HEK293T-Δ*CAMKMT*::*egfp* and NIH-3T3-Δ*Lepr*::*egfp*. The junction sequences between the outside of UHA and UHA, UHA and *egfp*, *egfp* and DHA, and DHA and the outside of DHA are highlighted in Figure 4B.

As we expected, the results of all the analytical experiments, including fluorescence detection, DNA gel electrophoresis of the PCR products, and targeted sequencing, were correct in the DNA-guided genome editing (Figure 4, Table S2). These results suggest that in the *Svi*Cas3-based genome editing in the presence of t/g-DNA, D-loops and R-loops may exist simultaneously in different transformed cells, thus affecting the GEE. Given that the DNA-guided genome editing (without g-DNA, that is, without PAM) conducted by *Svi*Cas3 can be precisely achieved in the eukaryotic microorganism (*S. cerevisiae* LYC4) and mammalian cells (HEK293T and NIH-3T3 cell lines), we believe that the discovery of this function of *Svi*Cas3 will lead to the evolution of genome editing from protein-guided and RNA-guided to DNA-guided.

**DNA-guided, HDR-directed base and gene editing in HEK293T cells**

Approximately two-thirds of human genetic diseases are caused by point mutations (single nucleotide mutations, that is, single nucleotide polymorphisims (SNPs)) [27]. Base editors can directly install and correct single nucleotide mutations of interest in genomic DNA and have been successfully applied to a variety of systems, including prokaryotes, plants, animals, and human embryos [27-31]. Since in *Svi*Cas3-based, DNA-guided genome editing, the formation of D-loops and R-loops enables *Svi*Cas3 to bind to the bulge of the NTS, resulting in nicks or cleavage of the NTS, we inferred that it is possible for *Svi*Cas3 to be developed as a HDR-directed base editing tool (In fact, nucleotide substitution) for cellular genomic DNA. Accordingly, we selected four bases on exon 2 (with 173 nt) of the gene *CAMKMT* as target bases for base editing. Furthermore, we designed and constructed a simultaneous base editing and gene editing (gene editing used for control and monitoring) vector AIO-mCherry-*cas*3-$t_{g1}$/$t_{b2}$-Δ*CAMKMT*::*egfp* (Figure S5). The designed and constructed template donors of $t_b$-DNA and $t_g$-DNA (used for base editng of exon 2 and gene editng of exon 1 in the CAMKMT gene, respectively) are shown in Figure 5B. After transfecting HEK293T cells with the vector AIO-mCherry-*cas*3-$t_{g1}$/$t_{b2}$-Δ*CAMKMT*::*egfp*, we obtained potential base-edited and gene-edited cells with green fluorescence (similar to Figure 2A): HEK293T-Δ*CAMKMT*::*egfp*-g1/b2. We further detected and analysed them through DNA gel electrophoresis and DNA sequencing of the PCR products of the edited bases and the edited sequences, following the same procedures described above.

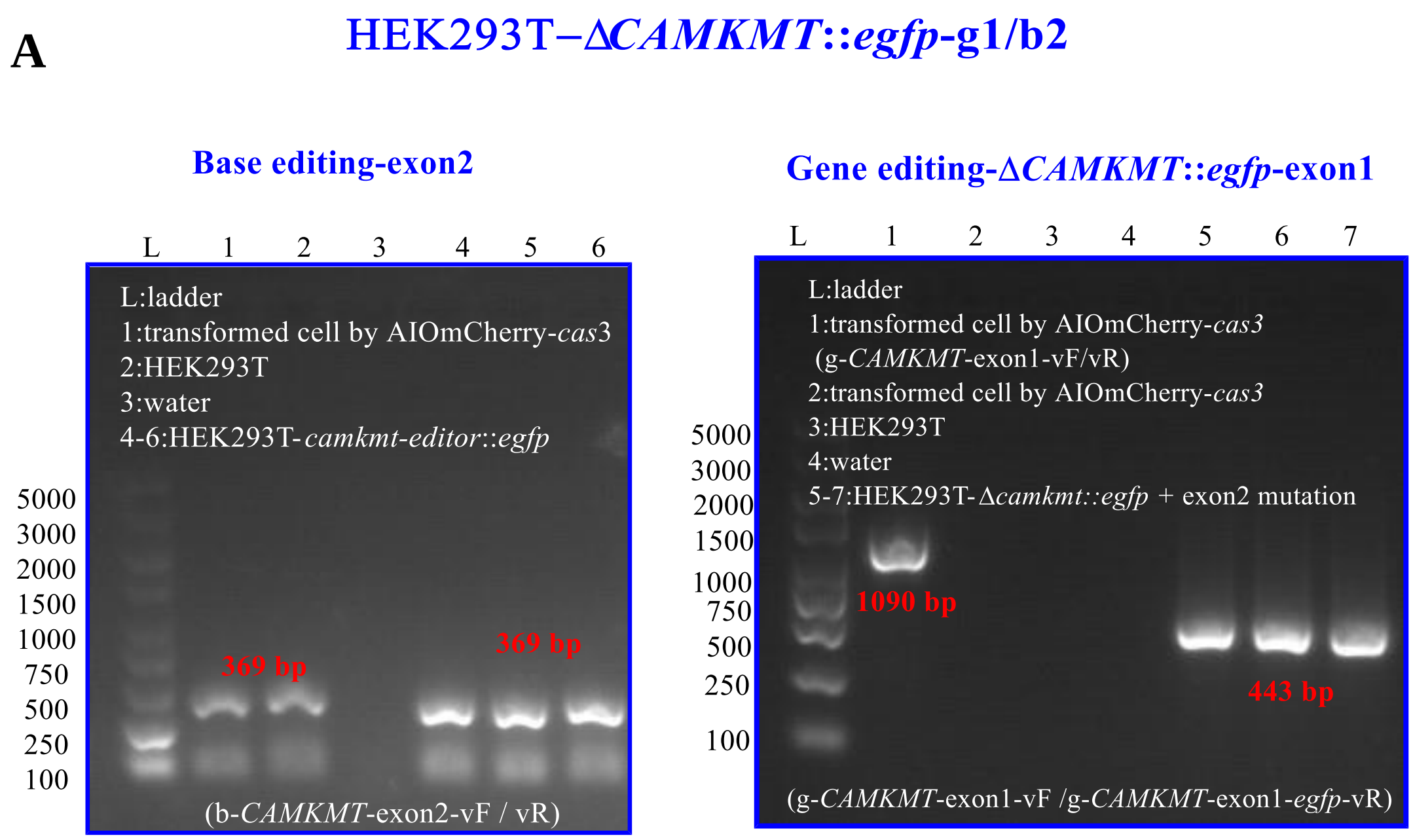

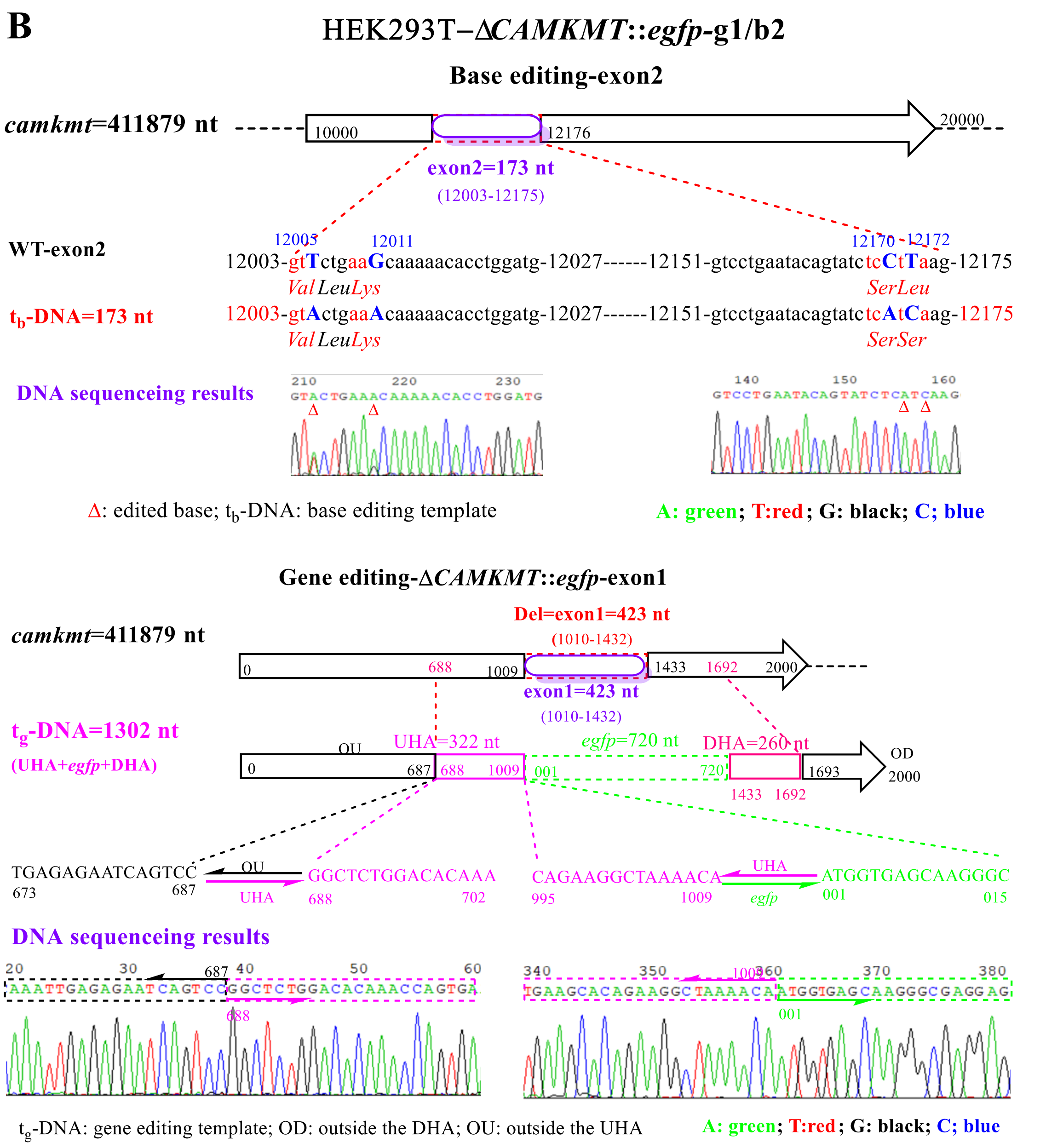

**Figure 5. Mutant verification in the DNA-guided, HDR-directed base and gene editing of HEK293T cells. (A)** The DNA gel electrophoretogram of the PCR products of target sequences in base editing and gene editing in HEK293T cell genomes conducted by DNA-guided *Svi*Cas3. **(B)** The designed and constructed $t_b$-DNA (base editing template for exon 2) and $t_g$-DNA (gene editing template for exon1), as well as the DNA sequencing results

of the PCR products of both the edited bases and the edited sequences in the genome of the variant HEK293T-Δ*CAMKMT*::*egfp*-g1/b2 (The four edited bases: $T_{12005}$→$A_{12005}$/transversion, $G_{12011}$→$A_{12011}$/transition, $C_{12170}$→$A_{12170}$/transversion and $T_{12172}$→$C_{12172}$/transition, are highlighted in Figure 5B).

The results of all analytical experiments (fluorescence dectection, DNA gel electrophoresis and sequencing) were consistent with the designed template donor in DNA-guided base editing and gene edting (Figure 5 and Table S2), demonstrating that the DNA-guided *Svi*Cas3 can simultaneously perform both base editing and gene editing at two different loci over 10000 nt (exon 1: 1010 to 1432, exon 2: 12003 to 12175). Notably, as seen in Figure 5B, there are reasonable, obvious ($T_{12005}$→$A_{12005}$ and $G_{12011}$→$A_{12011}$) and ambiguous ($C_{12170}$→$A_{12170}$ and $T_{12172}$→$C_{12172}$) double peaks of base overlap in the DNA sequencing map, indicating that the base editing conducted by the single *Svi*Cas3 in mammalian cells is simple and effective, but in each batch of base editing experiments, the editing efficiency of each base is different. Given that the DNA-guided, *Svi*Cas3-based, HDR-directed base editing does not require the involvement of cytidine deaminase, adenine deaminase and reverse transcriptase [27-34], it is reasonable to infer that all transition and transversion mutations between A, T, G and C in the genomic DNA of mammalian cells can be achieved through the *Svi*Cas3-based base editor.

**Discussion**

Genome editing tools based on *Sp*Cas9 of *Streptococcus pyogene* [33,35-38] and base editors based on nucleotide deamination (cytosine base editor (CBE) and adenine base editor (ABE)) of ssDNA loops created by RNA-guided CRISPR proteins and reverse transcriptase (prime editing (PE)) [27,29-31,33,34] are currently the most extensively used gene and base editing tools. Gene editing based on Cas3 from type I CRISPR-Cas systems has also been widely reported [26,39-43]. However, all experimental results show that template-based genome editing on the basis of type II CRISPR-Cas systems is an RNA-guided mechanism involving tracrRNA and hydrolysis by Cas9 alone [17,44-46], and template-based genome editing on the basis of type I CRISPR-Cas systems is an RNA-guided mechanism implemented by the Cas3-Cascade complex [4,19,20,26,42,47]. In short, there have been no reports of gene editing conducted by Cas3 alone, let alone successful DNA-guided gene editing and base editng.

Our experimental results showed that the codon optimized gene *Svicas*3 could be developed as RNA-guided genome editing tools through the HDR mechanisms of eukaryotic host cells in the absence of Cascade, with relatively high genome editing efficiencies (GEE) of $4.4\times10^{-2}$~$7.3\times10^{-2}$ (Table S3) ($8.6\times10^{-7}$~$9.0\times10^{-5}$ in prokaryotic genome editing, see the sister article: Prokaryotic genome editing based on the subtype I-B-*Svi* CRISPR-Cas system). Particularly, we found the *Svi*Cas3 enzyme guided by DNA could be developed as HDR-directed gene and base editing tools for eukaryotic genomic editing, with the GEE (5.5-7.8%) similar to that (4.4-7.3%) of RNA-guided genome editing (Table S3). Notably, in contrast to the RNA-guided genome editing conducted by *Svi*Cas3, the design and preparation of g-DNA was not required in the DNA-guided genome editing conducted by *Svi*Cas3. Furthermore, these distinctive features in the RNA-guided genome editing mediated by *Svi*Cas3, including flexible PAMs, no off-target cases and indel formation detected in the analysis of potential off-target sites, can be interpreted to show that *Svi*Cas3 is an RNA- and DNA-guided endonuclease that identifies R-loops and D-loops without the help of Cascade. In short, the formation of R-loop and D-loop enables the *Svi*Cas3 to bind to the bulge of NTS, resulting in nicks or cleavage of the NTS, thus enabling gene and base editing.

*Svi*Cas3-based genome editing tools based on DNA-guided mechanisms could be easily, reliably and effectively applied for the genome editing of prokaryotic (see the sister article: Prokaryotic genome editing based on the subtype I-B-*Svi* CRISPR-Cas system) and eukaryotic cells (Figures 1-4), so we envision that the developed *Svi*Cas3-based, HDR-directed gene and base editing tools based on DNA-guided principles will soon be applied to all fields related to basic and applied biology, including clinical medicine and biopharmaceuticals, industry, agriculture, environmental modification, and more. Obviously, the molecular mechanism of *Svi*Cas3-based genome editing remains to be studied. A variety of *Svi*Cas3-based tools, including chromosome imaging, gene expression control, genome-wide functional screening, etc., also remain to be developed.

**Materials and Methods**

**Strains, plasmids and gene editing elements**

*Saccharomyces cerevisiae* LYC4, human embryonic kidney 293 / HEK293T cells and NIH-3T3 mouse embryonic fibroblast cells / NIH-3T3 were used as hosts of interest for non-self-targeting genome editing and *E. coli* DH5α and *E. coli* TOP10 were used as hosts for plasmid amplification and storage.

Plasmid pYES2-NTA (used in genome editing of *S. cerevisiae* LYC4) harboring a gene *ura*3 encoding orotidine-5’-phosphate decarboxylase required for uracil biosynthesis

(http://www.addgene.org/) was selected as vectors for construction of gene editing plasmids (named plasmid-t/g- or t-*target gene abbreviation*) that carry an engineered g-DNA for crRNA transcription and an engineered t-DNA for editing template in the genome editing of *S. cerevisiae* LYC4.

Commercialized plasmid pRS415 (rooted from pRS415_pGal-nCas9, Addgene, Beijing Zhongyuan, Ltd, Beijing, China ) bearing the codon optimized *cas*9 and *leu*2 encoding 3-isopropylmalate dehydrogenase required for leucine biosynthesis (http://www.addgene.org/) was used for construction of Cas expression plasmids carrying a *cas*(es) encoding a Cas protein(s) in *S. cerevisiae* LYC4.

One All-in-One expression vector (containing three required elements for genome editing mediated by CRISPR-Cas system: crRNA generation, t-DNA duplication and Cas expression), AIO-mCherry (encoding dual U6 promoter-driven sgRNAs and mCherry-coupled Cas9-D10A nickase to enhance efficient and accurate genome editing), was used for the construction of gene editing tools in genome editing of HEK293T and NIH-3T3 cells. All strains and the plasmids used in this study are listed in Table S1.

All the details of the primers, g-DNAs and t-DNAs used in this study are summarized in Table S2.

## Solutions, media and chemicals

An LDTE buffer [1.02 g LiAc•H2O, 0.12 g Tris / Tris(hydroxymethyl)aminomethane and 0.029 g EDTA / ethylenediaminetetraacetic acid added in 80 ml $ddH_2O$, pH7.5, 115$^o$C, 30 min, supplemented with 20 ml filter-sterilized solution of 0.771% (W/V) DL-Dithiothreitol (DTT)] was prepared for the preparation of *S. cerevisiae* LYC4 competent cells. Buffer 1SM for electroporation of NIH-3T3 cells was prepared as follows: (i) respectively preparing 200 mM KCl, 150 mM $MgCl_2$ and 600 mM $NaH_2PO_4.2H_2O$ solutions [0.746 g KCl, 0.714 g $MgCl_2$ and 4.68 g $NaH_2PO_4.2H_2O$, each dissolved in 50 mL $ddH_2O$, pH 7.2], and (ii) preparing buffer 1SM containing sodium succinate, mannitol, KCl, $MgCl_2$ and $NaH_2PO_4$ [0.203 g sodium succinate, 0.227 g mannitol, 1.25 ml 200 mM KCl, 5 mL 150 mM $MgCl_2$, 10 mL 600 mM $NaH_2PO_4.2H_2O$, supplemented with $ddH_2O$ to 50 ml, filter-sterilization]. All solutions prepared above were stored at 4 $^oC$ in a refrigerator.

Yeast extract peptone dextrose medium (YPD) (1 liter: 10 g yeast extract, 20 g peptone and 20 g glucose added in $ddH_2O$, 115$^oC$, 30 min; 20 g agar supplemented for solid medium) was prepared for the preparation of *S. cerevisiae* LYC4 competent cells. Synthetic dropout medium-leucine (SD-leu) [1 liter: 1.7 g YNB / yeast nitrogen base medium, 5 g $(NH_4)_2SO_4$ and 20 g glucose added in $ddH_2O$, 115$^oC$, 30 min and then 4 ml filter-sterilized mixture of 0.5% (W/V) methionine, 0.5% (W/V) histidine and 0.5% (W/V) uracil added; 20 g agar and 182 g sorbitol supplemented for solid medium**], i**nduced synthetic dropout medium-leucine (ISD-leu) [1 liter: 1.7 g YNB, 5 g $(NH_4)_2SO_4$, 20 g galactose and 10 g raffinose added in $ddH_2O$, 115$^oC$, 30 min and then 4 ml filter-sterilized mixture of 0.5% (W/V) methionine, 0.5% (W/V) histidine and 0.5% (W/V) uracil added], synthetic dropout medium-uracil (SD-ura) [1 liter: 1.7 g YNB, 5 g $(NH_4)_2SO_4$ and 20 g glucose added in $ddH_2O$, 115$^oC$, 30 min and then 4 ml filter-sterilized mixture of 1% (W/V) leucine, 0.5% (W/V) methionine and 0.5% (W/V) histidine added; 20 g agar and 182 g sorbitol supplemented for solid medium] and synthetic dropout medium-leucine/uracil (SD-leu/ura) [1 liter: 1.7 g YNB, 5 g $(NH_4)_2SO_4$ and 20 g glucose added in $ddH_2O$, 115$^oC$, 30 min and then 4 ml filter-sterilized mixture of 0.5% (W/V) methionine and 0.5% (W/V) histidine added; 20 g agar and 182 g sorbitol supplemented for solid medium] were prepared for transformation of Cas expression plasmid (named plasmid-*cas gene abbreviation*) pRS415-*cas*3, preparation of *S. cerevisiae* LYC4-pRS415-*cas*3 competent cells, transformation of plasmid pYES2-NTA-t/g-Δ*crtE* and selection of *S. cerevisiae* LYC4-Δ*crtE*, respectively.

Dulbecco's modified eagle medium (DMEM) (cat: #ATCC® 30-2002™, Manassas, VA, USA) was purchased from BioVector NTCC Inc, Beijing, China and 10% serum cell culture medium (SCCM) [1 liter: 100 ml fetal bovine serum, 100 mg penicillin, 100 mg streptomycin, 10 g glutamine added in DMEM] was prepared for the cell cultivation of HEK293T and NIH-3T3.

All reagents used in this study were of biochemical grade, analytical grade or higher, and most of them were purchased from Sangon Biotech Co., Shanghai, China.

**Construction of gene editing tools for genome editing in *S. cerevisiae* LYC4**

In the construction of Cas expression plasmid pRS415-*cas*3 for genome editing of *S. cerevisiae* LYC4, the codon optimized gene *cas*3 (43% GC) was directly chemically synthesized by General Biosystems, Inc., Chuzhou, Anhui province, China. The linear skeleton of plasmid pRS415 without gene *cas*9 was prepared by a basic PCR (template: pRS415 DNA, primers: pRS415-*cas*3-F / pRS415-*cas*3-R, reaction conditions: pre-denaturation at 95 $^{o}$C for 5 min, 30 cycles of denaturation at 95 $^{o}$C for 30 s, annealing at 62 $^{o}$C for 30 s and extension at 72 $^{o}$C 2 min, and a final extension at 72 $^{o}$C for 10 min, TransStart FastPfu DNA Polymerase, TransGen Biotech, Beijing, China) and then purified and collected by a DNA electrophoresis in 1.5 % agarose gels and an AxyPrep Gel DNA Isolation Kit (Corning Incorporated, Shanghai, China). Further, the purified linear skeleton of pRS415 was ligated with the codon optimized *cas*3 using a ClonExpress® Entry One Step Cloning Kit (Sangon Biotech Co., Shanghai, China) according to the instruction provided by the supplier (2 μl 5×CE MultiS Buffer, 1 μl Exnase MultiS, 6 μl insert *cas*3 and 1 μl skeleton of pRS415, 37 $^{o}$C for 30 min and on ice for 5 min). The ligated product pRS415-*cas*3 was finally transformed, tested (pRS415-*cas*3-Fsc / pRS415-*cas*3-Fsc) and preserved in *E.coli* DH5α (Figures 1B and 1C, Tables S1 and S2). In the construction of pRS415-*cas*7-5-3, all the construction procedures excepting codon optimized DNA fragments of *cas*5 and *cas*7) and corresponding primers were the same as described in the construction steps of pRS415-*cas*3. To be more precise, the chemically synthesized codon optimized *cas*5 and *cas*7 fragments (General Biosystems, Inc., Chuzhou, Anhui province, China) are respectively composed of ENO2 promoter, *cas*5, SV40NLS and ENO2 terminator, and CDC19 promoter, *cas*7, SV40NLS and CDC19 terminator, in proper order.

The construction procedures of gene editing plasmids pYES2-NTA-t/g-Δ*crtE* and pYES2-NTA-t-Δ*crtE* in genome editing of *S. cerevisiae* LYC4 are the same as those described in the sister article (Prokaryotic genome editing based on the subtype I-B-*Svi* CRISPR-Cas system) excepting plasmid, promoter, spacer, terminator, template and primer (Figures 1A, 1C and 3, Tables S1 and S2).

## Construction of gene editing tools for genome editing in mammalian cells

In the construction of gene editing vector AIO-mCherry-*cas*3-t/g-Δ*DROSHA*::*egfp* for genome editing of HEK293T cells, the codon optimized *cas*3 fragment with 65% GC was chemically synthesized by General Biosystems, Inc., Chuzhou, Anhui province, China. After amplification through *E. coli* TOP10 cultivation, extracted by an AxyPrep Easy-96 Plasmid Kit (Axygen Biosciences, Hangzhou, China), digestion by *Xba*I and *Hind*III and purification, the purified linear skeleton of vector AIO-mCherry-#74120 (*37*) and the purified *cas*3 fragment were ligated using T4 ligase under the conditions of 16 $^{o}$C for 12 h and further 65 $^{o}$C for 10 min (inactivation). After DNA sequencing, the resultant product AIO-mCherry-*cas*3 was transformed and preserved in *E. coli* TOP10. The further construction of the gene editing vector AIO-mCherry-*cas3*-t/g-Δ*DROSHA*::*egfp* followed the same procedures as described in construction of gene editing plasmid pYES2-NTA-t/g-Δ*crtE* above excepting the plasmid (AIO-mCherry-*cas*3), spacer, template and primer (Figure S2, Tables S1 and S2).

Excepting target genes, corresponding g-DNAs, t-DNAs and primers, the construction procedures of the six gene editing vectors AIO-mCherry-*cas*3-t/g-Δ*CAMKMT*::*egfp*, AIO-mCherry-*cas*3-t/g-Δ*Lepr*::*egfp*, AIO-mCherry-*cas3*-t-Δ*DROSHA*::*egfp* , AIO-mCherry-*cas*3-t-Δ*CAMKMT*::*egfp*, AIO-mCherry-*cas*3-t-Δ*Lepr*::*egfp* and AIO-mCherry-*cas*3-$t_{g1}$/$t_{b2}$-Δ*CAMKMT*::*egfp* for mammalian cell genome editing are the same as those described above in the construction of gene editing vector AIO-mCherry-*cas*3-t/g-Δ*DROSHA*::*egfp* (Figures 2 and 4, Tables S1 and S2).

**Construction of gene-edited mutants of *S. cerevisiae* LYC4**

The construction steps of gene-edited mutants of *S. cerevisiae* LYC4 were as follows: (i) Preparation of *S. cerevisiae* LYC4 competent cells: A single colony of *S. cerevisiae* LYC4 from a YPD plate was inoculated into a flask containing 50 ml YPD and incubated in a rotory shaker at 200 rpm at 30°C for 12-14 h ($OD_{600}$: 0.9-1.2). The culture was totally transferred into a pre-cooled centrifugal tube and then centrifuged at 5000 rpm for 5 min at 4°C. The pellet was washed twice respectively by 25 ml pre-cooled dd$H_2O$ and 25 ml pre-cooled 1 M sorbitol (18.217 g sorbitol in 100 ml dd$H_2O$, 115°C, 30 min) through a combination of re-suspension and centrifugation. The washed pellet was re-suspended with 25 ml LDTE buffer and incubated in a rotory shaker at 200 rpm at 30°C for 30 min. The incubated cells were collected by centrifugation and transferred into a 1.5 ml-eppendorf tube for centrifugal washing twice (1 ml dd$H_2O$ and 1 ml 1 M sorbitol, respectively). After the final washed pellet was re-suspended with 200 µl 1 M sorbitol, about 200 µl of *S. cerevisiae* LYC4 competent cells was obtained and stored in a refrigerator at -80°C. (ii) Transformation of Cas expression plasmid pRS415-*cas*3 (or pRS415-*cas*7-5-3): In the transformation process, 5 µl pRS415-*cas*3 (or pRS415-*cas*7-5-3) and 50 µl of *S. cerevisiae* LYC4 competent cells were transferred into a pre-cooled electroporation cuvette. After cooling on ice for 5 min, the competent cell mixture was treated by electro-transformation at 1.5 kV for 5 ms and then transferred to a pre-cooled eppendorf tube containing 945 µl 1 M sorbitol. The transformed cell mixture was spread evenly to three SD-leu plates (100 µl for each) and the plates were placed into an incubator to incubate at 30°C for 3 days. Three red single colonies as potential transformants of *S. cerevisiae* LYC4-pRS415-*cas*3 (or *S. cerevisiae* LYC4-pRS415-*cas*7-5-3) were randomly selected and the pRS415-*cas*3 plasmids (or pRS415-*cas*7-5-3) from the three red colonies were respectively extracted and then verified by the basic PCR described above (primers: pRS415-*cas*3-Fsc / pRS415-*cas*3-Rsc or pRS415-*cas*7-5-Fsc / pRS415-*cas*7-5-Rsc). Correct transformants of *S. cerevisiae* LYC4-pRS415-*cas*3 and *S. cerevisiae* LYC4-pRS415-*cas*7-5-3 were respectively stored in a refrigerator at -20°C. (iii) Transformation of gene editing plasmid pYES2-NTA-t/g-Δ*crtE* (or pYES2-NTA-t-Δ*crtE*): The preparation of *S. cerevisiae* LYC4-pRS415-*cas*3 competent cells or *S. cerevisiae* LYC4-pRS415-*cas*7-5-3 competent cells followed the same procedures as described in Preparation of *S. cerevisiae* LYC4 competent cells, but using ISD-leu instead of YPD. Also the transformation of plasmid pYES2-NTA-t/g-Δ*crtE* (or pYES2-NTA-t-Δ*crtE*) followed the same procedures as described in Transformation of Cas expression plasmid pRS415-*cas*3 but using SD-leu/ura instead of SD-leu. Three white single colonies as potential Δ*crtE*-edited mutants propagated on the SD-leu/ura plates were randomly selected and the genomes of randomly selected three white colonies were extracted and verified through DNA gel electrophoresis analysis and DNA sequencing analysis to the PCR products of edited sequences (primers: *crtE*-vF / vR and *crtE*-UF / DR ).

**Construction of gene-edited variants of HEK293T and NIH-3T3**

In the construction of gene-edited mutant HEK293T-Δ*DROSHA*::*egfp*, the supernatant of 5 ml HEK293T cultures (~$1.0\times10^6$ /ml) as seed in a 25 ml cell culture flask was removed by using a pipette and 5 ml PBS was used to wash the HEK293T cells. After 1 ml Trypsin-EDTA solution (0.25%) was added into the flask, the flask was placed into a $CO_2$ incubator and incubated at 37°C for about 3 min until the cells disperse totally.

Subsequently, 4 ml SCCM was added into the flask and the mix in this flask was transferred to a 15 ml centrifuge tube and centrifuged at 1000 rpm at room temperature for 3 min. The collected cells were suspended with 1 ml SCCM and transferred to a new 25 ml cell culture flask with 4 ml SCCM. Finally, the cell mix were placed in a $CO_2$ incubator again and incubated at 37°C for about 48 h (until cell numbers reach over $1.0\times10^6$ /ml).

About $5.0\times10^5$ HEK293T cells from the HEK293T culture were transferred into a well of a six-well cell culture plate and SCCM was added into the well (a total volume of 2 ml). After the plate was incubated at 37°C for about 8-12 h in a $CO_2$ incubator (until the cell confluence was up to 60-80%), a 500 µl Lip2000-Vector mixture (containing a 250 µl Lip2000-DMEM mix: 10 µl Lip2000 Transfection Reagent plus 240 µl DMEM, and a 250 µl vector-DMEM mix: 10 µl / 4 µg AIO-mCherry-*cas3*-t/g-Δ*DROSHA*::*egfp* or AIO-mCherry-*cas3*-t-Δ*DROSHA*::*egfp* plus about 240 µl DMEM) pre-incubated at 37°C for about 20 min was added into each well. After incubated at 37°C for about 8 h in a $CO_2$ incubator, about 2 ml of the consumed SCCM in each well of the plate was replaced by the same volume of fresh SCCM. The plate was further incubated at 37°C for about three days in the incubator and during the period the consumed SCCM in each well was replaced by the same volume of fresh SCCM once a day. Finally, about 2 ml of the transfected cells in each well was totally transferred into a 25 ml cell culture flask and 3 ml SCCM was supplemented. After the flask was incubated for about 2-4 days under the same conditions (until the number of cells reach about $1.0\times10^7$ /ml, mCherry and EGFP were respectively detected by Leica Microsystems CMS Gmbh-DMi8, Germany), the enomes of potential gene-edited cells of HEK293T-Δ*DROSHA*::*egfp* in the flasks were extracted and verified through DNA gel electrophoresis and DNA sequencing analysis to the PCR products of edited sequences (primers: *DROSHA-egfp*-vF / *DROSHA*-vR and *DROSHA-egfp*-$vF_a$ / *DROSHA*-vR). The construction procedures of the gene-edited variants HEK293T-Δ*CAMKMT*::*egfp* and HEK293T-Δ*CAMKMT*::*egfp*-g1/b2 are the same as those described in construction of HEK293T-Δ*DROSHA*::*egfp* above, excepting the corresponding gene editing vector (AIO-mCherry-*cas3*-t/g-Δ*CAMKMT*::*egfp* or AIO-mCherry-*cas3*-t-Δ*CAMKMT*::*egfp* or AIO-mCherry-*cas*3-$t_{g1}/t_{b2}$-Δ*CAMKMT*::*egfp*) and the corresponding primers (*CAMKMT-egfp*-vF / *CAMKMT*-vR or *CAMKMT-egfp*-$vF_a$ / *CAMKMT*-vR or g-*CAMKMT*-exon1-vF / g-*CAMKMT*-exon1-*egfp*-vR and b-*CAMKMT*-exon2-vF / vR) (Tables S1 and S2).

The construction process of the gene-edited mutant NIH-3T3-Δ*Lepr*::*egfp* was similar to that of the gene-edited mutant HEK293T-Δ*DROSHA*::*egfp* described above, except for the transformation method of gene editing vectors. In the transformation process, about $2.5\times10^5$ NIH-3T3 cells, 100 μl buffer 1SM and 6 μg gene editing vector were transferred into a pre-cooled electroporation cuvette. After mixed well, the vector was delivered into NIH-3T3 cells by a LONZA Nucleofector™ 2b (Made in Germany) according to the program U-030. The gene editing vectors were AIO-mCherry-*cas3*-t/g-Δ*Lepr*::*egfp* and AIO-mCherry-*cas3*-t-Δ*Lepr*::*egfp*, respectively and the verifying primers were *Lepr*-vF / *Lepr-egfp*-vR and *Lepr*-vF / *Lepr-egfp*-$vR_a$ , rescettively (Tables S1 and S2).

**Off-target analysis, DNA electrophoresis and DNA sequencing analysis**

Similar to those described in the sister article (Prokaryotic genome editing based on the subtype I-B-*Svi* CRISPR-Cas system)

**Data reporting**

No statistical methods were used to predetermine sample size. The experiments were not randomized and the investigators were not blinded to allocation during experiments and outcome assessment.

## Acknowledgments

We are grateful to Zhi-Nan Xu, Wei Liu and Chang Dong, College of Chemical and Biological Engineering, Zhejiang University, Hangzhou, China, for generous gift of some plasmids and strains.

## Additional information

### Competing interests

A patent application (CN107557373A / WO2019056848A1 / EP3556860A1 / US11286506 B2) has been filed for the content disclosed in this study.

### Funding

The work was not funded by any agency.

### Author contributions

Wang-Yu Tong contributed to the research design, the result interpretation and the paper writing. Yong Li, Shou-Dong Ye, An-Jing Wang, Yan-Yan Tang, Mei-Li Li, Zhong-Fan Yu, Ting-Ting Xia, Qing-Yang Liu and Si-Qi Zhu carried out the experiments and participated in the result interpretation.

## Additional files

**Supplementary Information for Template-based eukaryotic genome editing directed by the *Svi*Cas3**

Table of contents

**Data availability**

Almost all relevant data for this study are included in the article and the supplementary Information. The target reference sequences of the genomic DNA of the three cells (*Saccharomyces cerevisiae* S288C; Homo sapiens (human)-HEK293T; Mus musculus (house mouse)-NIH-3T3) in the off-target analysis are accessible to the NCBI database (https://www.ncbi.nlm.nih.gov/genome/).

# Table of contents

# Supplementary Information for

# Template-based eukaryotic genome editing directed by *Svi*Cas3

## Authors:

Wang-Yu Tong*, Yong Li, Shou-Dong Ye, An-Jing Wang, Yan-Yan Tang, Mei-Li Li, Zhong-Fan Yu, Ting-Ting Xia, Qing-Yang Liu and Si-Qi Zhu

## Affiliations:

*Integrated Biotechnology Laboratory, School of Life Sciences, Anhui University, 111 Jiulong Road, Hefei 230601, China*

** Corresponding author: tongwy@ahu.edu.cn*

*Tel.: +86-551-63861282*

*Fax: +86-551-63861282*

# Table of contents

## Supplementary discussion

### 1. Genome editing efficiency

In RNA-guided genome editing directed by *Svi*Cas3, a gene editing plasmid, in addition to the essential components of the plasmid itself, should contain at least one *Svicas*3 gene, one guide-DNA (g-DNA) fragment and one template-DNA (t-DNA) frgment for homology-directed repair (HDR) mechanisms. However, in DNA-guided genome editing conducted by *Svi*Cas3, g-DNA is not needed because t-DNA serves as both an editing template and a DNA recognition guide. After a gene editing plasmid (carrying *Svicas*3 and t-DNA) enters host cells, the consequent molecular events that occur are: (i) duplication, transcription and translation of the *Svicas*3 gene, (ii) site-specific cleavage of genomic DNA by *Svi*Cas3 and (iii) homologous recombination between target DNA and t-DNA by HDR machinery. Events (i) and (iii) are wholly controlled by molecular components of plasmid(s) and host cells, and only event (ii) should be performed by *Svi*Cas3.

In previous RNA-guided, template-based prokaryotic microbial genome editing based on the subtype I-B-*Svi* CRISPR-Cas system, we found the GEEs were significantly dependent on the corresponding TEs (up to four orders of magnitude) rather than HREs, which was consistent with the molecular principles of genome editing discussed (see the sister article: Prokaryotic genome editing based on the subtype I-B-*Svi* CRISPR-Cas system) [1,2]. To further verify this view, a series of transformation experiments were carried out in the template-based eukaryotic genome editing conducted by *Svi*Cas3, respectively.

As can be seen from Table S3, the change of GEE values in the eukaryotic genome editing directed by the *Svi*Cas3 is not significant (From 4.4% to 7.8%, the same order of magnitude), whether through RNA-guide (4.4%-7.3%) or DNA-guided mode (5.5%-7.8%). Further analysis shows that in template-based genome editing directed by the *Svi*Cas3, the changes among TEs (6.6-14.8%) are limited, nor were changes among HREs (36.0%-98.0%) (all in the same order of magnitude). It is worth noting that in the eukaryotic genome editing directed by the *Svi*Cas3, the GEEs by the DNA-guide mode (5.5%-7.8%) is not lower than the GEEs by the RNA-guide mode (4.4%-7.3%), implying that in genome editing conducted by the *Svi*Cas3 in the presence of t/g-DNA, R-loops derived from g-DNA and D-loops from t-DNA (UHA and DHA) may coexist in transformed cells, thus affecting GEEs. In the eukaryotic genome editing directed by the *Svi*Cas3, all the HREs are higher than the TEs (Table S3), indicating again that to improve GEE, the optimization of experimental strategies and operational conditions is indispensable.

**2 Off-target analysis in RNA-guided genome editing**

Off-target mutations are a major concern in genome editing in mammalian cells, because they can lead to genomic instability and the functional failure of some normal genes, especially in biomedical and clinical applications [3-5]. In the RNA-guided, template-based prokaryotic microbial genome editing directed by *Svi*Cas3, we performed off-target analysis on gene-edited mutants, and did not find off-target cases and indel-formation (see the sister article: Prokaryotic genome editing based on the subtype I-B-*Svi* CRISPR-Cas system). To further investigate off-target effects in the RNA-guided, template-based eukaryotic genome editing conducted by *Svi*Cas3, four potential off-target sites for each gene-edited mutant were randomly selected (87 potential off-target sites in *S. cerevisiae* LYC4-Δ*crtE* based on the motif: PAM+seed=3+6=TTC+tagtgc; 315 potential off-target sites in HEK293T-Δ*DROSHA*::*egfp* based on the motif: PAM+seed=3+11=TTC+tttcaacagtg; 11 potential off-target sites in HEK293T-Δ*CAMKMT*::*egfp* based on the motif: PAM+seed=3+12=TCC+ttgaatgttgaa; and 17 potential off-target sites in NIH-3T3-Δ*Lepr*::*egfp* based on the motif: PAM+seed=3+10=TAC+gttcctgagt, respectively). Further, we designed and chemically synthesized 16 primer pairs for off-target-analysis (Table S2) based on the principle that the targeting specificity of Cas9 in genome editing could be tightly controlled by suitable selection of both spacer and PAM [3,6], and preformed the off-target analytical experiments.

In common with the results of the RNA-guided, template-based microbial genome editing directed by *Svi*Cas3, no off-target cases and indel-formation were detected in the RNA-guided, template-based genome editing of eukaryotic cells conducted by *Svi*Cas3 (Figure S6), which can be directly explained by the fact that the *Svi*Cas3 is also a DNA-guided endonuclease (locating D-loop), so the specificity of the *Svi*Cas3 can be further enhanced by forming two big stable D-loops or multiple dynamic intermediate D-loops between t-DNA (because both UHA and DHA can be much longer than a spacer of about 40 bp) and target sequceces [7,8]. Also, the fact that in the DNA-guided genome editing couducted by the single *Svi*Cas3, no PAM-selection is needed and the view that a directional R-loop formation (D-loop in DNA-guided genome editing) provides efficient off-target site rejection [7-11] are indirectly supported the results of "no off-target cases and indel-formation".

### 3. RNA-guided Cas9-based gene editing in mammalian cells

To clearly compare the the effectiveness between *Svi*Cas3 and *Sp*Cas9 in genome editing of mammalian cells, we randomly selectced two tarege genes, *DROSHA* (encoding an RNase III enzyme, a core nuclease that initiates miRNA processing in the nucleus) and *Vegfa* (encoding vascular endothelial growth factor A that specifically acts on endothelial cells) (https://www.ncbi.nlm.nih.gov) respectively in the two cell lines HEK293T and NIH-3T3, for the *Sp*Cas9-based gene editing. In the gene editing of the two genes *DROSHA* (HEK293T, Homo sapiens-chr V) and *Vegfa* (NIH-3T3, Mus musculus-chr XVII), we firstly selected the expression vector pX459 VRER (addgene: plasmid # 101716) as the original vector of gene editing tools, and then designed and constructed the two gene editing vectors: pX459M-*cas*9-t/g-Δ*DROSHA*::*egfp* and pX459M-*cas*9-t/g-Δ*Vegfa*::*egfp* (Figure S7A, Tables S1 and S2), following the procedures described in the section "Construction of gene editing tools in genome editing of mammalian cells". The arbitrarily designed fragment sizes of both t-DNA (UHA plus *egfp* plus DHA) and deletion were: 417+726+355=1498 nt and 303 nt (Δ*DROSHA*::*egfp*) and 385+726+333=1444 nt and 283 nt (Δ*Vegfa*::*egfp*), respectively. The selected PAMs respectively were ggg (Δ*DROSHA*::*egfp*) and agg (Δ*Vegfa*::*egfp*) (Figure S7D, Table S2). After transfecting each of the two gene editing vectors (pX459M-*cas*9-t/g-Δ*DROSHA*::*egfp* and pX459M-*cas*9-t/g-Δ*Vegfa*::*egfp*) into corresponding cells, we obtained the potential gene-edited variants with green fluorescence (Figure S7B) (Note: Each UHA of the two gene editing vectors has no promoter, so the EGFP protein can only be expressed by inserting the *egfp* gene into the framework of the two target genes, and green fluorescence can be detected.): HEK293T-Δ*DROSHA::egfp*-Cas9 and NIH-3T3-Δ*Vegfa*::*egfp*-Cas9, and tested them using a basic PCR (primers: *DROSHA*-vF-9 / and *DROSHA-egfp*-vR-9 and *Vegfa*-vF-9/*Vegfa-egfp*-vR-9) and DNA sequencing analysis (Figures S7C and S7D, Table S2). All the analytical results were consistent with rational expectations and designed t-DNA (Figure S7 and Table S2) showed that the RNA-guided genome editing mediated by *Sp*Cas9 was successful. As can be seen from Figure 7B and Table S3, the GEE of RNA-guided genome editing directed by the *Svi*Cas9 (0.5-0.7%) was significantly lower than that (4.4-7.3%) of RNA-guided genome editing guided by the *Svi*Cas3, which resulted mainly from the decrease in HREs rather than the decrease in transformation / transfection efficiencies (TEs). We inferred that the decrease of HRE may be due to the significant activation of non-homologous end joining (NHEJ) mechanisms (resulting in numerous off-target effects, see Figure S6), which inhibits the activity of homologous recombination mechanisms.

Furthermore, as a comparsion, we conducted a study on the off-target effects of genome editing mediated by *Sp*Cas9. From Figure S6, we can see that in the RNA-guided genome editing mediated by the *Svi*Cas3, no off-target changes or indel formation were detected, while in the RNA-guided genome editing mediated by *Sp*Cas9, two out of four randomly selected potential off-target sites were found to show off-target changes (8468 potential off-target sites in HEK293T-Δ*DROSHA*::*egfp*-Cas9 based on the motifs: seed+PAM=10+3=agactttgta+TGG, seed+PAM=12+3=ccagactttgta+AGG, seed+PAM=13+3=tacaagactttgt+AGG, seed+PAM=14+3=gaccagactttgta+GGG). The results show that the off-target frequency of the *Svi*Cas3-based gene editing is at least much lower than that of Cas9-based gene editing [6,12].

## 4. Main features of typical gene editing techniques

The targeting approaches of genome engineering based on site-specific recombination processes mainly include: recombinases (e.g., phi C31 integrase) [13,14], transposons (e.g., the ISY100 transposase) [15], homologous recombination / HR [16,17] and non-homologous end joining / NHEJ [18], among which the homologous gene targeting reported in yeast is one of the first methods for rational genome engineering (with an efficiency of only $10^{-6}$ to $10^{-9}$) [19,20]. Generally, endonucleases used in genome engineering can be divided into five groups [21,22]: meganucleases (also known as homing endonucleases) [21,23,24], chemical nucleases (synthetic endonucleases resulting from the fusion of DNA-reactive agents to a DNA-binding polymer) [25], zinc finger nucleases (ZFNs) [26-30], transcription activator like effector nucleases (TALENs) [31-33] and CRISPR-Cas enzymes [11,34-37], which mainly introduce double strand break / DSB or single strand break / SSB at the target DNA site to stimulate the cell's own DNA repair pathways to make DNA modifications (HDR: conservative homology-directed repair in the presence of t-DNA; NHEJ: error-prone non-homologous end joining in the absence of t-DNA).

Meganucleases (I-SceI meganuclease, the first reported for meganuclease used in mammalian cells) [23,38] can be used as a tool for targeted genome engineering, but their disadvantages, such as site-sequence degeneracy, homodimer nature, the conversion of sequences adjacent to a DNA break, and the need to first introduce a known cleavage site into the region of interest, hinder their widespread use in genome editing [21,23,38]. Although its off-target effect and relatively large molecular weight limit its practical application, the most commonly used genome editing tool is still the Cas9-based gene editing system based on RNA-guided mechanisms [39,40]. The main features of typical gene editing techniques in three classes of genome editing based on the principle of DNA-orientation are summarized in Table S4.

As seen from Table S4, genome editing technology based on *Svi*Cas3 has obvious advantages over other technologies, such as free selection of target sequences, simple operation and no off-target changes detected. In addition to these advantages and features mentioned above, the fragment ranges of t-DNA from 807 to 2001 nt, insertion from 0 to 726 nt, and deletion from 84 to 1623 nt in the template-based eukaryotic genome editing are at least acceptable (Table S2). In particular, t-DNA without the PAM requirement acts as both a template and a guide, suggesting that CRISPR, as the basis of crRNA design, is unnecessary in genome editing directed by *Svi*Cas3. Furthermore, we demonstrated that the DNA-guided *Svi*Cas3 enzyme can simultaneously perform both gene editing and base editing in the genomic DNA of mammalian cells.

## Figures S1 to S7

**Figure S1. Construction of gene editing tools in the RNA-guided genome editing of *S. cerevisiae* LYC4.**

**Figure S2. Construction of gene editing vector in the RNA-guided genome editing of HEK293T and NIH-3T3 cells.**

**Figure S3. Construction of gene editing tools in the DNA-guided genome editing of *S. cerevisiae* LYC4.**

**Figure S4. Construction of gene editing tools in the DNA-guided genome editing of mammalian cells.**

**Figure S5. Construction of the DNA-guided gene and base editing vector AIO-mCherry-*cas*3-$t_{g1}$/$t_{b2}$-Δ*CAMKMT*::*egfp* for the *CAMKMT* gene in HEK293T cells.**

**Figure S6. Off-target analyses in the template-based genome editing of eukaryotic cells conducted by the *Svi*Cas3.**

**Figure S7. Construction of Cas9-based gene editing tools and mutant verification in mammalian cell genome editing mediated by Cas9.**

**A**

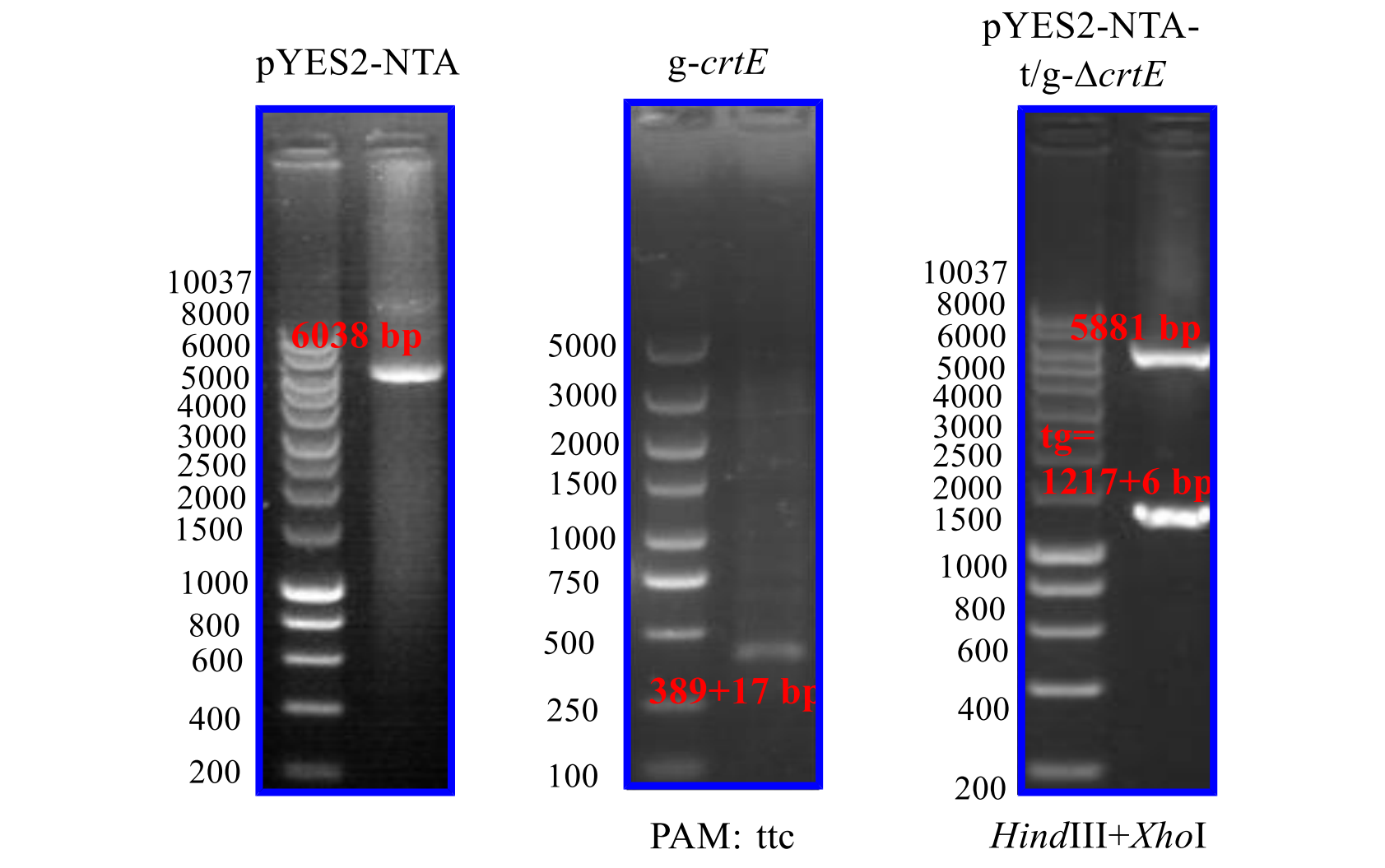

**B**

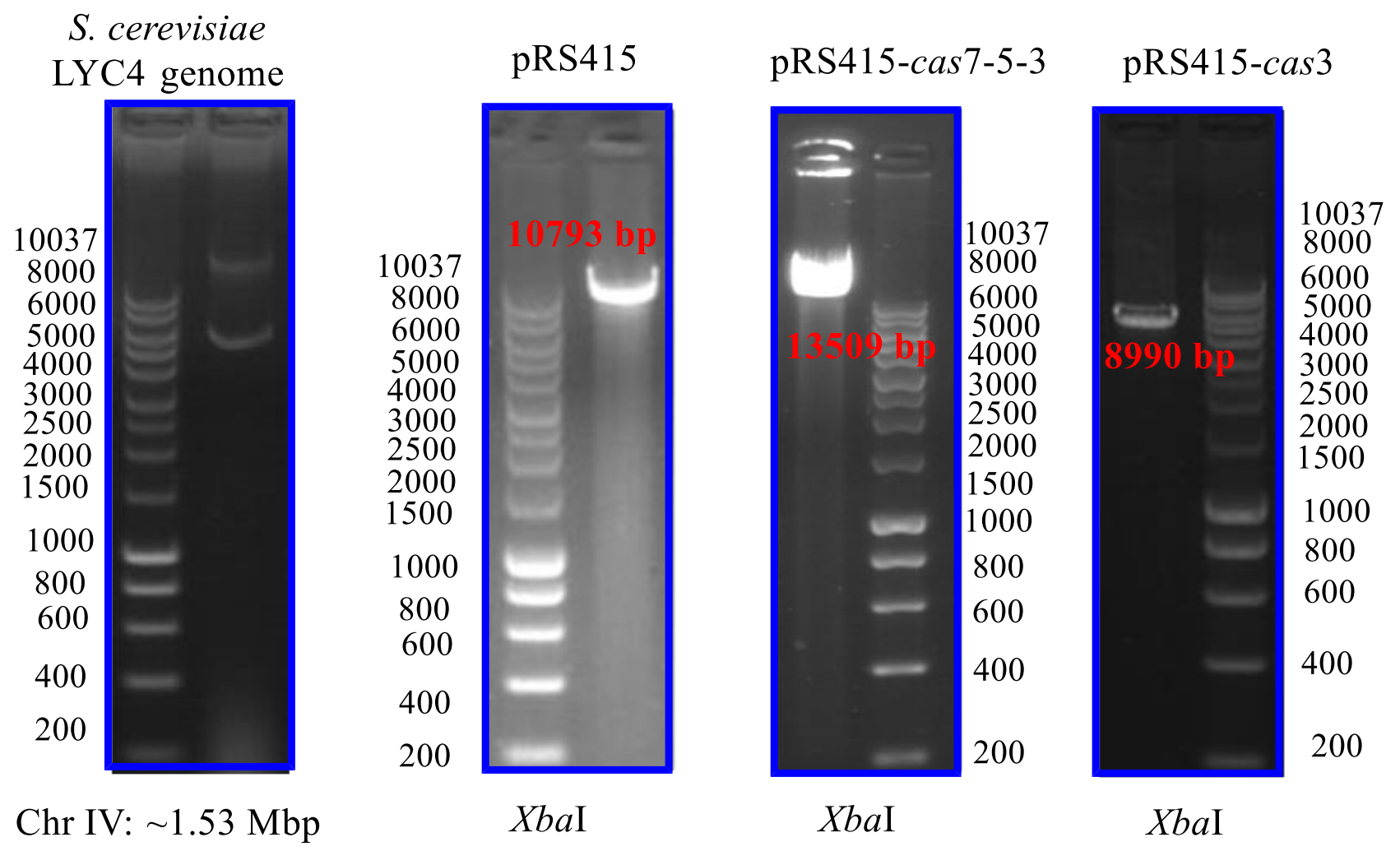

**C**

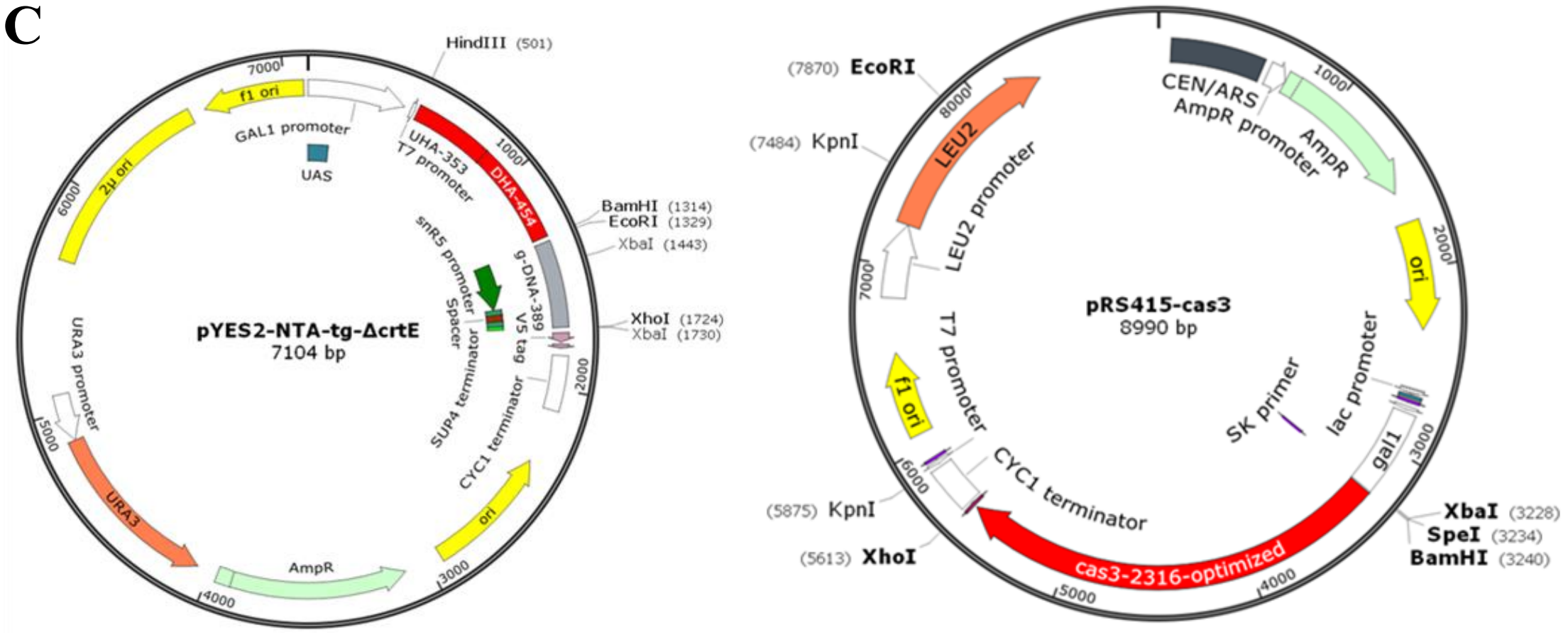

**Figure S1. Construction of gene editing tools in the RNA-guided genome editing of *S. cerevisiae* LYC4. (A)** The extracted pYES2-NTA, chemically synthesized g-*crtE* and the result of pYES2-NTA-t/g-Δ*crtE* digested by *Hind*III and *Xho*I. **(B)** The extracted *S. cerevisiae* LYC4 genome and the linearized results of plasmids pRS415, pRS415-*cas*7-5-3 and pRS415-*cas*3 digested by *Xba*I. **(C)** The maps of gene editing plasmid pYES2-NTA-t/g-Δ*cr*tE and Cas expression plasmid pRS415-*cas*3 (The map of plasmid pRS415-*cas*7-5-3 was the same as this map of plasmid pRS415-*cas*3, except for the codon optimized genes of *cas*5 and *cas*7).

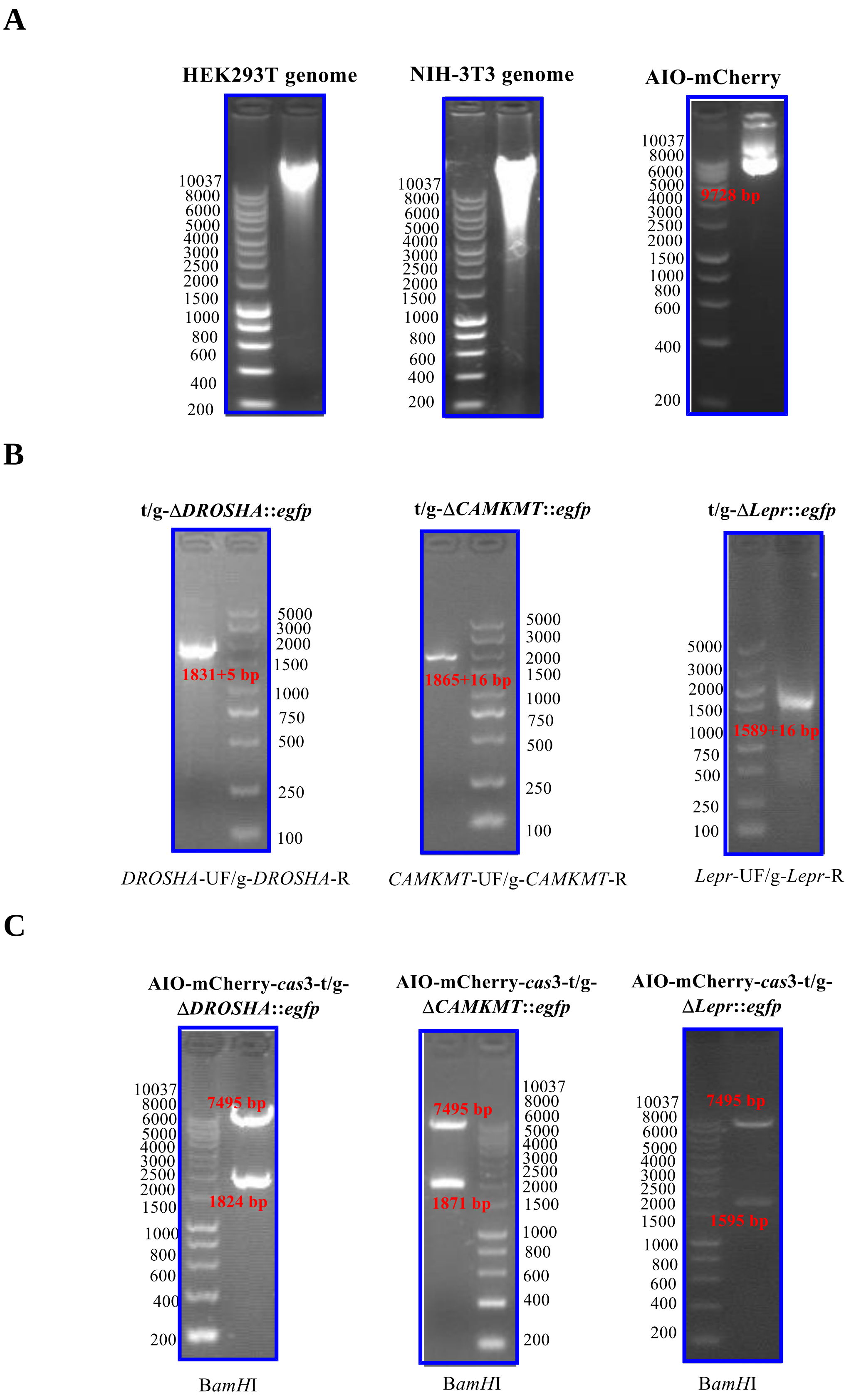

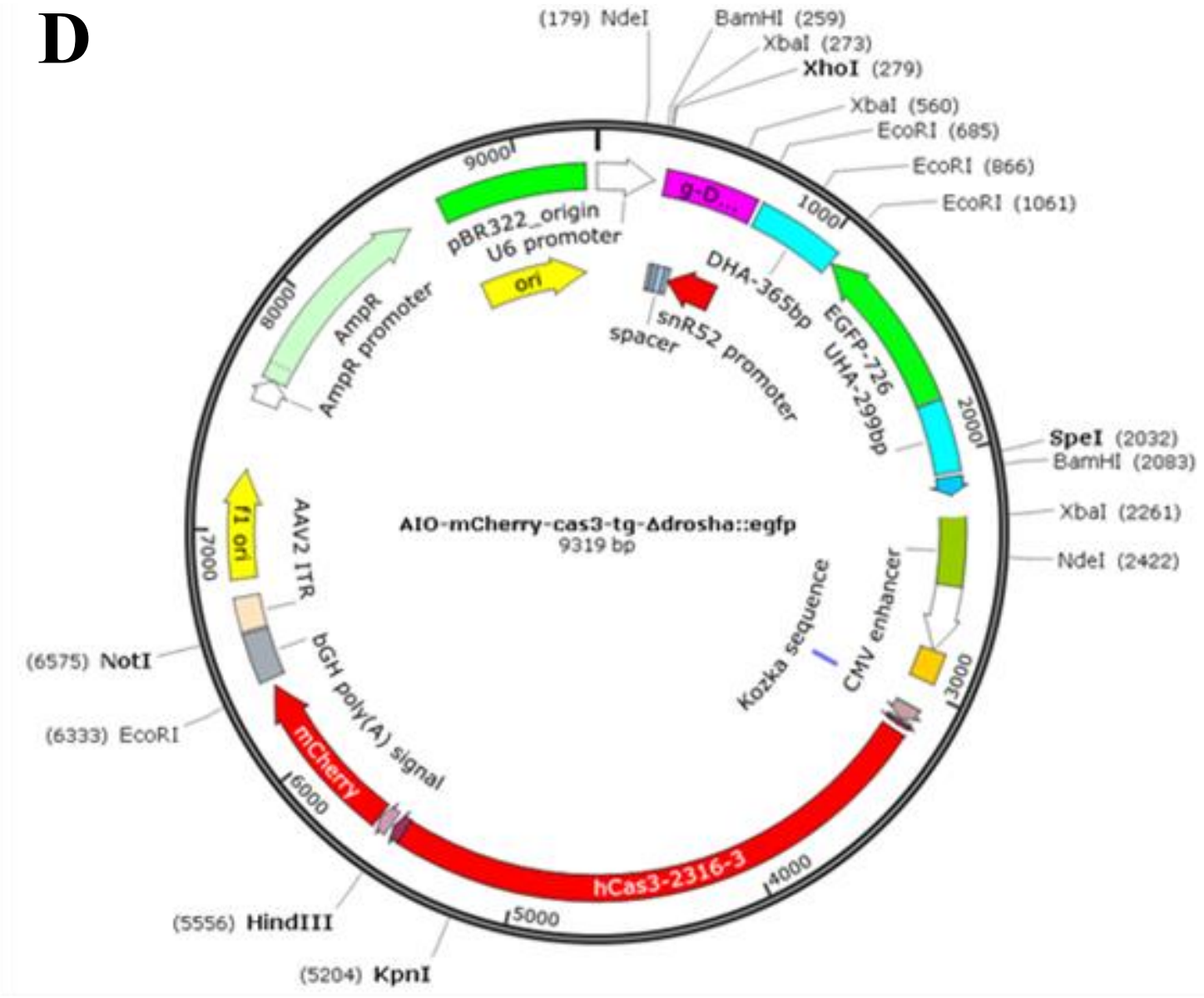

**Figure S2. Construction of gene editing vector in the RNA-guided genome editing of HEK293T and NIH-3T3 cells. (A)** The extracted HEK293T cell genome, extracted NIH-3T3 cell genome and extracted vector AIO-mCherry. **(B)** The overlap PCR amplified t/g-Δ*DROSHA*::*egfp*, overlap PCR amplified t/g-Δ*CAMKMT*::*egfp*, overlap PCR amplified t/g-Δ*Lepr*::*egfp*. **(C)** The linearized results of the three gene editing vectors AIO-mCherrym-*cas*3-t/g-Δ*DROSHA*::*egfp*, AIO-mCherrym-*cas*3-t/g-Δ*CAMKMT*::*egfp* and AIO-mCherrym-*cas*3-t/g-Δ*Lepr*::*egfp* digested by *Bam*HI (each vector harboring two *Bam*HI restriction sites). **(D)** The map of gene editing plasmid AIO-mCherry-*cas*3-t/g-Δ*DROSHA*::*egfp* (In the RNA-guided genome editing of mammalian cells, the maps of the other two genome editing vectors AIO-mCherry-*cas*3-t/g-Δ*CAMKMT*::*egfp* and AIO-mCherry-*cas*3-t/g-Δ*Lepr*::*egfp* were the same as this map, except for the corresponding g-DNAs and t-DNAs).

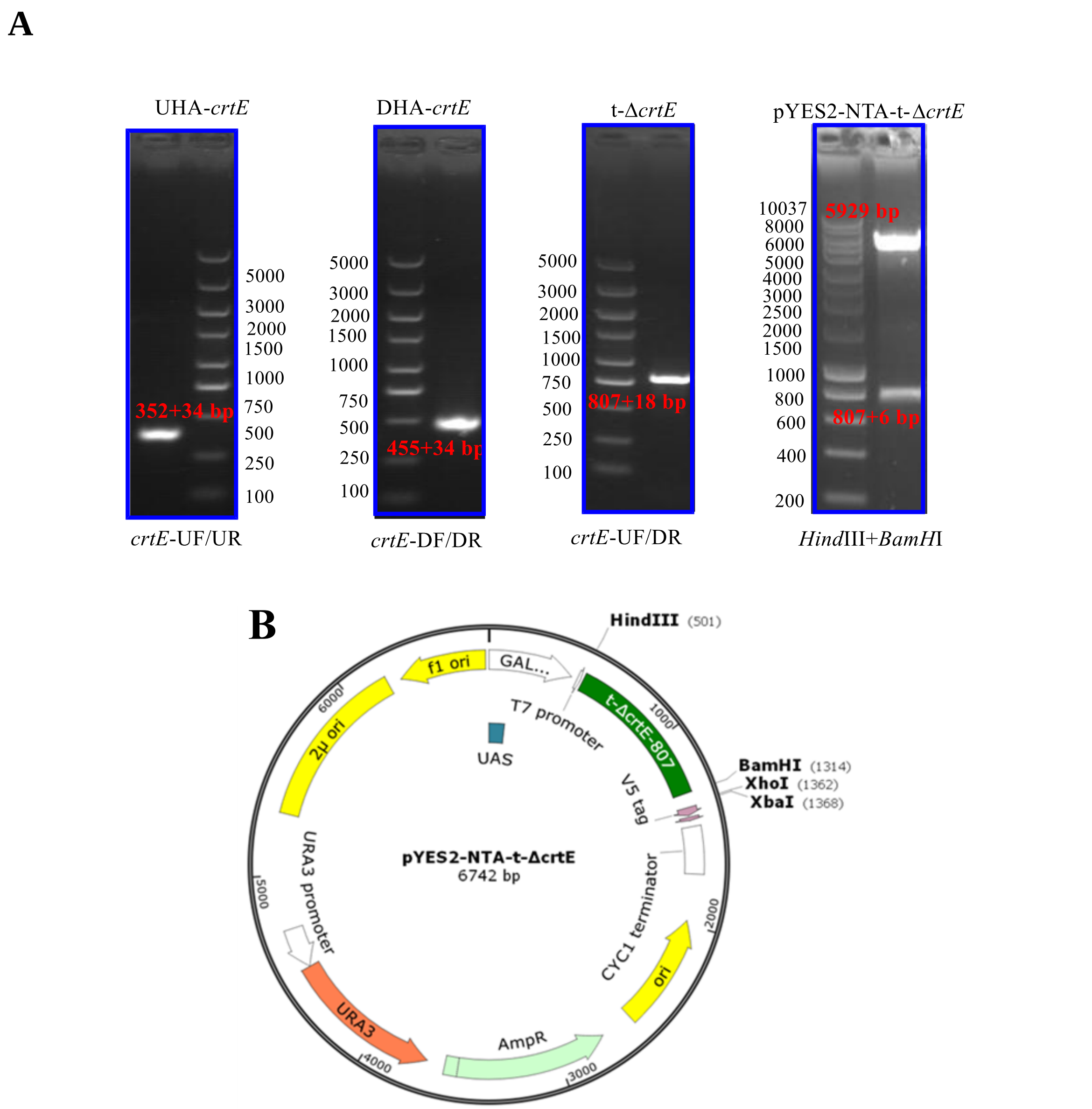

**Figure S3. Construction of gene editing tools in the DNA-guided genome editing of *S. cerevisiae* LYC4. (A)** The PCR amplified UHA-*crtE* and DHA-*crtE*, overlap PCR amplified t-*crtE* and the result of gene editing plasmid pYES2-NTA-t-Δ*crtE* digested by *Hind*III+*BamH*I, respectively (UHA+DHA=353+454=807 nt, deletion=252 nt). **(B)** The map of gene editing plasmid pYES2-NTA-t-Δ*crt*E.

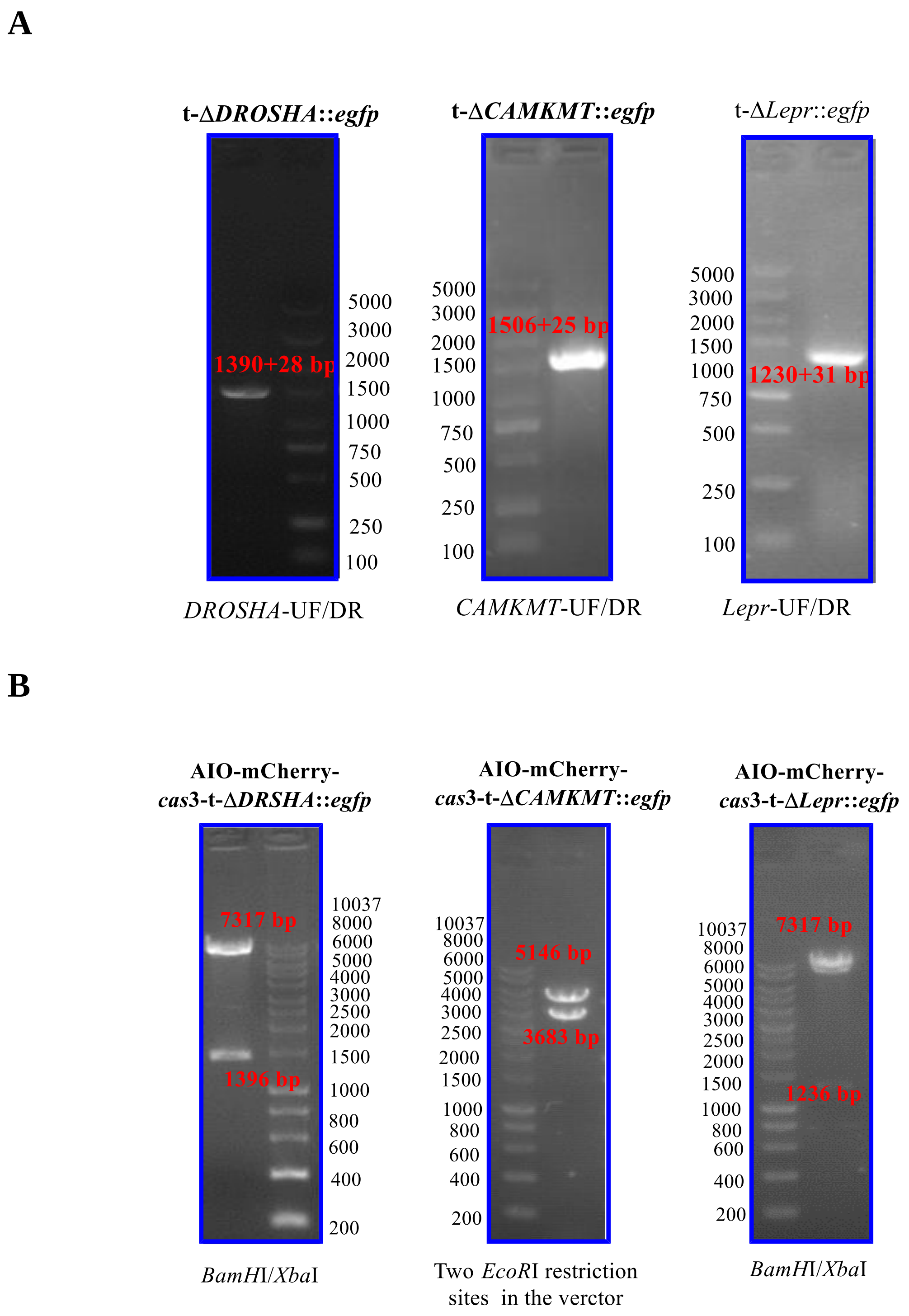

C

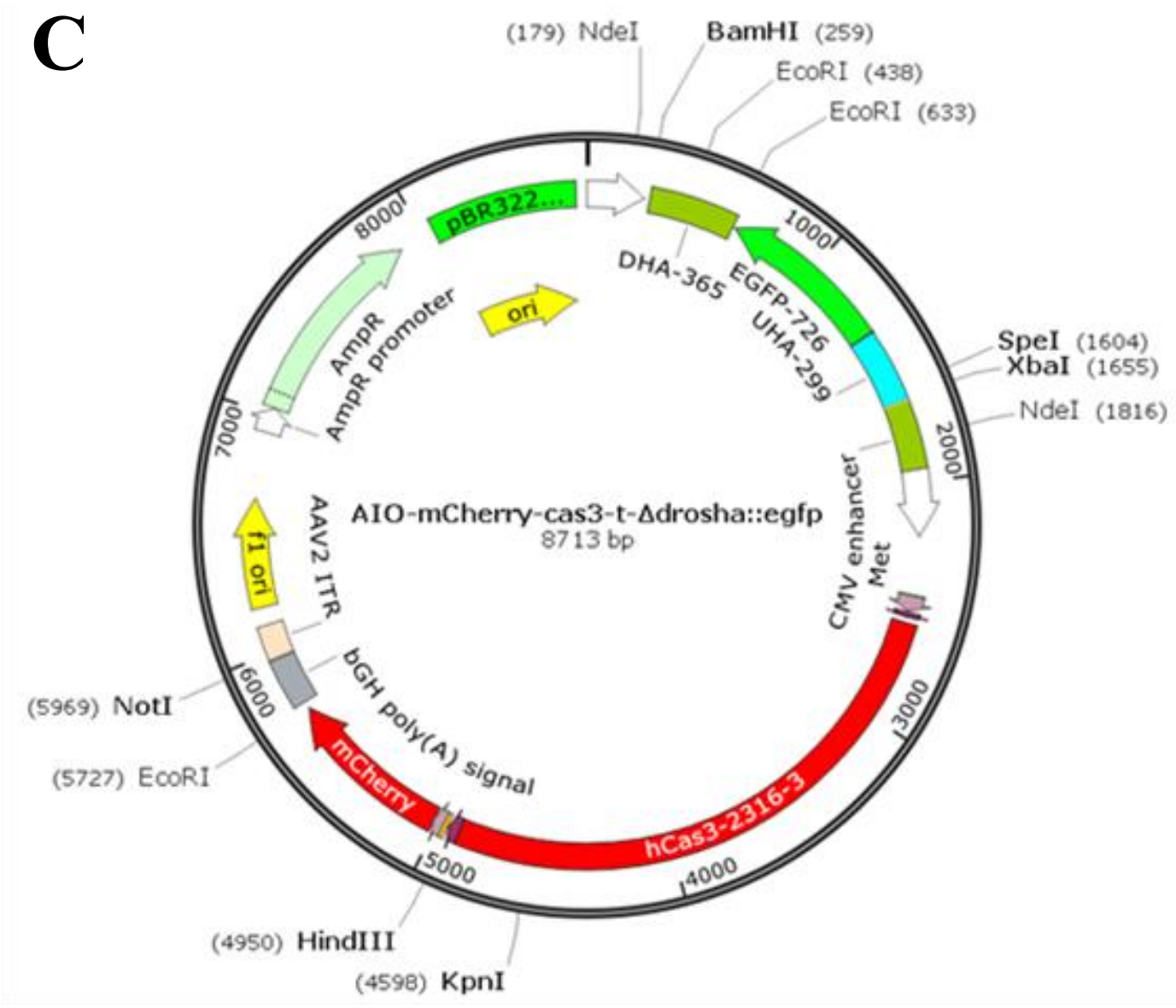

**Figure S4. Construction of gene editing tools in the DNA-guided genome editing of mammalian cells. (A)** The PCR amplified fragments of t-Δ*DROSHA*::*egfp*, t-Δ*CAMKMT*::*egfp* and t-Δ*Lepr*::*egfp*. **(B)** The results of the three vectors (AIO-mCherry-*cas*3-t-Δ*DROSHA*::*egfp*, AIO-mCherry-*cas*3-t-Δ*CAMKMT*::*egfp* and AIO-mCherry-*cas*3-t-Δ*Lepr*::*egfp*) digested by *Bam*HI+*Xba*I, *EcoR*I and *Bam*HI+*Xba*I, respectively. **(C)** The map of gene editing plasmid AIO-mCherry-*cas*3-t-Δ*DROSHA*::*egfp* (In the DNA-guided genome editing of mammalian cells, the maps of the other two genome editing vectors AIO-mCherry-*cas*3-t-Δ*CAMKMT*::*egfp* and AIO-mCherry-*cas*3-t-Δ*Lepr*::*egfp* were identical to this map, except for the corresponding t-DNAs).

A

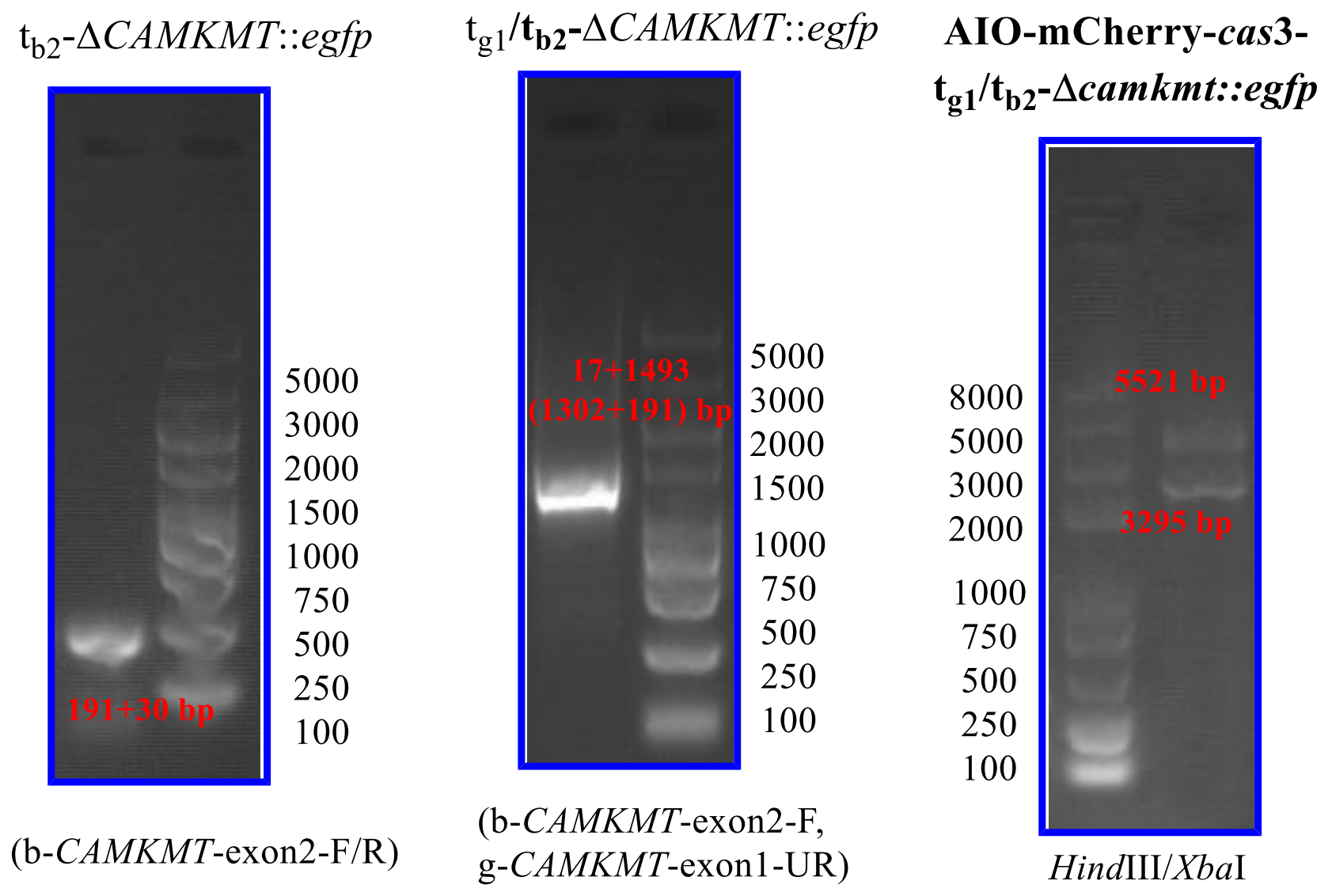

B

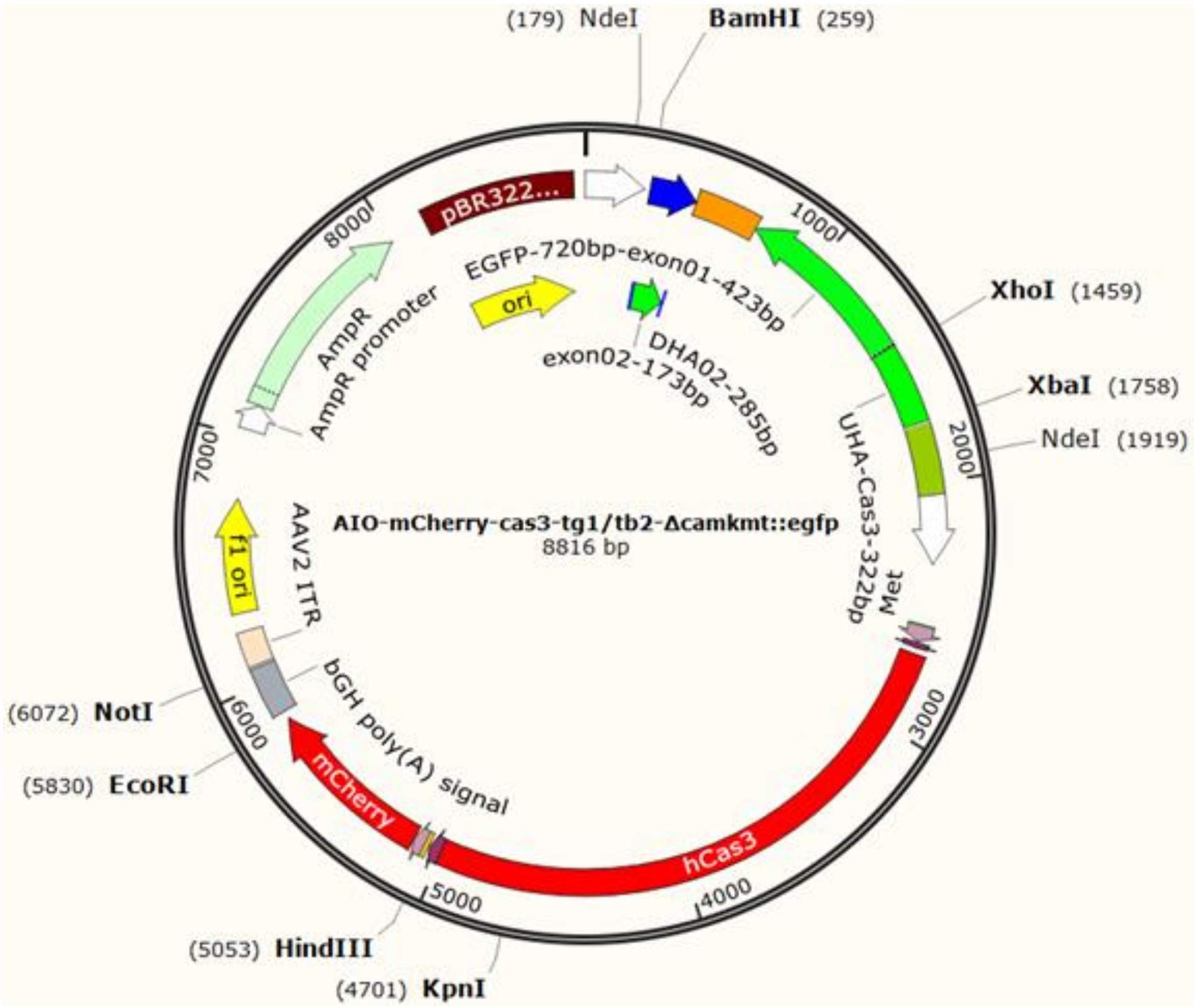

**Figure S5. Construction of the DNA-guided gene and base editing vector AIO-mCherry-*cas*3-$t_{g1}$/$t_{b2}$-Δ*CAMKMT*::*egfp* for the *CAMKMT* gene in HEK293T cells. (A)** The PCR amplified fragments of $t_{b2}$-Δ*CAMKMT*::*egfp* and $t_{g1}$/$t_{b2}$-Δ*CAMKMT*::*egfp*, and the result of the vector AIO-mCherry-*cas*3-$t_{g1}$/$t_{b2}$- Δ*CAMKMT*::*egfp* digested by *Hind*III plus *Xba*I. **(B)** The map of the vector AIO-mCherry-*cas*3-$t_{g1}$/$t_{b2}$- Δ*CAMKMT*::*egfp* ("b- or b2-" representing base editing of the exon2, "g- or g1-" representing gene editing of the exon1).

**A**

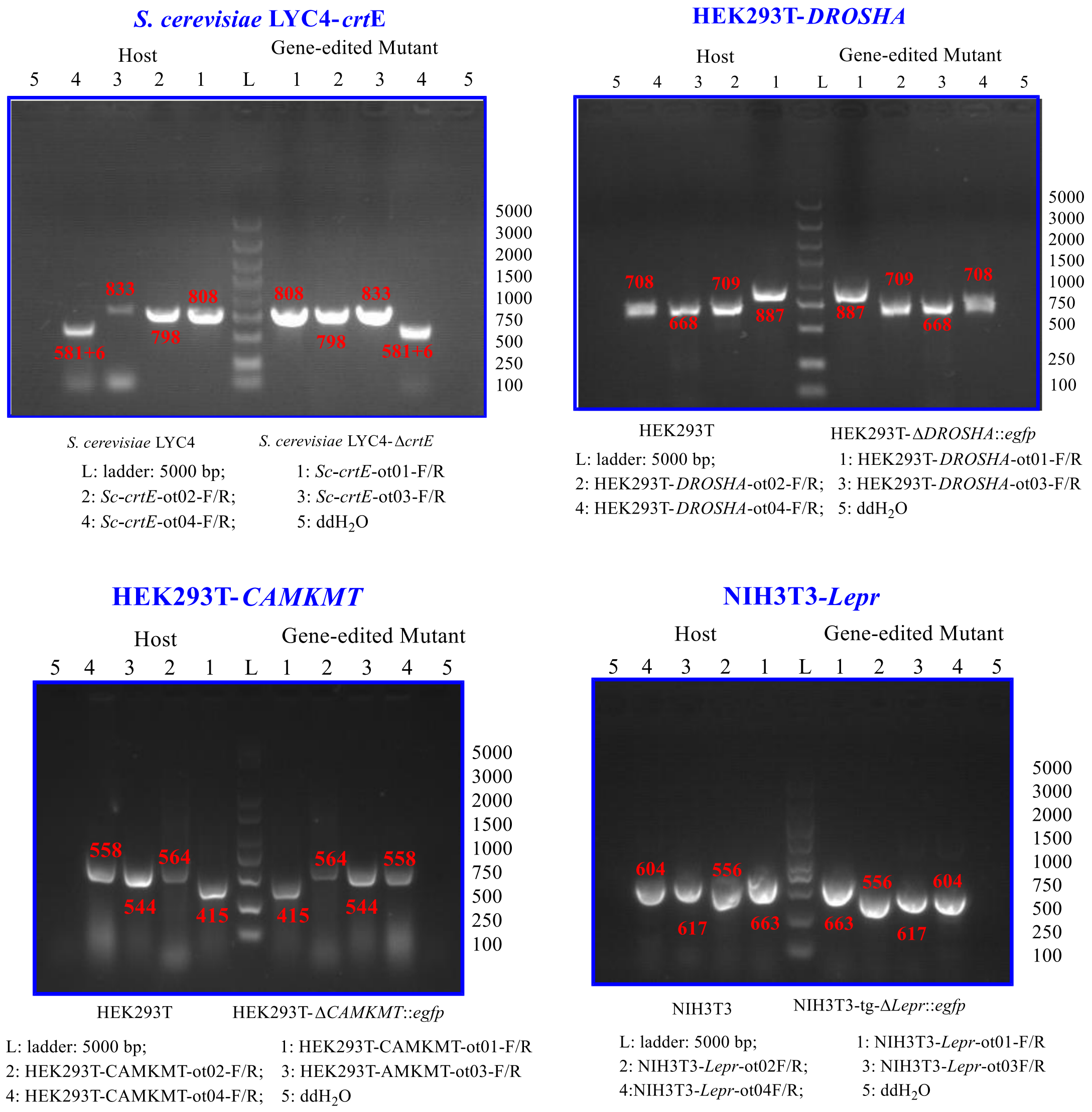
S. cerevisiae LYC4-crtE
Host
Gene-edited Mutant
5 4 3 2 1 L 1 2 3 4 5
5000
3000
2000
1500
1000
750
500
250
100
833
808
808
833
581+6
798
798
581+6
S. cerevisiae LYC4
S. cerevisiae LYC4-ΔcrtE
L: ladder: 5000 bp;
1: Sc-crtE-ot01-F/R
2: Sc-crtE-ot02-F/R;
3: Sc-crtE-ot03-F/R
4: Sc-crtE-ot04-F/R;
5: ddH2O
HEK293T-DROSHA
Host
Gene-edited Mutant
5 4 3 2 1 L 1 2 3 4 5
5000
3000
2000
1500
1000
750
500
250
100
708
709
709
708
668
887
887
668
HEK293T
HEK293T-ΔDROSHA::egfp
L: ladder: 5000 bp;
1: HEK293T-DROSHA-ot01-F/R
2: HEK293T-DROSHA-ot02-F/R;
3: HEK293T-DROSHA-ot03-F/R
4: HEK293T-DROSHA-ot04-F/R;
5: ddH2O
HEK293T-CAMKMT
Host
Gene-edited Mutant
5 4 3 2 1 L 1 2 3 4 5
5000
3000
2000
1500
1000
750
500
250
100
558
564
564
558
544
415
415
544
HEK293T
HEK293T-ΔCAMKMT::egfp
L: ladder: 5000 bp;
1: HEK293T-CAMKMT-ot01-F/R
2: HEK293T-CAMKMT-ot02-F/R;
3: HEK293T-AMKMT-ot03-F/R
4: HEK293T-CAMKMT-ot04-F/R;
5: ddH2O
NIH3T3-Lepr
Host
Gene-edited Mutant
5 4 3 2 1 L 1 2 3 4 5
5000
3000
2000
1500
1000
750
500
250
100
604
556
556
604
617
663
663
617
NIH3T3
NIH3T3-tg-ΔLepr::egfp
L: ladder: 5000 bp;
1: NIH3T3-Lepr-ot01-F/R
2: NIH3T3-Lepr-ot02F/R;
3: NIH3T3-Lepr-ot03F/R
4:NIH3T3-Lepr-ot04F/R;
5: ddH2O

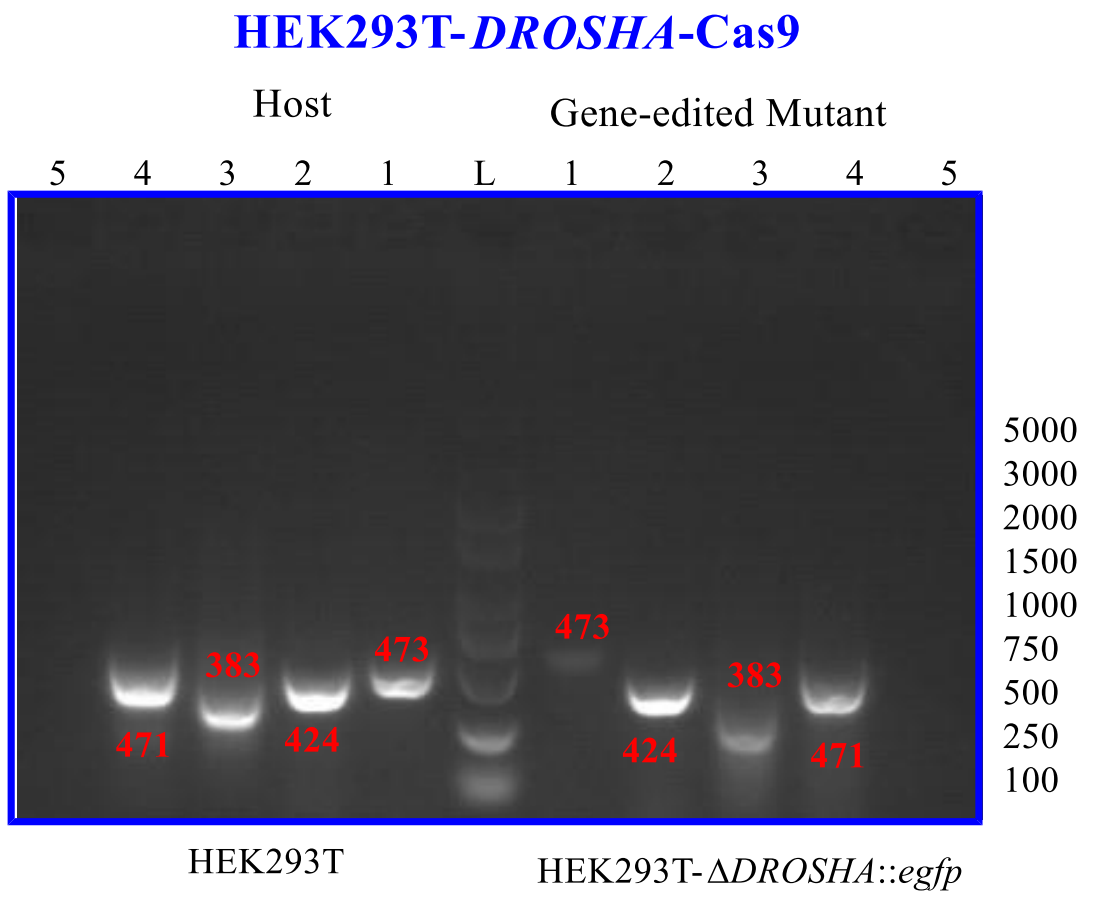

L: ladder: 5000 bp; 1: HEK293T-*DROSHA*-9-ot01-F/R
2: HEK293T-*DROSHA*-9-ot02-F/R; 3: HEK293T-*DROSHA*-9-ot03-F/R
4: HEK293T-*DROSHA*-9-ot04-F/R; 5: $ddH_2O$

**B**

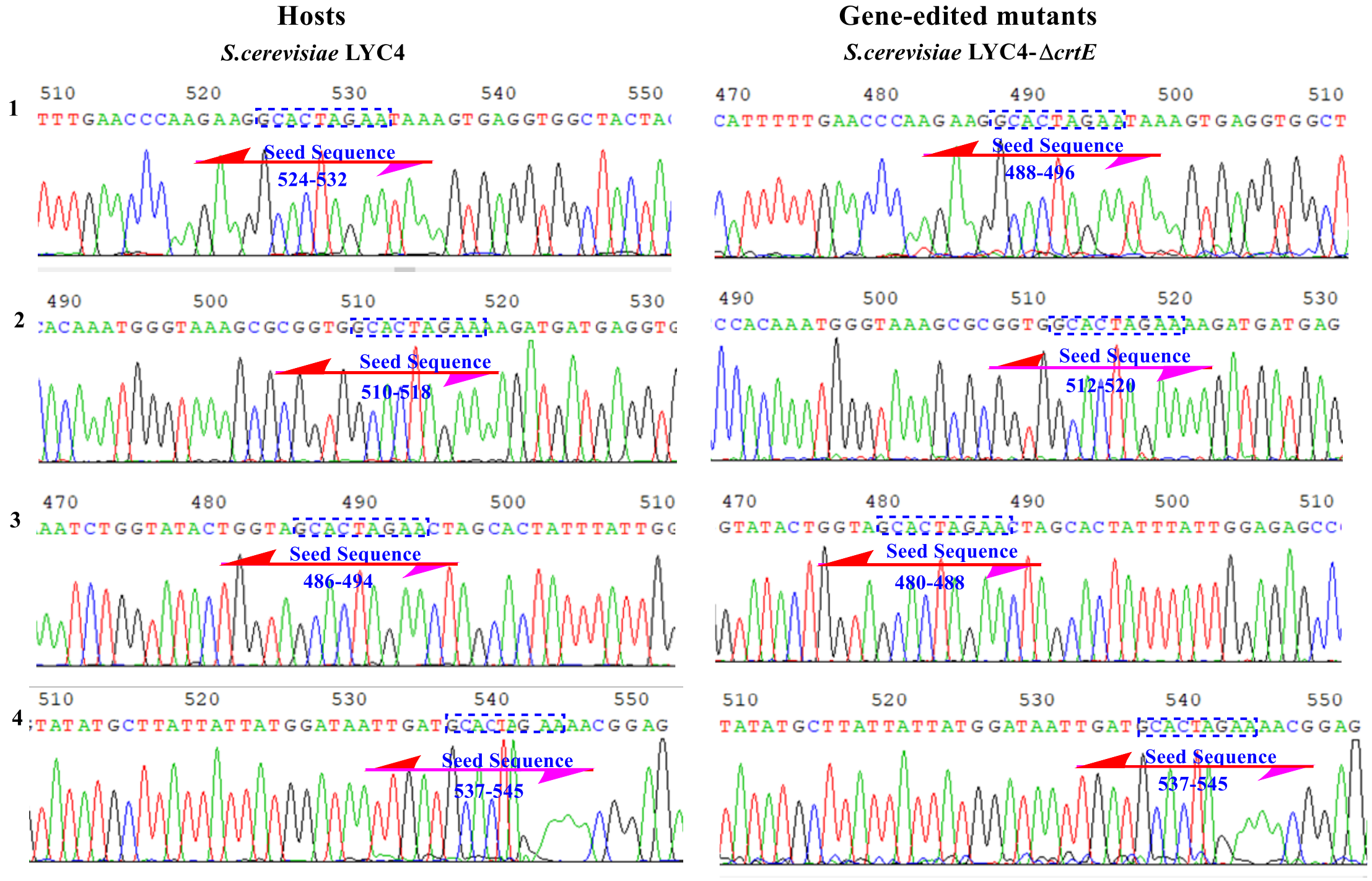

**A: green; T:red; G: black; C; blue**

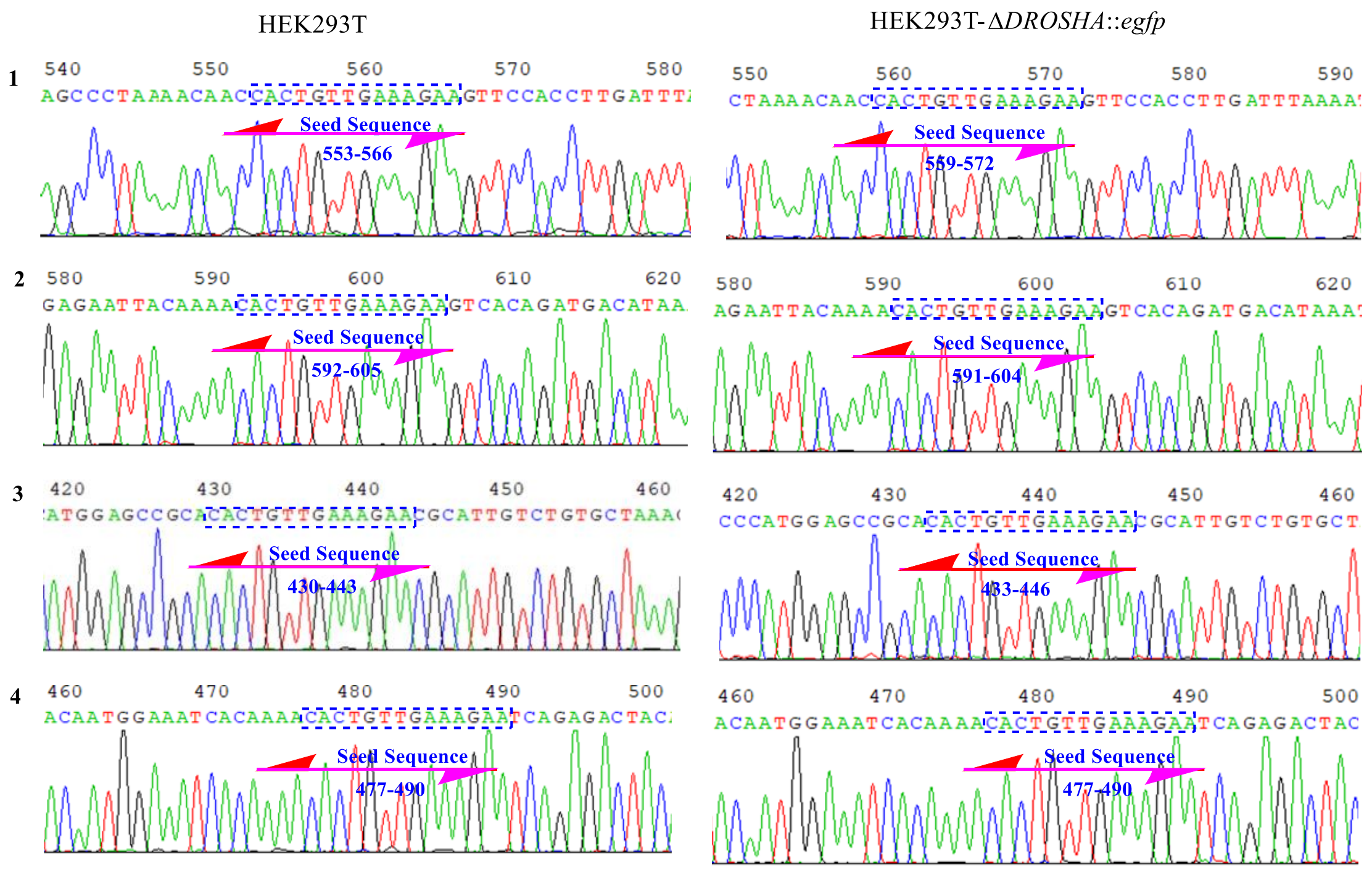

A: green; T:red; G: black; C; blue

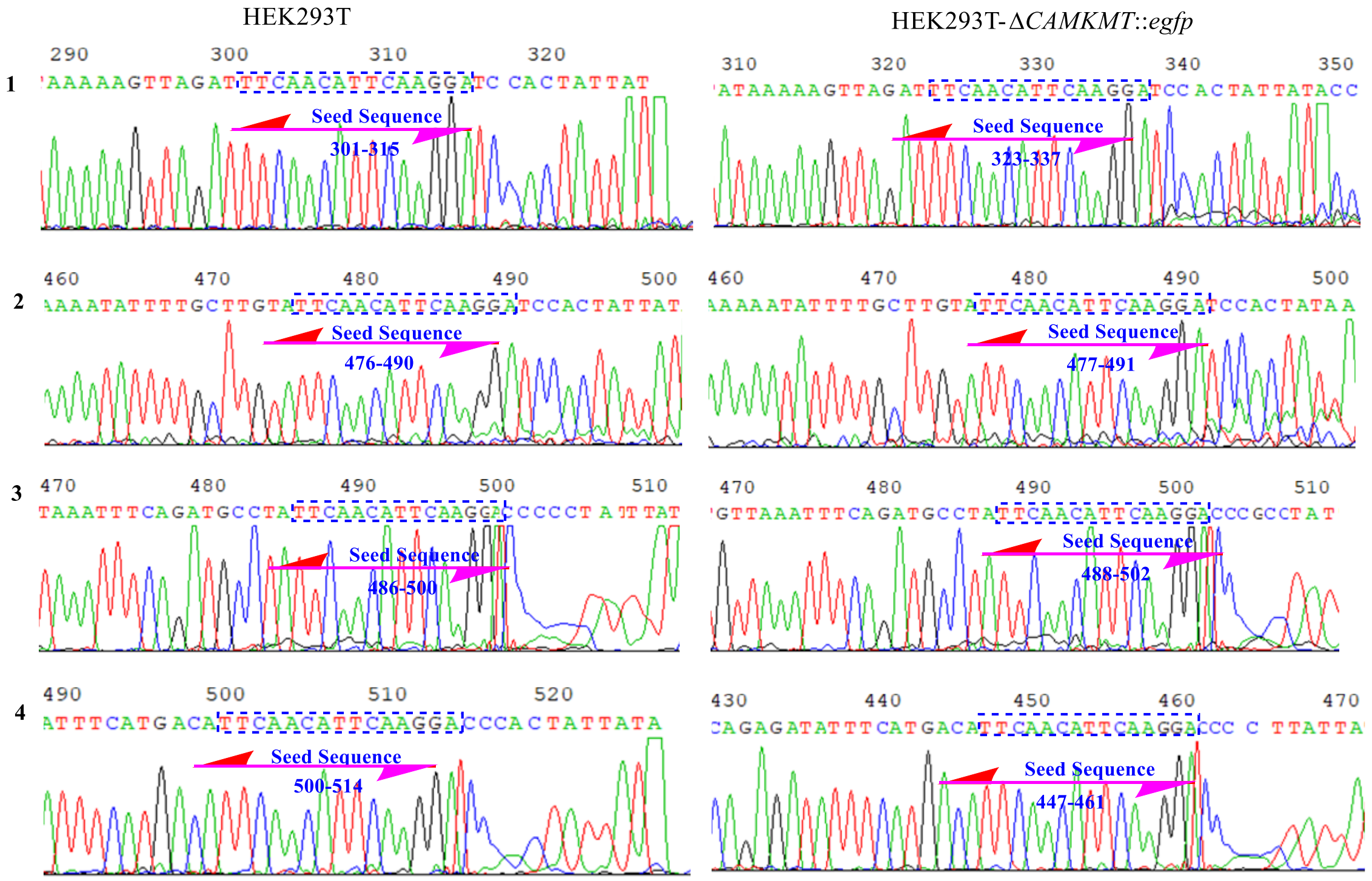

A: green; T:red; G: black; C; blue

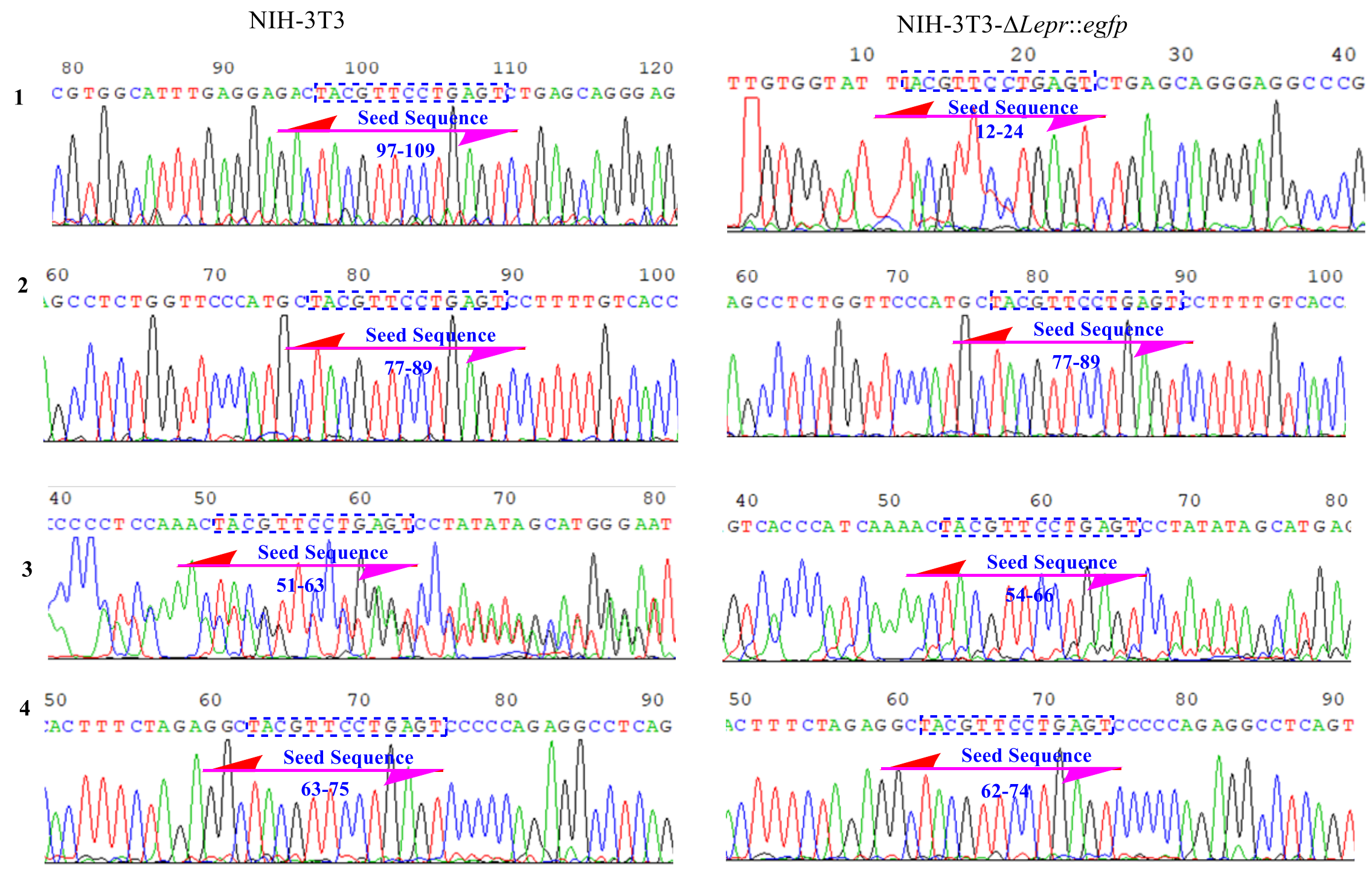

A: green; T:red; G: black; C; blue

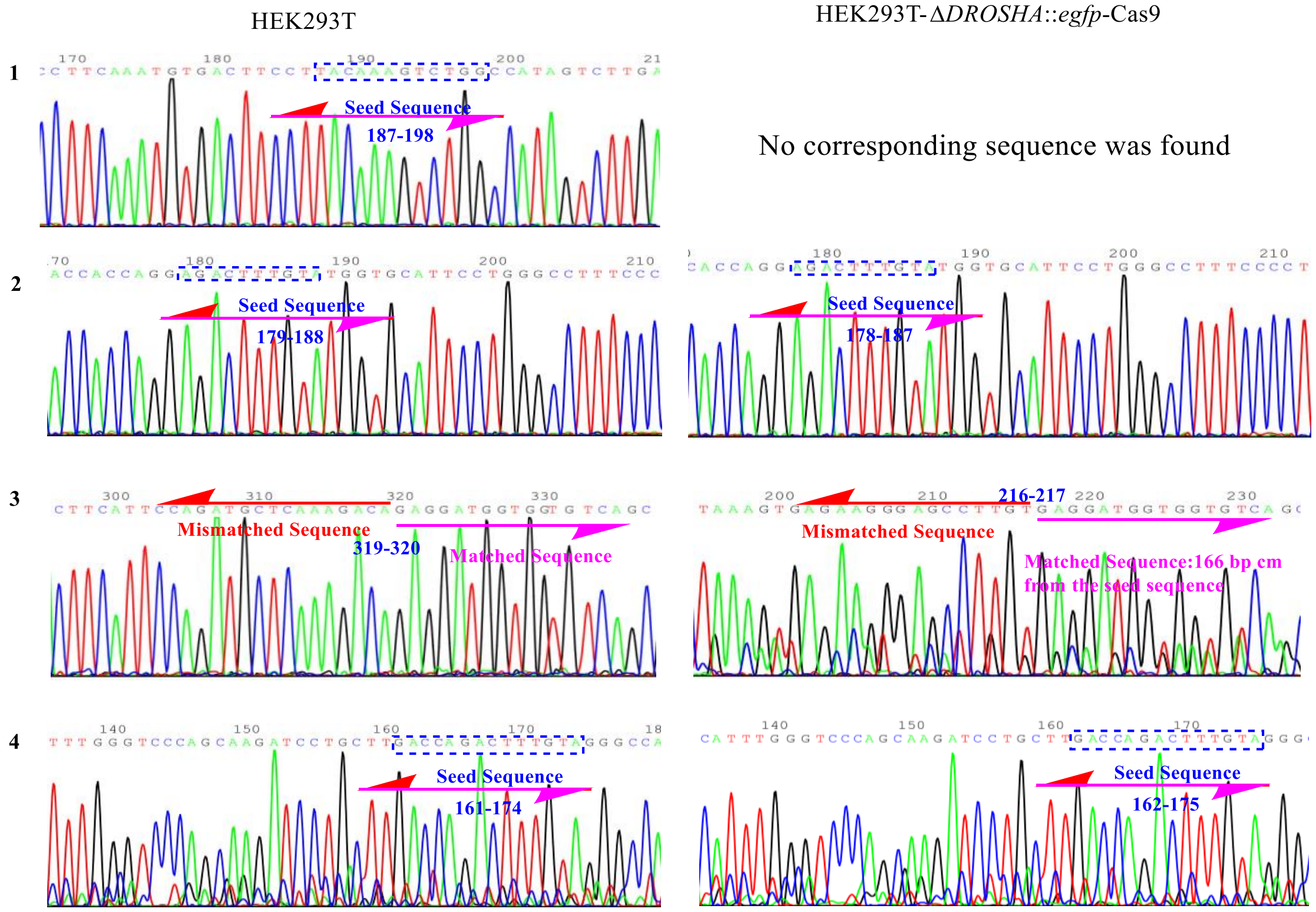

A: green; T:red; G: black; C; blue

**Figure S6. Off-target analysis of the RNA-guided genome editing conducted by the *Svi*Cas3 and *Sp*Cas9 in eukaryotic cells. (A)** DNA gel electrophoresis of the PCR products of potential off-target sites in the genomes of gene-edited mutants. **(B)** The DNA sequencing results of the PCR products of potential off-target sites in the gene-edited mutants (As can be seen from Figure S6, in the RNA-guided genome editing mediated by the *Svi*Cas3, no off-target cases and indel formation were detected, while in the RNA-guided genome editing mediated by *Sp*Cas9, off-target cases were detected in HEK293T-Δ*DROSHA*::*egfp*-Cas9 cells).

**A**

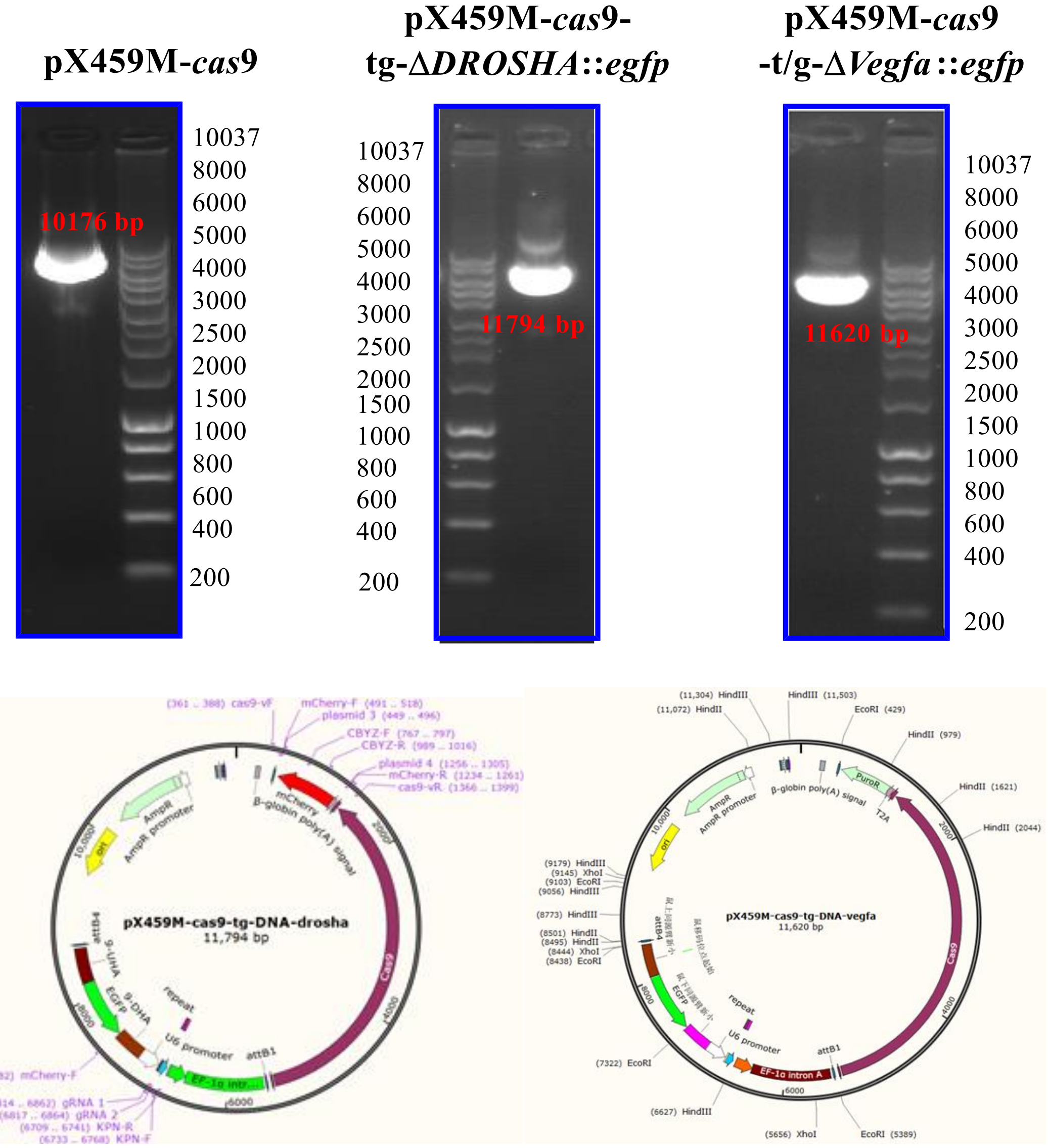

**B**

**pX459M-*cas*9-t/g-Δ*DROSHA*::*egfp* (with mCherry)**

Bright field mCherry EGFP

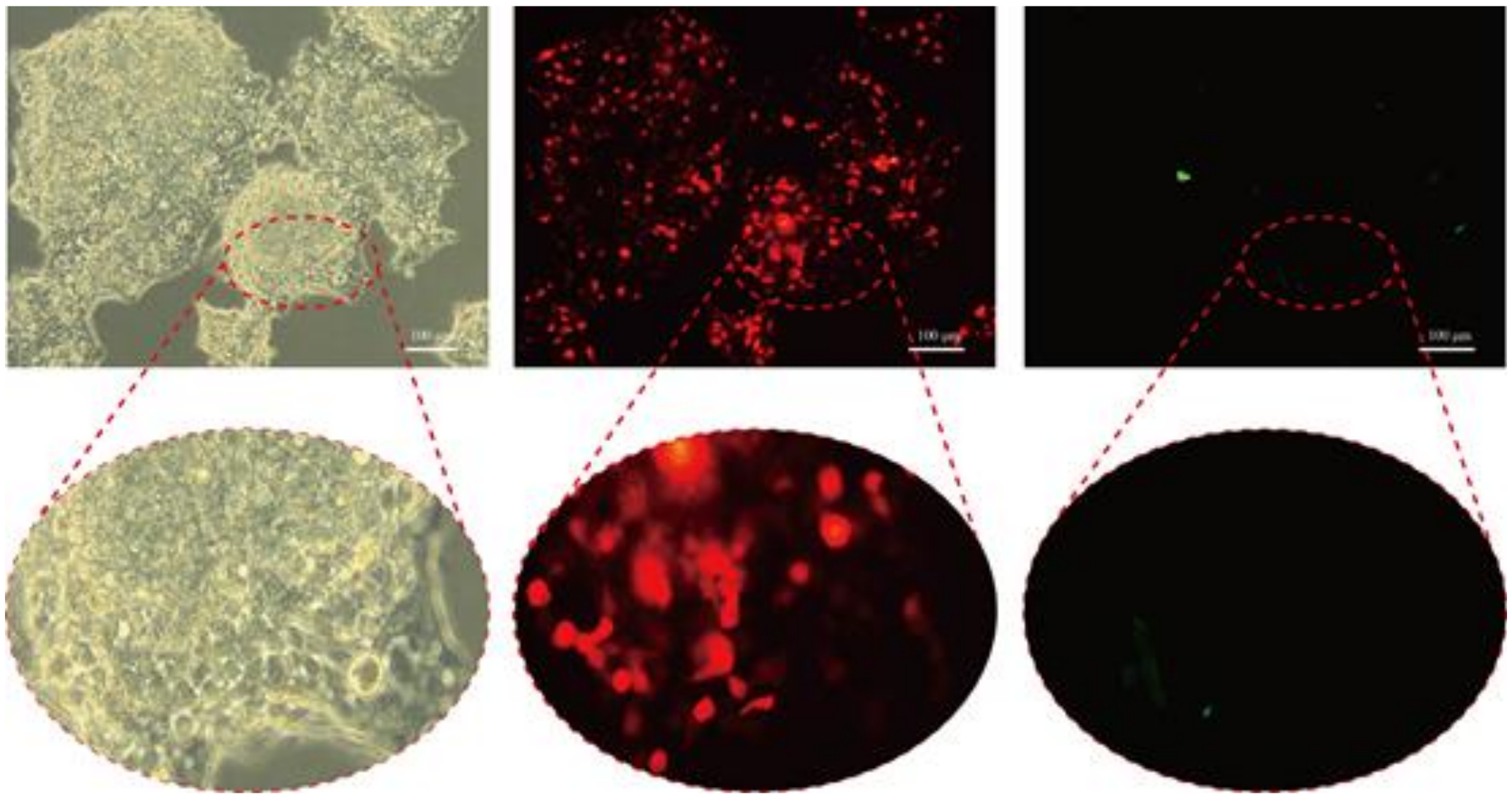

HEK293T
HEK293T-t/g-Δ*DROSHA*::*egfp*
HEK293T-Δ*DROSHA*::*egfp*

HEK293T-t/g-Δ*DROSHA*::*egfp*

HEK293T-Δ*DROSHA*::*egfp*

**pX459M-cas9-Δ*Vegfa*::*egfp* (without mCherry)**

Bright field EGFP

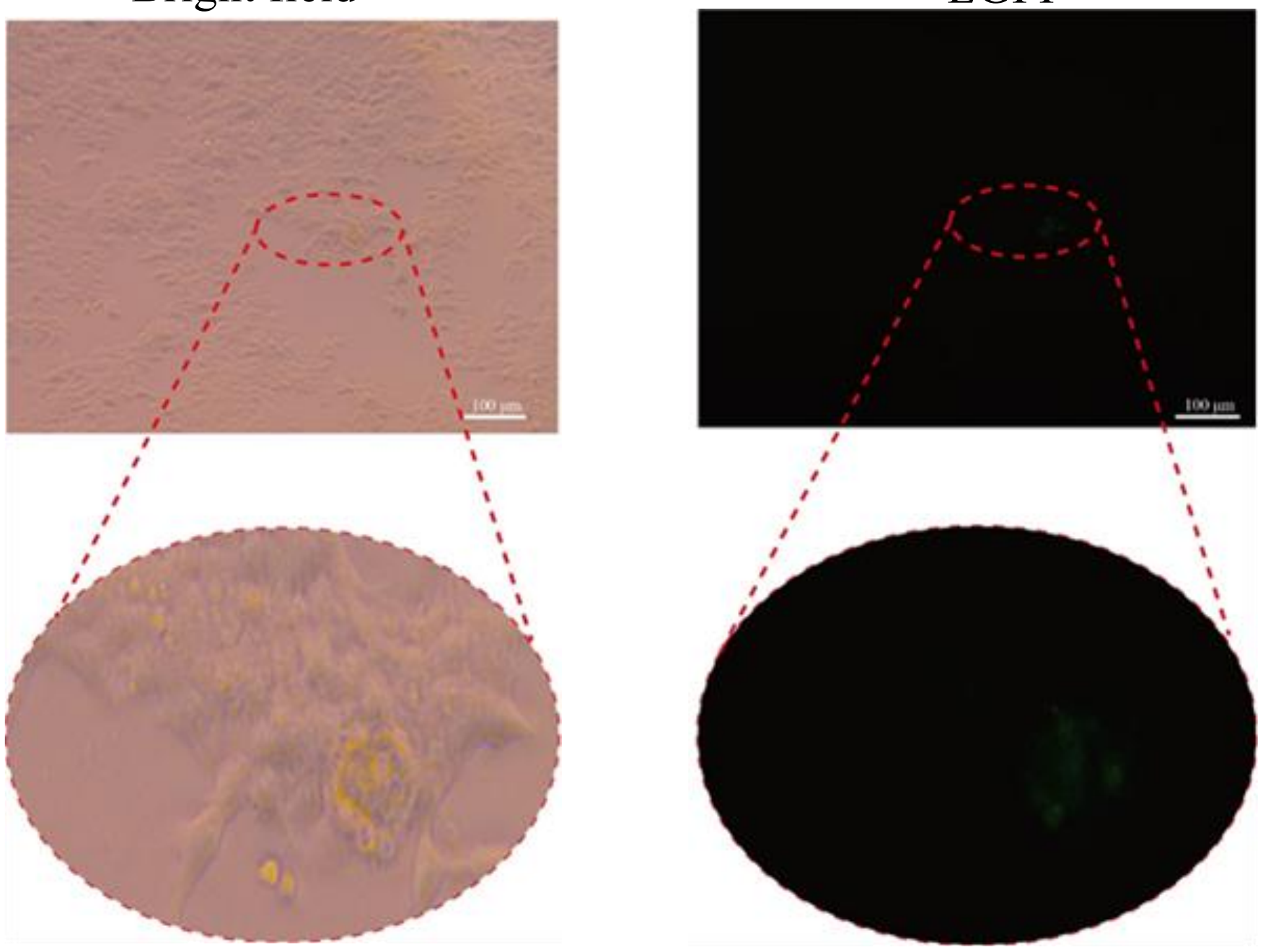

NIH-3T3
NIH-3T3-t/g-Δ*Vegfa*::*egfp*
NIH-3T3-Δ*Vegfa*::*egfp*

NIH-3T3-Δ*Vegfa*::*egfp*

C

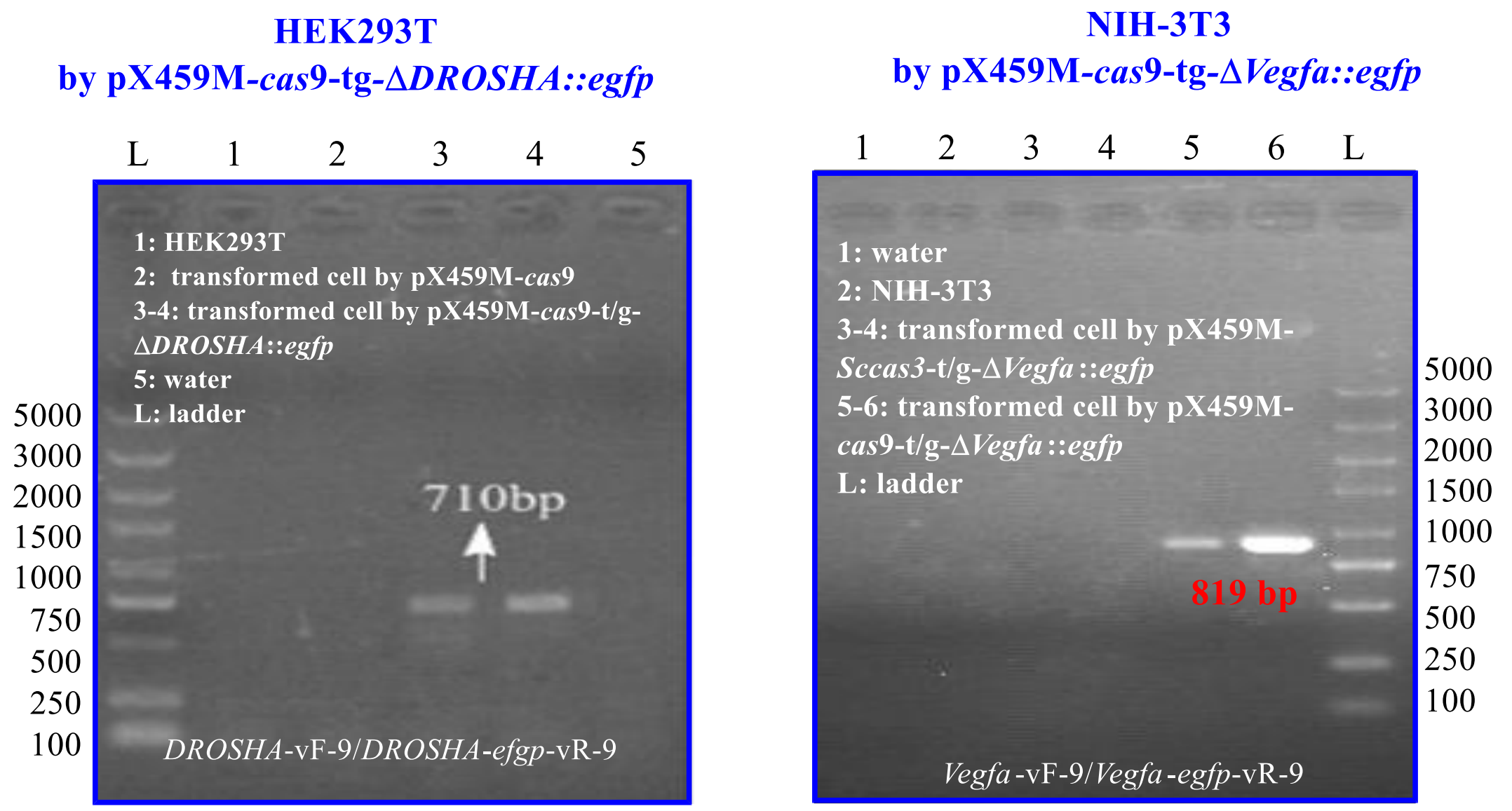

***Sccas3*: codon optimized gene *Svicas3* based on the base preference of *Saccharomyces cerevisiae***

D

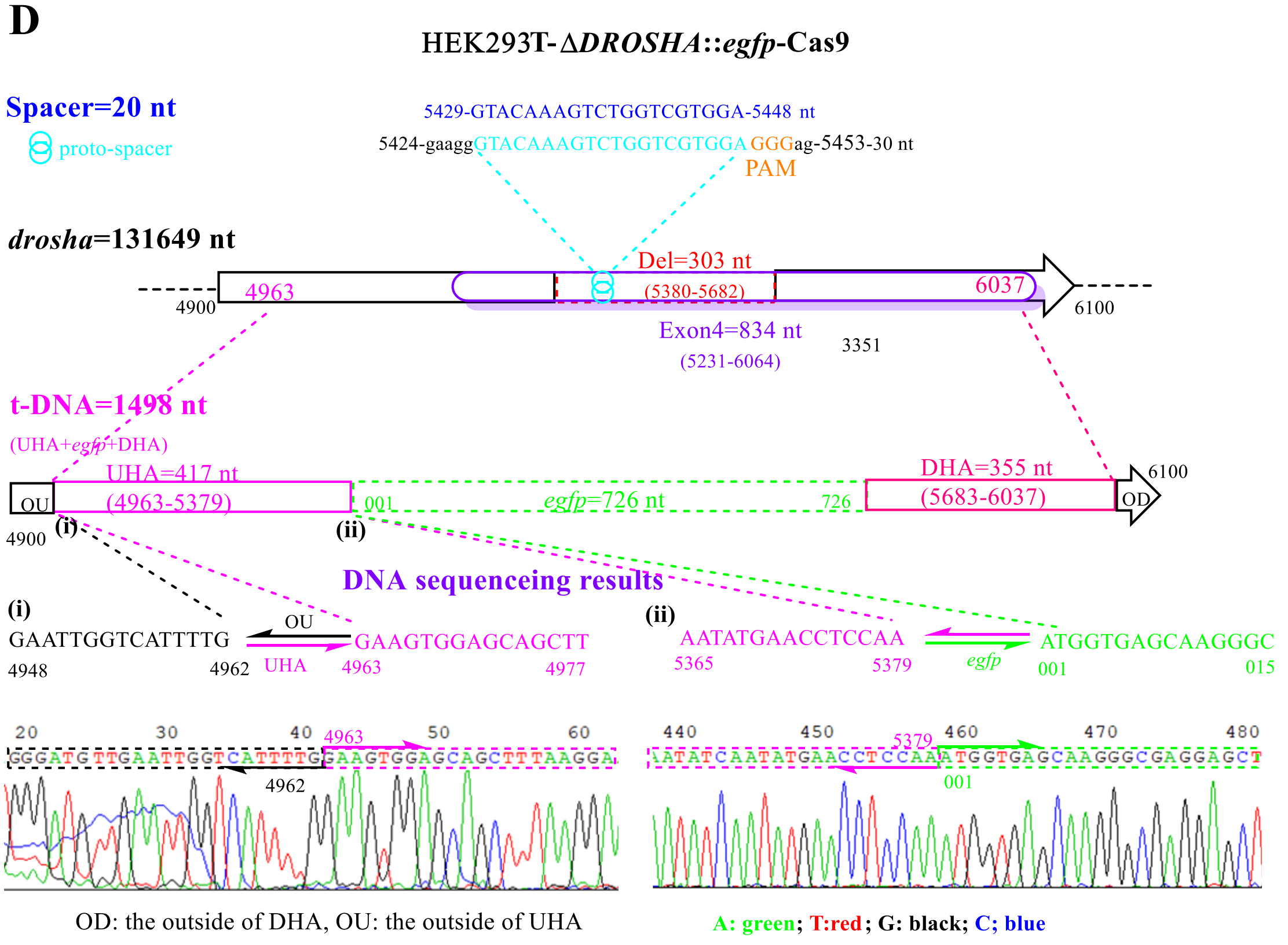

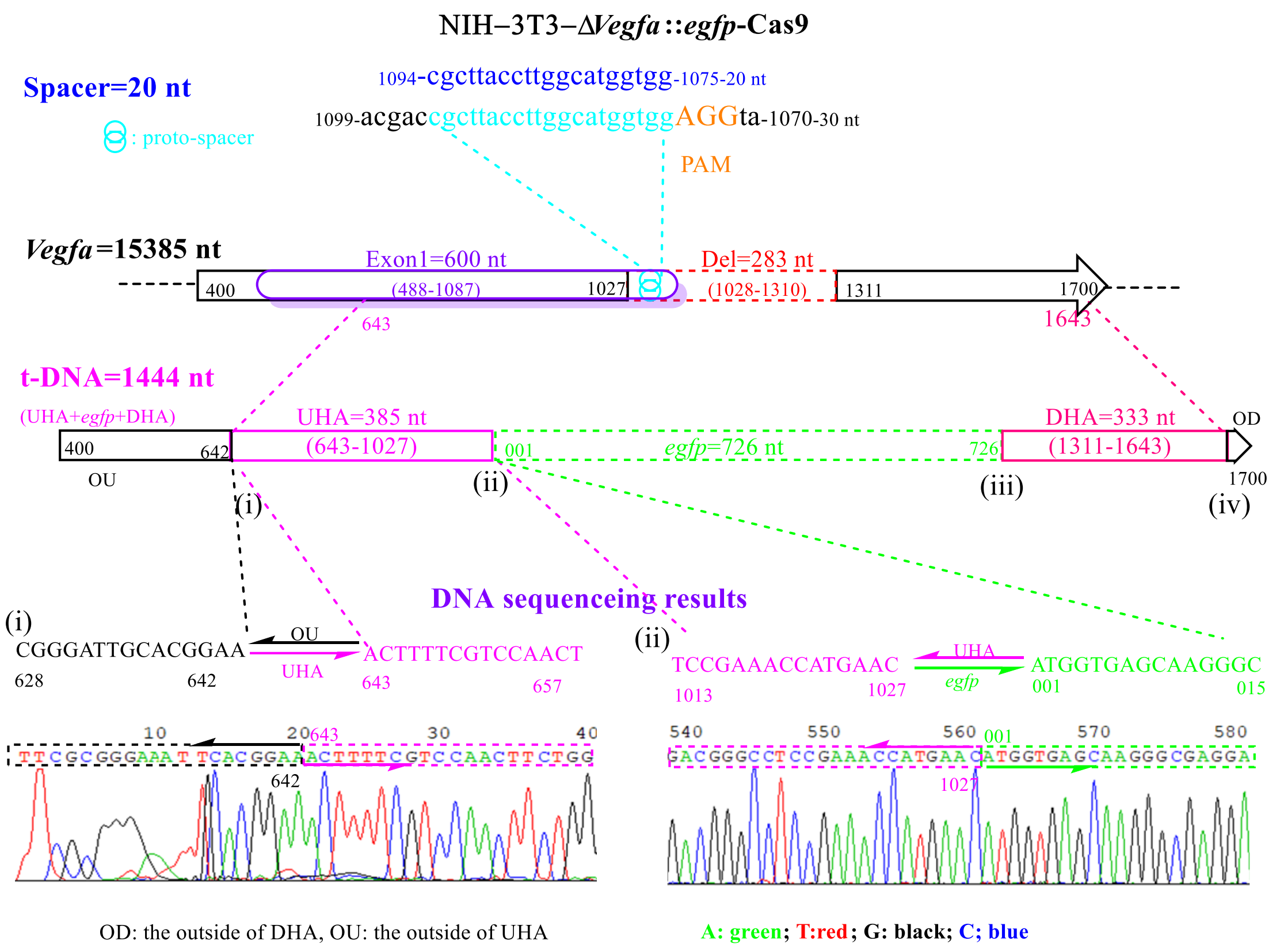

**Figure S7. Construction of Cas9-based gene editing tools and mutant verification in mammalian cell genome editing mediated by Cas9. (A)** The construction of Cas9-based genome editing tools (The original plasmid pX459M-*cas*9, the two constructed Cas9-based gene editing plasmids pX459M-*cas*9-tg-Δ*DROSHA*::*egfp* with mCherry gene and pX459M-*cas*9-tg-Δ*Vegfa*::*egfp* without mCherry gene as well as their maps). **(B)** Typical micrographs of *Sp*Cas9-based genome editing process in mammal cells. **(C)** The results of DNA gel electrophoresis of the PCR product of target sequences in the Cas9-based mammalian cell genome editing by pX459M-*cas*9-tg-Δ*DROSHA*::*egfp* and pX459M-*cas*9-tg-Δ*Vegfa*::*egfp*, respectively. **(D)** The engineered t-DNA structure diagrams and the results of DNA sequencing of the PCR product of target sequences in the Cas9-based mammalian

cell genome editing by pX459M-*cas*9-tg-Δ*DROSHA*::*egfp* and pX459M-*cas*9-tg-Δ*Vegfa*::*egfp* (PCR products only occur in gene-edited variants HEK293T-Δ*DROSHA*::*egfp*-9 and NIH-3T3-Δ*Vegfa*::*egfp*-9 because the primers *DROSHA*-*egfp*-vR-9 and *Vegfa*-*egfp*-vR-9 are a portion of gene *egfp* in t-Δ*DROSHA*::*egfp* and t-Δ*Vegfa*::*egfp* and the primers *DROSHA*-vF-9 and *Vegfa*-vF-9 are a portion of genes *DROSHA* and *Vegfa* that are not contained in t-Δ*DROSHA*::*egfp* and t-Δ*Vegfa*::*egfp*, respectively.).

## Tables S1 to S4

**Table S1.** Strains and plasmids used in this study

**Table S2.** Primers, g-DNAs and t-DNAs used in this study

**Table S3.** Genome editing efficiency in the template-based eukaryotic genome editing directed by the *Svi*Cas3

**Table S4.** The main features of typical gene editing techniques

**Table S1.** Strains and plasmids used in this study (http://www.addgene.org/; https://www.ncbi.nlm.nih.gov)

| Item | Description | Application | Source |
|---|---|---|---|
| **Strain** | | | |
| *E. coli* DH5α | $F^-$ , φ80d*lac*ZΔM15 , Δ(*lac*ZYA-*arg*F)U169 , *deo*R , *rec*A1 , *end*A1 , *hsd*R17(rk$^-$，mk$^+$)，*pho*A，*sup*E44，λ-， *thi*-1，*gyr*A96，*rel*A1 | | Our lab |
| *E. coli* DH5α-pRS415 | pRS415 | *S. cerevisiae* LYC4 | This study |
| *E. coli* DH5α-pRS415-*cas*3 | pRS415-*cas*3 | *S. cerevisiae* LYC4 | This study |
| *E. coli* DH5α-pRS415-*cas*7-5-3 | pRS415-*cas*7-5-3 | *S. cerevisiae* LYC4 | This study |
| *E. coli* DH5α-pYES2-NTA | pYES2-NTA | *S. cerevisiae* LYC4 | This study |
| *E. coli* DH5α-pYES2-NTA-t/g-Δ*crtE* | pYES2-NTA-t/g-Δ*crtE* | *S. cerevisiae* LYC4 | This study |
| *E. coli* DH5α-pYES2-NTA-t-Δ*crtE* | pYES2-NTA-t-Δ*crtE* | *S. cerevisiae* LYC4 | This study |
| *E. coli* DH5α-pYES2-NTA-g-Δ*crtE* | pYES2-NTA-g-*crtE* | *S. cerevisiae* LYC4 | This study |
| *E. coli* TOP10 | $F^-$, *mcr*AΔ(*mrr-hsd* RMS-*mcr*BC), *φ*80, *lac*ZΔM15, Δ*lac*X74, *rec*A1, *ara*Δ139Δ(*ara-leu*)7697, *gal*U, *gal*K, *rps*L, (*Strr*) *end*A1, *nup*G | HEK293T / NIH-3T3 | Our lab |
| *E. coli* TOP10-AIO-mCherry | AIO-mCherry | HEK293T / NIH-3T3 | This study |
| *E. coli* TOP10-AIO-mCherry-*cas*3 | AIO-mCherry-*cas*3 | HEK293T / NIH-3T3 | This study |
| *E. coli* TOP10-AIO-mCherry-t/g-Δ*DROSHA*::*egfp* | AIO-mCherry-t/g-Δ*DROSHA*::*egfp* | HEK293T | This study |
| *E. coli* TOP10-AIO-mCherry- *cas*3-t/g-Δ*DROSHA*::*egfp* | AIO-mCherry- *cas*3-t/g-Δ*DROSHA*::*egfp* | HEK293T | This study |
| *E. coli* TOP10-AIO-mCherry-t/g-Δ*CAMKMT*::*egfp* | AIO-mCherry-t/g-Δ*CAMKMT*::*egfp* | HEK293T | This study |
| *E. coli* TOP10-AIO-mCherry- *cas*3-t/g-Δ*CAMKMT*::*egfp* | AIO-mCherry-*cas*3-t/g-Δ*CAMKMT*::*egfp* | HEK293T | This study |
| *E. coli* TOP10-AIO-mCherry-t/g-Δ*Lepr*::*egfp* | AIO-mCherry-t/g-Δ *Lepr*::*egfp* | NIH-3T3 | This study |
| *E. coli* TOP10-AIO-mCherry-*cas*3-t/g-Δ*Lepr*::*egfp* | AIO-mCherry-*cas*3-t/g-Δ*Lepr*::*egfp* | NIH-3T3 | This study |
| *E. coli* TOP10-AIO-mCherry-t-Δ*DROSHA*::*egfp* | AIO-mCherry-t-Δ*DROSHA*::*egfp* | HEK293T | This study |
| *E. coli* TOP10-AIO-mCherry- *cas*3-t-Δ*DROSHA*::*egfp* | AIO-mCherry-*cas*3-t-Δ*DROSHA*::*egfp* | HEK293T | This study |
| *E. coli* TOP10-AIO-mCherry- t-Δ*CAMKMT*::*egfp* | AIO-mCherry-t-Δ*CAMKMT*::*egfp* | HEK293T | This study |
| *E. coli* TOP10-AIO-mCherry- *cas*3-t-Δ*CAMKMT*::*egfp* | AIO-mCherry-*cas*3-t-Δ*CAMKMT*::*egfp* | HEK293T | This study |
| *E. coli* TOP10-AIO-mCherry-$t_{g1}$/$t_{b2}$-Δ*CAMKMT*::*egfp* | AIO-mCherry-$t_{g1}$/$t_{b2}$-Δ*CAMKMT*::*egfp* | HEK293T | This study |
| *E. coli* TOP10- AIO-mCherry-*cas*3-$t_{g1}$/$t_{b2}$-Δ*CAMKMT*::*egfp* | AIO-mCherry-*cas*3-$t_{g1}$/$t_{b2}$-Δ*CAMKMT*::*egfp* | HEK293T | This study |
| *E. coli* TOP10-AIO-mCherry-t-Δ*Lepr*::*egfp* | AIO-mCherry-t-Δ *Lepr*::*egfp* | NIH-3T3 | This study |
| *E. coli* TOP10-AIO-mCherry-*cas*3-t-Δ*Lepr*::*egfp* | AIO-mCherry-*cas*3-t-Δ*Lepr*::*egfp* | NIH-3T3 | This study |
| *E. coli* TOP10-pX459M-*cas*9 | pX459M-*cas*9 | HEK293T / NIH-3T3 | This study |
| *E. coli* TOP10-pX459M-*cas*9-t/g-Δ*DROSHA*::*egfp*-9 | pX459M-*cas*9-t/g-Δ*DROSHA*::*egfp*-9 | HEK293T | This study |
| *E. coli* TOP10-pX459M-*cas*9-t/g-Δ*Vegfa*::*egfp*-9 | pX459M-*cas*9-t/g-Δ*Vegfa*::*egfp*-9 | NIH-3T3 | This study |
| human embryonic kidney 293 /HEK293T | wild-type | HEK293T | Our lab |
| HEK293T-AIO-mCherry | AIO-mCherry | HEK293T | This study |
| HEK293T-AIO-mCherry-*cas*3 | AIO-mCherry-*cas*3 | HEK293T | This study |
| HEK293T-AIO-mCherry-t/g-Δ*DROSHA*::*egfp* | AIO-mCherry-t/g-Δ*DROSHA*::*egfp* | HEK293T | This study |
| HEK293T-AIO-mCherry-t-Δ*DROSHA*::*egfp* | AIO-mCherry-t-Δ*DROSHA*::*egfp* | HEK293T | This study |
| HEK293T-Δ*DROSHA*::*egfp* | Δ*DROSHA*::*egfp* | HEK293T | This study |
| HEK293T-pX459M-Cas9-t/g-Δ*DROSHA*::*egfp*-9 | pX459M-Cas9-t/g-Δ*DROSHA*::*egfp*-9 | HEK293T | This study |
| HEK293T-Δ*DROSHA*::*egfp*-Cas9 | Δ*DROSHA*::*egfp*-9 | HEK293T | This study |
| HEK293T-AIO-mCherry-t/g-Δ*CAMKMT*::*egfp* | AIO-mCherry-t/g-Δ*CAMKMT*::*egfp* | HEK293T | This study |
| HEK293T-AIO-mCherry-t-Δ*CAMKMT*::*egfp* | AIO-mCherry-t-Δ*CAMKMT*::*egfp* | HEK293T | This study |
| HEK293T-AIO-mCherry-$t_{g1}$/$t_{b2}$-Δ*CAMKMT*::*egfp* | AIO-mCherry-$t_{g1}$/$t_{b2}$-Δ*CAMKMT*::*egfp* | HEK293T | This study |
| HEK293T-Δ*CAMKMT*::*egfp* | Δ*CAMKMT*::*egfp* | HEK293T | This study |
| HEK293T-Δ*CAMKMT*::*egfp*-g1/b2 | Δ*CAMKMT*::*egfp*-g1/b2 | HEK293T | This study |
| NIH-3T3 mouse fetus fibroblast cells/NIH-3T3 | wild | NIH-3T3 | Our lab |
| NIH-3T3-AIO-mCherry | AIO-mCherry | NIH-3T3 | This study |
| NIH-3T3-AIO-mCherry-*cas*3 | AIO-mCherry-*cas*3 | NIH-3T3 | This study |
| NIH-3T3-AIO-mCherry-t/g-Δ*Lepr*::*egfp* | AIO-mCherry-t/g-Δ*Lepr*::*egfp* | NIH-3T3 | This study |
| NIH-3T3-AIO-mCherry-t-Δ*Lepr*::*egfp* | AIO-mCherry-t-Δ*Lepr*::*egfp* | NIH-3T3 | This study |
| NIH-3T3-Δ*Lepr*::*egfp* | Δ*Lepr*::*egfp* | NIH-3T3 | This study |
| NIH3T3-pX459M-Cas9-t/g-Δ*Vegfa*::*egfp*-9 | pX459M-Cas9-t/g-Δ*Vegfa*::*egfp*-9 | NIH3T3 | This study |
| NIH-3T3-Δ*Vegfa*::*egfp*-Cas9 | Δ*Vegfa*::*egfp*-9 | NIH-3T3 | This study |
| *S. cerevisiae* LYC4 | carrying genes *crtE*, *crtB*, *crtB* and *zds* in chromosome IV in *S. cerevisiae* S288C | *S. cerevisiae* LYC4 | Our lab |
| *S. cerevisiae* LYC4-Δ*crtE* | Δ*crtE* | *S. cerevisiae* LYC4 | This study |
| *S. cerevisiae* LYC4-pYES2-NTA-t/g-Δ*crtE* | pYES2-NTA-t/g-Δ*crtE* | *S. cerevisiae* LYC4 | This study |
| *S. cerevisiae* LYC4-pRS415-*cas*3 | pRS415-*cas*3 | *S. cerevisiae* LYC4 | This study |
| *S. cerevisiae* LYC4-pRS415-*cas*7-5-3 | pRS415-*cas*7-5-3 | *S. cerevisiae* LYC4 | This study |
| *S. cerevisiae* LYC4-pYES2-NTA-t-Δ*crtE* | pYES2-NTA-t-Δ*crtE* | *S. cerevisiae* LYC4 | This study |
| *S. cerevisiae* LYC4-pYES2-NTA-g-Δ*crtE* | pYES2-NTA-g-Δ*crtE* | *S. cerevisiae* LYC4 | This study |
| **Plasmid** | | | |
| AIO-mCherry (9728 6bp) | Mammalian Expression (high copy), CRISPR, AmpR, f1 ori, U6 promoter, mCherry, Cas9n | HEK293T / NIH-3T3 | Addgene |
| AIO-mCherry-*cas*3 | *cas*3 | HEK293T / NIH-3T3 | This study |
| AIO-mCherry-t/g-Δ*DROSHA*::*egfp* | t/g-Δ*DROSHA*::*egfp* | HEK293T | This study |
| AIO-mCherry-t-Δ*DROSHA*::*egfp* | t-Δ*DROSHA*::*egfp* | HEK293T | This study |
| AIO-mCherry-*cas*3-t/g-Δ*DROSHA*::*egfp* | *cas*3-t/g-Δ*DROSHA*::*egfp* | HEK293T | This study |
| AIO-mCherry-*cas*3-t-Δ*DROSHA*::*egfp* | *cas*3-t-Δ*DROSHA*::*egfp* | HEK293T | This study |
| AIO-mCherry-t/g-Δ*CAMKMT*::*egfp* | t/g-Δ*CAMKMT*::*egfp* | HEK293T | This study |
| AIO-mCherry-t-Δ*CAMKMT*::*egfp* | t/g-Δ*CAMKMT*::*egfp* | HEK293T | This study |
| AIO-mCherry-$t_{g1}$/$t_{b2}$-Δ*CAMKMT*::*egfp* | $t_{g1}$/$t_{b2}$-Δ*CAMKMT*::*egfp* | HEK293T | This study |
| AIO-mCherry-*cas*3-t/g-Δ*CAMKMT*::*egfp* | *cas*3-t/g-Δ*CAMKMT*::*egfp* | HEK293T | This study |
| AIO-mCherry-*cas*3-t-Δ*CAMKMT*::*egfp* | *cas*3-t-Δ*CAMKMT*::*egfp* | HEK293T | This study |
| AIO-mCherry-*cas*3-$t_{g1}$/$t_{b2}$-Δ*CAMKMT*::*egfp* | *cas*3-$t_{g1}$/$t_{b2}$-Δ*CAMKMT*::*egfp* | HEK293T | This study |
| AIO-mCherry-t/g-Δ*Lepr*::*egfp* | t/g-Δ*Lepr*::*egfp* | NIH-3T3 | This study |
| AIO-mCherry-t-Δ*Lepr*::*egfp* | t-Δ*Lepr*::*egfp* | NIH-3T3 | This study |
| AIO-mCherry-*cas*3-t/g-Δ*Lepr*::*egfp* | *cas*3-t/g-Δ*Lepr*::*egfp* | NIH-3T3 | This study |
| AIO-mCherry-*cas*3-t-Δ*Lepr*::*egfp* | *cas*3-t-Δ*Lepr*::*egfp* | NIH-3T3 | This study |
| pCas(12545 bp) | *cas*9, lambda-Red recombinase expression plasmid, $Kan^R$, ParaB promoter | *E. coli* JM109 (DE3) | Addgene |
| pRS415 (rooted from pRS415_pGal-nCas9,10793 bp) | Yeast Expression, $Leu2^+$, , $Amp^R$, *cas*9 | *S. cerevisiae* LYC4 | Addgene |
| pRS415-*cas*3 | *cas*3 | *S. cerevisiae* LYC4 | This study |
| pRS415-*cas*7-5-3 | *cas*7-5-3 | *S. cerevisiae* LYC4 | This study |
| pX459M-*cas*9 (10176 bp) | codon optimized *cas*9, without *mCheery* gene | HEK293T / NIH-3T3 | This study |
| pX459M-*cas*9-t/g-Δ*DROSHA*::*egfp* | t/g-Δ*DROSHA*::*egfp*-9, with *mCheery* gene | HEK293T | This study |
| pX459M-*cas*9-t/g-Δ*Vegfa*::*egfp* | t/g-Δ*Vegfa*::*egfp*-9, without *mCheery* gene | NIH-3T3 | This study |
| pYES2-NTA (6038 bp) | Galactose-inducible expression plasmid in Yeast, $Ura3^+$, $Amp^R$, *Dcr*1 | *S. cerevisiae* LYC4 | Addgene |
| pYES2-NTA-t/g-Δ*crtE* | t/g-Δ*crtE* | *S. cerevisiae* LYC4 | This study |
| pYES2-NTA-t-Δ*crtE* | t-Δ*crtE* | *S. cerevisiae* LYC4 | This study |

**Table S2.** Primers, g-DNAs and t-DNAs used in this study*

| Items | Sequence (5´ to 3´) | Note |
|---|---|---|
| **Primers** | | |
| ***S. cerevisiae* LYC4** | | |
| *crtE*-F | tccattggagtttactccacaag | *crtE* |
| *crtE*-R | ttttccctgtacgaccttcggatt | *crtE* |
| *crtE*-UF | cccAAGCTTcagccattccattggagtttactcc | UHA |
| *crtE*-UR | aacggcaatacgaaacaaaccaccaactggcataggaataggtgttgga c | UHA |
| *crtE*-DF | gtccaacacctattcctatgccagttggtggtttgtttcgtattgccgtt | DHA |
| *crtE*-DR | cgcGGATCCttttccctgtacgaccttcggatt | DHA |
| *crtE*-vF | cagcatacacctcactagggtag | Δ*crtE* |
| *crtE*-vR | ccccttccagtgcattatgcaa | Δ*crtE* |
| *cas*3-Fsc | aaaaaaatgggtagattggatgccgttg | codon *cas*3 |
| *cas*3-Rsc | gtcagccctgcttaaaacttcaccggccctatatgcag | codon optimized *cas*3 |
| pRS415-Fsc | tatagggccggtgaagttttaagcagggctgaccccaag | pRS415 without *cas*9 |
| pRS415-Rsc | caacggcatccaatctacccatttttttcccgggggatccactagttc | pRS415 without *cas*9 |
| pRS415-*cas*3-Fsc | ccggattctagaactagtggatc | pRS415 with codon optimized *cas*3 |
| pRS415-*cas*3-Rsc | cttttcggttagagcggatgtgg | pRS415 with codon optimized *cas*3 |
| *cas*7-Fsc | ataaaacatcatcacagaattcatggttgccggtgcacc | codon optimized *cas*7 |
| *cas*7-Rsc | aaaccttggatggatcacaccttcctcttcttcttgggg | codon optimized *cas*7 |
| pRS415-*cas*7-Fsc | gaagaagaggaaggtgtgatccatccaaggtttcaaggccg | pRS415 with codon optimized *cas*7 |
| pRS415-*cas*7-Rsc | ggcaaccatgaattctgtgatgatgttttatttgttttgattgg | pRS415 with codon optimized *cas*7 |
| *cas*5-Fsc | atacaataatagaattcatgacaggtactgaagttactgcc | codon optimized *cas*5 |
| *cas*5-Rsc | aattcttagttaaaagcacttcacaccttcctcttcttcttgggg | codon optimized *cas*5 |
| pRS415-*cas*5-Fsc | aagaggaaggtgtgaagtgcttttaactaagaattattagtcttttctgc | pRS415-*cas*7-3 |
| pRS415-*cas*5-Rsc | tcagtacctgtcatgaattctattattgtatgttatagtattagttgcttgg | pRS415-*cas*7-3 |
| pRS415-*cas*7-5-Fsc | tgattgtttatttacaggatcccctctccccgcgcgttg | pRS415 with codon optimized *cas*7-5 |
| pRS415-*cas*7-5-Rsc | gtagtggaatcaaggatccgcggtttgcgtattgggcg | pRS415 with codon optimized *cas*7-5 |
| pRS415-*cas*7-5-3-Fsc | gctccggtgcccgtcagtggatccttgattccactacagttac | pRS415-*cas*7-5-3 |
| pRS415-*cas*7-5-3-Rsc | gacgctttgagccgccaccactagtgcaaattaaagccttcgagcgtcc | pRS415-*cas*7-5-3 |
| *Sc*-*crtE*-ot-F | ctccgt+TTC+tagtgc (xN+PAM+seed: 6+3+6=15 nt ) | number of potential off-target site: 87 |
| *Sc*-*crtE*-ot01-F/R | cctatcgaacagaagattagactac / cagccgtatgttcatcg | ChrII, off-target analysis |
| *Sc*-*crtE*-ot02-F/R | tttactacccatgcgacttgtc / gtacaaatagctacaccac | ChrII, off-target analysis |
| *Sc*-*crtE*-ot03-F/R | gttgatgatgggtcactattgaag / ggatttgcattgttcaatcc | ChrVII, off-target analysis |
| *Sc*-*crtE*-ot04-F/R | ctccgtttctagtgc / cacacttaaacgtatacgag | ChrXVI, off-target analysis |
| **HEK293T** | | |
| AIO-mCherry-F | cccAAGCTTggctccggagccacgaacttc | AIO-Mcherry |
| AIO-mCherry-R | gcTCTAGAgccatttgtctgcagaattgg | AIO-Mcherry |
| g-*DROSHA*-F | aggaattcggcaggaagagtctttgaaaagataatgtatg | g-*DROSHA* |
| g-*DROSHA*-R | cgcGGATCCaagggccctctagactcgagagac | g-*DROSHA* |
| *DROSHA*-vF | ctgtttttaccatgaaggaaacgagagtatgaag | Δ*DROSHA-egfp* |
| *DROSHA*-vR | catttgctcttttcctagtatttcctatctc | Δ*DROSHA-egfp* |
| *DROSHA*-*egfp*-vF | cacaacgtctatatcatggccgacaagc | *egfp* |
| *DROSHA*-*egfp*-vR | acttgaagaagtcgtgctgcttcatgtggtcg | *egfp* |
| *DROSHA*-UF | cgcGGATCCatgcctataagtctttttagctgctg | UHA |
| *DROSHA*-U*egfp*-R | ctcgcccttgctcaccatcatgatgttccgcctggatatgtcac | UHA-*egfp* |
| *DROSHA*-U*egfp*-F | gtgacatatccaggcggaacatcatgatggtgagcaagggcgag | UHA-*egfp* |
| *DROSHA*-D*egfp*-R | ccaccttctaaggacctaattccacctttaatttagaattccttgtacagctc | DHA-*egfp* |
| *DROSHA*-D*egfp*-F | gagctgtacaaggaattctaaattaaaggtggaattaggtccttagaaggt gg | DHA-*egfp* |
| *DROSHA*-DR | ctcttcctgccgaattcctgacaccattaatagtccttaagacatgtaacagg | DHA |
| *DROSHA*-*egfp*-vF$_a$ | gcaccatcttcttcaaggacga | *egfp* |
| HEK293T-*DROSHA*-ot-F | tgctca+TTC+tttcaacagtg (xN+PAM+seed: 6+3+11=20 nt ) | number of potential off-target site: 315 |
| HEK293T-*DROSHA*-ot01-F/R | agatgtagcctcactcctgagatc / cattaagagctcaggtcaagtgtatg | ChrVI, off-target analysis |
| HEK293T-*DROSHA*-ot02-F/R | gatattgattcttccaagtcatgagc / ccacatacacgaatcaataagtg | ChrV, off-target analysis |
| HEK293T-*DROSHA*-ot03-F/R | caaagccaggctttggtcatcatgc / gttgtattgtcctcagatggttc | ChrV, off-target analysis |
| HEK293T-*DROSHA*-ot04-F/R | tattggttccatatgaaatgtggaatag / ctccaaataataagagctagctatg | ChrV, off-target analysis |
| g-*DROSHA*-F-9 | gagggcctatttcccatgattc | g-*DROSHA*-9 |
| g-*DROSHA*-R-9 | ccggagccaagcttaaaaaaaagcaccgactcggtgccac | g-*DROSHA*-9 |
| *DROSHA*-vF-9 | tggtgaacaatgggatgttgaattggtc | Δ*DROSHA*-*egfp*-9 |
| *DROSHA*-vR-9 | gtttacctgctccgttcgtagc | Δ*DROSHA*-*egfp*-9 |
| *DROSHA*-UF-9 | ctcgaggaattcggcaggaagaagtggagcagctttaaggaatggtcg | UHA-9 |
| *DROSHA*-U*egfp*-R-9 | cgcccttgctcaccatttggaggttcatattgatattgcacagg | UHA-*egfp*-9 |
| *DROSHA*-U*egfp*-F-9 | cctgtgcaatatcaatatgaacctccaaatggtgagcaagggcg | UHA-*egfp*-9 |
| *DROSHA*-D*egfp*-R-9 | ctgataattaacctgctgcggcatgactgttagaattccttgtacag | DHA-*egfp*-9 |
| *DROSHA*-D*egfp*-F-9 | ctgtacaaggaattctaacagtcatgccgcagcaggttaattatcag | DHA-*egfp*-9 |
| *DROSHA*-DR-9 | tcatgggaaataggccctcgatggtgttctccctcggtcataatcag | DHA-9 |
| *DROSHA*-*egfp*-vF-9 | cacaacgtctatatcatggccgacaagc | *egfp* |
| *DROSHA*-*egfp*-vR-9 | cttgccggtggtgcagatgaacttc | *egfp* |
| pX459M-*cas*9-*DROSHA*-F | gctctagaagatttctttttcttagcttgaccagctttcttagtagcagcag | linearization of pX459M-*cas*9-F |
| pX459M-*cas*9-*DROSHA*-R | cgcggatcctgcatcgagactagataactgatctacccagctttcttgtac | linearization of pX459M-*cas*9-R |
| pX459M-*cas*9-*mCherry*-F | cgcggatccttagaattccttgtacagctcg | *mCherry* |
| pX459M-*cas*9-*mCherry*-R | gctctagatctctgttaaagcaagcaggag | *mCherry* |
| HEK293T-*DROSHA*-9-ot | seed+PAM=10+3=agactttgta+ TGG,<br>seed+PAM=12+3=ccagactttgta+AGG,<br>seed+PAM=13+3=tacaagactttgt+AGG,<br>seed+PAM=14+3=gaccagactttgta+GGG | Potential off-target sites: 8468, PAM: NGG |
| HEK293T-*DROSHA*-9-ot01-F/R | ccaataatgtcagttcagatattgtg / gttaacatttatatttggttaactag | ChrII, off-target analysis |
| HEK293T-*DROSHA*-9-ot02-F/R | cacagaacaggaaggatttcaatg / cgaaaatacaggtctccttg | ChrVIII, off-target analysis |
| HEK293T-*DROSHA*-9-ot03-F/R | caaggcaggcagagtatacaag / cccgctgacaccaccatcctc | ChrVIII, off-target analysis |
| HEK293T-*DROSHA*-9-ot04-F/R | ctcactcccaggaaatctgcc / ctttgaacctcattttccgtg | ChrX, off-target analysis |
| g-*CAMKMT*-F | ccatcataggagctatggactctttgaaaagataatgtatgattatg | g-*CAMKMT* |
| g-*CAMKMT*-R | cgGGATCCagacataaaaaacaaaaaaacccctcg | g-*CAMKMT* |
| *CAMKMT*-vF | ctgtagatggaatagcagacctgaag | Δ*CAMKMT::egfp* |
| *CAMKMT*-vR | gctcatatttcattgacaattcactatcttaagag | Δ*CAMKMT::egfp* |
| *CAMKMT*-*egfp*-vF | gatcactctcggcatggacgag | *egfp* |
| *CAMKMT*-UF | cgGGATCCtcagcttgattctctaag | UHA |
| *CAMKMT*-U*egfp*-R | tcgcccttgctcaccattccactattatgcctagaaaggag | UHA-*egfp* |
| *CAMKMT*-U*egfp*-F | atggtgagcaagggcgag | UHA-*egfp* |
| *CAMKMT*-D*egfp*-R | ttagaattccttgtacagctcgtcc | DHA-*egfp* |
| *CAMKMT*-D*egfp*-F | ctgtacaaggaattctaagagatggagagagtaggtcagaaag | DHA-*egfp* |
| *CAMKMT*-DR | cattatcttttcaaagagtccatagctcctatgatggtatg | DHA |
| *CAMKMT*-vF$_a$ | tgctgttatagagtaaatttatgtattgttgagtgtgg | Δ*CAMKMT::egfp* |
| *CAMKMT*-vR$_a$ | cctaaattcagccaaaagttaacagttctg | Δ*CAMKMT::egfp* |
| g-*CAMKMT*-exon1-vF | caccccccatacatacgcaaattgag | Δ*CAMKMT::egfp*-exon1 |
| g-*CAMKMT*-exon1-vR | cagagactggcgagaaggggaaaag | Δ*CAMKMT::egfp*-exon1 |
| g-*CAMKMT*-exon1-*egfp*-vR | gaacttgtggccgtttacgtcg | *egfp* |
| g-*CAMKMT*-exon1-UR | gctctagaggctctggacacaaaccagtg | g-exon1-UHA |
| g-*CAMKMT*-exon1-U*egfp*-R | aagcacagaaggctaaaacaatggtgagcaagggcgaggag | g-exon1-UHA-*egfp* |
| g-*CAMKMT*-exon1-U*egfp*-F | tcgcccttgctcaccattgttttagccttctgtgcttcatctc | g-exon1-UHA-*egfp* |
| g-*CAMKMT*-exon1-D*egfp*-R | tggacgagctgtacaagtaagtaagggagaacctgctcgcc | g-exon1-DHA-*egfp* |
| g-*CAMKMT*-exon1-D*egfp*-F | gcgagcaggttctcccttacttacttgtacagctcgtccatgcc | g-exon1-DHA-*egfp* |
| b-*CAMKMT*-exon1-exon2-R | gaagtagaagaacccccttctaaatggttaccttGaTgagatactgtattc | g-b-exon1-exon2-DHA |
| b-*CAMKMT*-exon1-exon2-F | atctcAtCaaggtaaccatttagaaggggttcttctacttcgatctg | g-b-exon1-exon2-DHA |
| b-*CAMKMT*-exon2-F | cgcggatccatttcaaaggtActgaaAcaaaaacac | b-exon2 |
| b-*CAMKMT*-exon2-F’ | cgcggatccatttcaaaggtTctgaaGcaaaaacac | b-exon2 |
| b-*CAMKMT*-exon2-vF | tatttggctttatttcaaaggtActgaaAc | b-exon2 |
| b-*CAMKMT*-exon2-vR | atacagcaagacccagttaaaatagcag | b-exon2 |
| b-*CAMKMT*-exon2-vF$_a$ | gtctttgctctctgatgtcttcaagtaatg | b-exon2 |
| b-*CAMKMT*-exon2-vR$_a$ | tagaataatggttaccttGaTgagatactg | b-exon2 |
| HEK293T-*CAMKMT*-ot-F | ataatagtgga+TCC+ttgaatgttgaa (xN+PAM+seed: 6+3+11=20 nt ) | number of potential off-target site: 11 |
| HEK293T-*CAMKMT*-ot01-F/R | ataatagtggaTCCttgaatgttgaa / cctttgaggtagctgtctatgaaag | ChrI, off-target analysis |
| HEK293T-*CAMKMT*-ot02-F/R | ataatagtggaTCCttgaatgttgaa / gtacaatagtatccacagtgcac | ChrI, off-target analysis |
| HEK293T-*CAMKMT*-ot03-F/R | ataatagtggaTCCttgaatgttgaa / agaccaaaaagactgactgttatg | ChrI, off-target analysis |
| HEK293T-*CAMKMT*-ot04-F/R | ataatagtggaTCCttgaatgttgaa / cattctgtatccttgagtaggg | ChrIII, off-target analysis |
| **NIH-3T3** | | |
| g-*Lepr*-F | caatacatgttatgctgagtgatctttgaaaagataatgtatgattatg | g-*Lepr* |
| g-*Lepr*-R | cgGGATCCagacataaaaaacaaaaaaacccctcg | g-*Lepr* |
| *Lepr*-vF | tgaggaaaattgatgcaaggatttccag | Δ*Lepr*::*egfp* |
| *Lepr*-vR | catctaattgactcaattaagtcttctctgtcc | Δ*Lepr*::*egfp* |
| *Lepr*-*egfp*-vR | gaacttgtggccgtttacgtcg (reverse complement) | *egfp* |
| *Lepr*-*egfp*-vR$_a$ | cttgccggtggtgcagatgaacttc (reverse complement) | *egfp* |
| *Lepr*-UF | cgggatccgaaggtagacgctcagggttg | UHA |
| *Lepr*-U*egfp*-R | tcctcgcccttgctcaccatgataccacttgaattaaatttag | UHA-*egfp* |
| *Lepr*-U*egfp*-F | atggtgagcaagggcgag | UHA-*egfp* |
| *Lepr*-D*egfp*-R | ttagaattccttgtacagctcgtcc | DHA-*egfp* |
| *Lepr*-D*egfp*-F | ggacgagctgtacaaggaattctaaactgaagggaagacactggc | DHA-*egfp* |
| *Lepr*-DR | atcatacattatcttttcaaagatcactcagcataacatgtattgattg | DHA |
| NIH-3T3-*Lepr*-ot-F | gtggtatc+TAC+gttcctgagt (xN+PAM+seed: 8+3+10=21nt) | number of potential off-target site: 17 |
| NIH-3T3-*Lepr*-ot01-F/R | ctgcccacctcttaccatag/agtggtagcacaatccttaaatc | ChrI, off-target analysis |
| NIH-3T3-*Lepr*-ot02-F/R | gatgaagggtgtgaatggtg/catgatagaaagacacgtggaatac | ChrI, off-target analysis |
| NIH-3T3-*Lepr*-ot03-F/R | caccttttacatcagctacgg/gtatgctaggtgctgttgaag | ChrI, off-target analysis |
| NIH-3T3-*Lepr*-ot04-F/R | aagtagacatggtggcagag/gccatggaagtgttagctttc | ChrII, off-target analysis |
| g-*Vegfa*-F-9 | gagggcctatttcccatgattc | g-*Vegfa*-9 |
| g-*Vegfa*-R-9 | ccggagccaagcttaaaaaaaagcaccgactcggtgccac | g-*Vegfa*-9 |
| *Vegfa*-vF-9 | ttgcatcggaccagtcgcgctgacggac | Δ*Vegfa*::*egfp*-9 |
| *Vegfa*-vR-9 | gtgcaggtggcccaacaagctagagcggtg | Δ*Vegfa*::*egfp*-9 |
| *Vegfa*-UF-9 | gttggctccgaatttctcgacttttcgtccaacttctgggctcttctcgctc | UHA-9 |
| *Vegfa*-U*egfp*-R-9 | cgcccttgctcaccatgttcatggtttcggaggccc | UHA-*egfp*-9 |
| *Vegfa*-U*egfp*-F-9 | gggcctccgaaaccatgaacatggtgagcaagggcg | UHA-*egfp*-9 |
| *Vegfa*-D*egfp*-R-9 | gatcgtacgtgcggtgactctggtgggtggttagaattccttgtacag | DHA-*egfp*-9 |
| *Vegfa*-D*egfp*-F-9 | ctgtacaaggaattctaaccacccaccagagtcaccgcacgtacgatc | DHA-*egfp*-9 |
| *Vegfa*-DR-9 | ctcttcctgccgaattcctcagcaatccatcctaaaactttccccaaactcactgc | DHA-9 |
| *Vegfa*-*egfp*-*v*R-9 | acttgaagaagtcgtgctgcttcatgtggtcg | *egfp*-9 |
| pX459M-*cas*9-F | gaccttccgcttcttctttggcatgg | linearization of pX459M-*cas*9-F |
| pX459M-*cas*9-R | actagtggtggcggctcaaagcgtc | linearization of pX459M-*cas*9-R |
| pX459M-Cas9-t/g-Δ*Vegfa*::*egfp*-F | ttcctgccgaattcctcgag | linearization of pX459M-Cas9-t/g-Δ*Vegfa*::*egfp*-F |
| pX459M-Cas9-t/g-Δ*Vegfa*::*egfp*-R | tttttttaagcttggctccgg | linearization of pX459M-Cas9-t/g-Δ*Vegfa*::*egfp*–R |
| NIH3T3-*Vegfa*-ot-F | atgaact+TTC+tgctctcttg (xN+PAM+seed: 7+3+10=20 nt) | number of potential off-target site: 341 |
| NIH3T3-*Vegfa*-ot01-F/R | ggattacctcctggtatgttattc / cacacaaactccaaacaaaagac | ChrXVII, off-target analysis |
| NIH3T3-*Vegfa*-ot02-F/R | gccattcagatatactgctttctag / cactaaccaggaagaaatgtactg | ChrXV, off-target analysis |
| NIH3T3-*Vegfa*-ot03-F/R | cagccctgttctgctaatctc / gccagatgtggtaacacatatg | ChrXV, off-target analysis |
| NIH3T3-*Vegfa*-ot04-F/R | gtaacttgcggattattgactcg / tagaacagcctcagactttagc | ChrXVII, off-target analysis |
| **g-DNA** | | |
| g-*crtE* (389+17＝406 bp)<br>snR52 promoter-269 bp<br>SUP4 terminator-20 bp | cgGAATTCtctttgaaaagataatgtatgattatgctttcactcatatttatacagaaacttgatgttttctttcgagtatatacaaggtgattacatgtacgtttgaagtacaactctagattttgtagtgccctcttgggctagcggtaaaggtgcgcatttttcacaccctacaatgttctgttcaaaagattttggtcaaacgctgtagaagtgaaagttggtgcgcatgtttcggcgttcgaaacttctccgcagtgaaagataaatgatc**gtcctcatcgccccttcgaggggtcgcaac***tagtgcctcatcttcttcttcagcatcctccgaaaatggt***gtcctcatcgccccttcgaggggtcgcaac**ttttttgtttttatgtctCTCGAGcgg | PAM: ttc<br>R: 30+30 nt<br>S: 40 nt (*crtE*=1131 nt) |
| g-*DROSHA* (389 bp)<br>snR52 promoter-269 bp<br>SUP4 terminator-20 bp | tctttgaaaagataatgtatgattatgctttcactcatatttatacagaaacttgatgttttctttcgagtatatacaaggtgattacatgtacgtttgaagtacaactctagattttgtagtgccctcttgggctagcggtaaaggtgcgcatttttcacaccctacaatgttctgttcaaaagattttggtcaaacgctgtagaagtgaaagttggtgcgcatgtttcggcgttcgaaacttctccgcagtgaaagataaatgatc**gtcctcatcgccccttcgaggggtcgcaac***tttcaacagtgggcaca* | PAM: ttc<br>R: 30+30 bp<br>S: 40 bp (outside of the exon4) |
| | *atacaatagagggccatgcccag***gtcctcatcgccccttcgaggggtc**<br>**gcaac**tttttttgtttttatgtct | |
| g-*DROSHA*-9 (352 bp)<br>U6 promoter-241+8 bp<br>*Sp* terminator-41 bp | gagggcctatttcccatgattccttcatatttgcatatacgatacaaggctgtt<br>agagagataattagaattaatttgactgtaaacacaaagatattagtacaaa<br>atacgtgacgtagaaagtaataatttcttgggtagtttgcagttttaaaattat<br>gttttaaaatggactatcatatgcttaccgtaacttgaaagtatttcgatttcttg<br>gctttatatatcttgtggaaaggac**gaaacacc***gtacaaagtctggtcgtg*<br>*ga***gttttagagctagaaatagcaagttaaaataaggctagtccg**ttat<br>caacttgaaaaagtggcaccgagtcggtgcttttttt | PAM: ggg<br>R: 42 bp<br>S: 20 bp |
| g-*CAMKMT* (359 bp)<br>snR52 promoter-269 bp<br>SUP4 terminator-20 bp | tctttgaaaagataatgtatgattatgctttcactcatatttatacagaaacttg<br>atgttttctttcgagtatatacaaggtgattacatgtacgtttgaagtacaactc<br>tagattttgtagtgccctcttgggctagcggtaaaggtgcgcatttttcaca<br>ccctacaatgttctgttcaaaagattttggtcaaacgctgtagaagtgaaagt<br>tggtgcgcatgtttcggcgttcgaaacttctccgcagtgaaagataaatgat<br>c**tcgcaac***ttgaatgttgaagatgtccttaccagctttgacaatacag***gt**<br>**cctcatcgccccttcgagggg**tttttttgtttttatgtct | PAM: tcc<br>R: 7+23 bp<br>S: 40 bp (Exon3=64+1 nt) |
| g-*Lepr* (359 bp)<br>snR52 promoter-269 bp<br>SUP4 terminator-20 bp | tctttgaaaagataatgtatgattatgctttcactcatatttatacagaaacttg<br>atgttttctttcgagtatatacaaggtgattacatgtacgtttgaagtacaactc<br>tagattttgtagtgccctcttgggctagcggtaaaggtgcgcatttttcaca<br>ccctacaatgttctgttcaaaagattttggtcaaacgctgtagaagtgaaagt<br>tggtgcgcatgtttcggcgttcgaaacttctccgcagtgaaagataaatgat<br>c**tcgcaac***gttcctgagttatccaaaacagtcttccactgttgctttg***gtcc**<br>**tcatcgccccttcgagggg**tttttttgtttttatgtct | PAM: tac<br>R: 7+23 bp<br>S: 40 bp (Exon2=328+2 nt) |
| g-*Vegfa*-9 (352 bp)<br>U6 promoter-241+8 bp<br>*Sp* terminator-41 bp | gagggcctatttcccatgattccttcatatttgcatatacgatacaaggctgtt<br>agagagataattagaattaatttgactgtaaacacaaagatattagtacaaa<br>atacgtgacgtagaaagtaataatttcttgggtagtttgcagttttaaaattat<br>gttttaaaatggactatcatatgcttaccgtaacttgaaagtatttcgatttcttg<br>gctttatatatcttgtggaaaggac**gaaacacc***cgcttaccttggcatggt*<br>*gg***gttttagagctagaaatagcaagttaaaataaggctagtccg**ttat<br>caacttgaaaaagtggcaccgagtcggtgcttttttt | PAM: ggg<br>R: 42 bp<br>S: 20 bp |

| **t-DNA** | | |
|---|---|---|
| t-Δ*crtE*-807bp<br>(*crtE*=**<u>1131</u>** nt)<br>(*Saccharomyces cerevisiae* S288C, chr IV~1.53 Mbp) | Cagccattccattggagtttactccacaagatgatattgttttgttggaacca<br>tatcattatttgggaaaaaatcctggtaaagaaattcgttcacaattgattgaa<br>gcctttaattattggttggatgttaaaaaagaggaccttgaagttattcaaaat<br>gttgttggtatgttgcatactgcttctttgctcatggatgatgttgaagatagtt<br>ccgttttacgtcggggttcaccagttgctcatttgatctacggtattccacaa<br>actattaatacagccaattatgtttactttttagcctatcaagaaatttttaaatt<br>gcgtccaacacctattcctatgccagtt<br>Ggtggtttgtttcgtattgccgttcgtttgatgatggccaaatctgaatgtgat<br>attgattttgttcaattagttaatttgattagtatctattttcaaattcgtgatgatt<br>atatgaatcttcaatctagtgaatatgcacataacaaaaatttgccgaggac<br>cttactgaaggtaaattttcctttccaactattcatagtattcatactaatccaa<br>gtagtcgtttagttattaatacattacagaaaaaatctacatccccagaaattt<br>tgcatcattgtgttaattatatgcgtactgaaacacattcctttgaatatactcg<br>tgaagttttgaatactttgtccggtgccttggaacgtgaattgggtcgtttgc<br>aagaagaatttgccgaagctaatagtcgtatggatttgggcgacgttgaat<br>ccgaaggtcgtacagggaaaa | UHA: 353 bp<br>DHA: 454 bp<br>Del (**<u>1131</u>**-879 nt): 252 bp |
| t-Δ*DROSHA*::*egfp*-1390 bp<br>(Exon3=**<u>66</u>** nt)<br>(Homo sapiens,<br>chrV~181.54 Mbp) | Atgcctataagtcttttagctgctgatagttttgagttggcaaaactagtaat<br>aatgacattttaaaatttaaggacattaaaaatttagttggaaactttaacaca<br>gttcaacatgcttttaattttgtttgtaggctaagtgtattggaaactacaaaat<br>tagtgttatggtaattgcagaaatattttgtagcattggcagtttgggaaacat<br>gtttatgtctgttttattctttctgtttttttagagcttatattctctgtggaagatgt<br>gacatatccaggcggaacatcatg<br>Atggtgagcaagggcgaggagctgttcaccggggtggtgcccatcctg<br>gtcgagctggacggcgacgtaaacggccacaagttcagcgtgtccggc<br>gagggcgagggcgatgccacctacggcaagctgaccctgaagttcatct<br>gcaccaccggcaagctgcccgtgccctggcccaccctcgtgaccaccct<br>gacctacggcgtgcagtgcttcagccgctaccccgaccacatgaagcag<br>cacgacttcttcaagtccgccatgcccgaaggctacgtccaggagcgcac<br>catcttcttcaaggacgacggcaactacaagacccgcgccgaggtgaag | UHA: 299 bp<br>*egfp*: 717+9 bp<br>DHA: 365 bp<br>Del (**<u>66</u>**-49+247 nt): 264 bp |
| | ttcgagggcgacaccctggtgaaccgcatcgagctgaagggcatcgact<br>tcaaggaggacggcaacatcctggggcacaagctggagtacaactacaa<br>cagccacaacgtctatatcatggccgacaagcagaagaacggcatcaag<br>gtgaacttcaagatccgccacaacatcgaggacggcagcgtgcagctcg<br>ccgaccactaccagcagaacacccccatcggcgacggccccgtgctgct<br>gcccgacaaccactacctgagcacccagtccgccctgagcaaagaccc<br>caacgagaagcgcgatcacatggtcctgctggagttcgtgaccgccgcc<br>gggatcactctcggcatggacgagctgtacaaggaattctaa<br>Attaaaggtggaattaggtccttagaaggtggattgtaattattgtaggacc<br>tgctaagaagcagcacagagtgttgtgagagcaaggaacctaacaaagc<br>cctagggtcagtgacagtggttgtagaaaggccagcaaggatattattata<br>atagccatgtaagaggtgatgatgtgaagtagatgaattcaggaaatattta<br>tgaagcagaaacaattggatttttgtttatttttattaagggatagttgccata<br>taagtgacagcaattttaaagtaaattttagctaaggatggggggtgaggg<br>aggacaagtcagatgattccctgttacatgtcttaaggactattaatggtgtc | |
| t-Δ*DROSHA*::*egfp*-9 (1498 bp)<br>(Exon4=**834** nt)<br>(Homo sapiens,<br>chrV~181.54 Mbp) | Gaagtggagcagctttaaggaatggtcggtccagggactaaacatggca<br>gtgggctgggtggccaagaaaggggaagaggagtttattgagcatgag<br>ggtttcagatgttagcaggaacccagtattaaatgggtggcctattccagtg<br>gcttgactagggggtctttgagttgctcatcaagatggtcaggcatttatga<br>aaccctgtttacatagtaagaattatttttaaaaaaacttttcccttttttcttttct<br>gccatgaagtcacagaatgtcgttccacccgggacgagggtgtccccga<br>ggacgaggaggacatggagccagaccctcagcaccatcctttaggcccc<br>aaaatctgaggctgcttcaccctcagcagcctcctgtgcaatatcaatatga<br>acctccaa<br>Atggtgagcaagggcgaggagctgttcaccggggtggtgcccatcctg<br>gtcgagctggacggcgacgtaaacggccacaagttcagcgtgtccggc<br>gagggcgagggcgatgccacctacggcaagctgaccctgaagttcatct<br>gcaccaccggcaagctgcccgtgccctggcccaccctcgtgaccaccct<br>gacctacggcgtgcagtgcttcagccgctaccccgaccacatgaagcag<br>cacgacttcttcaagtccgccatgcccgaaggctacgtccaggagcgcac<br>catcttcttcaaggacgacggcaactacaagacccgcgccgaggtgaag | UHA: 417 bp<br>*egfp*: 717+9 bp<br>DHA: 355 bp<br>Del (**834**-159-382): 303 bp |
| | ttcgagggcgacaccctggtgaaccgcatcgagctgaagggcatcgact<br>tcaaggaggacggcaacatcctggggcacaagctggagtacaactacaa<br>cagccacaacgtctatatcatggccgacaagcagaagaacggcatcaag<br>gtgaacttcaagatccgccacaacatcgaggacggcagcgtgcagctcg<br>ccgaccactaccagcagaacacccccatcggcgacggccccgtgctgct<br>gcccgacaaccactacctgagcacccagtccgccctgagcaaagaccc<br>caacgagaagcgcgatcacatggtcctgctggagttcgtgaccgccgcc<br>gggatcactctcggcatggacgagctgtacaaggaattctaa<br>Cagtcatgccgcagcaggttaattatcagtaccctccgggctattctcacc<br>acaacttcccacctcccagttttaatagtttccagaacaaccctagttctttcc<br>tgcccagtgctaataacagcagtagtcctcatttcagacatctccctccata<br>cccactcccaaaggctcccagtgagagaaggtccccagaaaggctgaa<br>acactatgatgaccacaggcaccgagatcacagtcatgggcgaggtgag<br>aggcatcggtccctggatcggcgggagcgaggccgcagtcccgacag<br>gagaagacaagacagccggtacagatctgattatgaccgagggagaac<br>accatc | |
| $t_{g1}$-Δ*CAMKMT*::*egfp*-1302 bp<br>(Exon1=**423** nt)<br>(Homo sapiens,<br>chr II~242.19 Mbp) | Ggctctggacacaaaccagtgaaggcaggccggagaacctgaagctct<br>cagaggggagcaggtgtcaccgcaggcaagtccagccgaagtctgcgt<br>tccgcagcccacagaacgacaacttacccagagccgcctagagctggttt<br>gcagcacgccaatctacgtaacctcaatctagcacgagcaacaggcaga<br>tttcgccatctttgttgtggtcaaggagtcttcttgggttcctgggttctttagt<br>ctcgaaatatataagacttttcgttcttcgctagtcttctgagctcgagatgaa<br>gcacagaaggctaaaaca<br>Atggtgagcaagggcgaggagctgttcaccggggtggtgcccatcctg<br>gtcgagctggacggcgacgtaaacggccacaagttcagcgtgtccggc<br>gagggcgagggcgatgccacctacggcaagctgaccctgaagttcatct<br>gcaccaccggcaagctgcccgtgccctggcccaccctcgtgaccaccct<br>gacctacggcgtgcagtgcttcagccgctaccccgaccacatgaagcag<br>cacgacttcttcaagtccgccatgcccgaaggctacgtccaggagcgcac<br>catcttcttcaaggacgacggcaactacaagacccgcgccgaggtgaag<br>ttcgagggcgacaccctggtgaaccgcatcgagctgaagggcatcgact | $t_{g1}$: gene edit template for exon1<br>UHA: 322 bp<br>*egfp*: 717+3 bp<br>DHA: 260 bp<br>Del=Exon1=**423** bp |
| | tcaaggaggacggcaacatcctggggcacaagctggagtacaactacaa<br>cagccacaacgtctatatcatggccgacaagcagaagaacggcatcaag<br>gtgaacttcaagatccgccacaacatcgaggacggcagcgtgcagctcg<br>ccgaccactaccagcagaacacccccatcggcgacggccccgtgctgct<br>gcccgacaaccactacctgagcacccagtccgccctgagcaaagaccc<br>caacgagaagcgcgatcacatggtcctgctggagttcgtgaccgccgcc<br>gggatcactctcggcatggacgagctgtacaagtaa<br>Gtaagggagaacctgctcgcctcacctttgcctctggtcactcctctctcac<br>gtaccgggggagcgactgttcgcagttctttcctcttggcagctctgggag<br>acccttctcagtttgccgggagccacctgcgaagctccccctcgcttcacc<br>taagcgtcccatccaagaggaggaatcatgtggggttgggtggggagcg<br>ggacagccaaggctgagctcctgggaaactcagatcgaagtagaagaa<br>ccccttcta | |
| $t_{b2}$-Δ*CAMKMT*::*egfp*-191 bp<br>(Exon2=**173** nt)<br>(Homo sapiens,<br>chr II~242.19 Mbp) | ATTTCAAAGgt**A**ctgaa**A**caaaaacacctggatgattgcctgcg<br>acatgtatctgtaagaagatttgaatcatttaatctgttttcagtaacagaagg<br>caaagaaagggaaactgaagaggaggttggtgcatgggtccaatataca<br>agcatcttctgtcctgaatacagtatctc**A**t**C**aagGTAACCATT | $t_{b2}$: base edit template for exon2<br>Capital letters with dash-line denote complementary regions with the outer sides of the exon2<br>Bold capital letters indicate mutant bases (T→A, G→A,C→A and T→C, in proper order) |
| t-Δ*CAMKMT*::*egfp*-1506 bp<br>(Exon3=**64**+1 nt)<br>(Homo sapiens,<br>chr II~242.19 Mbp) | Tcagcttgattctctaagtatactgtgtgggtttttggctttaaaaaggttaag<br>aacttccctttacataatgaaaacccaacaagtaaatattatttgttagtacttt<br>aagatgtagattaatcatttaaaaagtcattatgttaaatattctgaggaccat<br>actgttatttctttccttctccctctagtgggcagaaaatattcttaaaatatag<br>catggaataatttattgctattgggtataccataatatacagtctcagtctcag<br>ctttttttgaatactaaaaatgttaacatttcagaatcattaaagcaaacgcttta<br>ctcctttctaggcataatagtgga<br>Atggtgagcaagggcgaggagctgttcaccggggtggtgcccatcctg<br>gtcgagctggacggcgacgtaaacggccacaagttcagcgtgtccggc<br>gagggcgagggcgatgccacctacggcaagctgaccctgaagttcatct<br>gcaccaccggcaagctgcccgtgccctggcccaccctcgtgaccaccct | UHA: 351 bp<br>*egfp*: 717+9 bp<br>DHA: 429 bp<br>Del (**64**-12+254 nt): 306 bp |
| | gacctacggcgtgcagtgcttcagccgctaccccgaccacatgaagcag cacgacttcttcaagtccgccatgcccgaaggctacgtccaggagcgcac catcttcttcaaggacgacggcaactacaagacccgcgccgaggtgaag ttcgagggcgacaccctggtgaaccgcatcgagctgaagggcatcgact tcaaggaggacggcaacatcctggggcacaagctggagtacaactacaa cagccacaacgtctatatcatggccgacaagcagaagaacggcatcaag gtgaacttcaagatccgccacaacatcgaggacggcagcgtgcagctcg ccgaccactaccagcagaacacccccatcggcgacggccccgtgctgct gcccgacaaccactacctgagcacccagtccgccctgagcaaagaccc caacgagaagcgcgatcacatggtcctgctggagttcgtgaccgccgcc gggatcactctcggcatggacgagctgtacaaggaattctaa Gagatggagagagtaggtcagaaagaacactggacactgaagtactgtc cctttatttttgtttccattttgacctcagttaattcagtagtggggctgtttcttc cttctttgctttgtttcgggtgctttctttgccattccctgcttctttactttcactg tagaataaacccagaacatatttttcttgatgctactatacattaataccagag tgtttttgtttgatgcagtattacaatattacagaatatattgccttttcatcaaa gcttctgattaattggtgatagatttgtccactgagaaaatcacttaaatatttc ttataattcttgtaattaatttcatgtgatgaaagacatcagccaagaatacctt tatgtcactgagttaaacagcataccatcataggagctatggac | |
| t-Δ*Lepr*::*egfp*-1230 bp (Exon02=**<u>328</u>**+2 nt) (Mus musculus, chr IV ~156.51 Mbp) | Gaaggtagacgctcagggttgtttattagactccgatccagcactctgaag gttaaatatttctacttgaaaatgctctaccatttaatctcaagtacttcctagtt aatcttcgtgacaatgatggcttgtttctctgtctttttgctgtcctagcagaat ttctttatgtgatagctgcacttaacctggcatatccaatctctccctggaaatt taagttgttttgtggaccaccgaacacaaccgatgactcctttctctcacctg ctggagccccaaacaatgcctcggctttgaagggggcttctgaagcaatt gttgaagctaaatttaattcaagtggtatc Atggtgagcaagggcgaggagctgttcaccggggtggtgcccatcctg gtcgagctggacggcgacgtaaacggccacaagttcagcgtgtccggc gagggcgagggcgatgccacctacggcaagctgaccctgaagttcatct gcaccaccggcaagctgcccgtgccctggcccaccctcgtgaccaccct gacctacggcgtgcagtgcttcagccgctaccccgaccacatgaagcag | UHA: 351 bp<br>*egfp*: 717+9 bp<br>DHA: 153 bp<br>Del (**<u>328</u>**-244 nt): 84 bp |
| | cacgacttcttcaagtccgccatgcccgaaggctacgtccaggagcgcac<br>catcttcttcaaggacgacggcaactacaagacccgcgccgaggtgaag<br>ttcgagggcgacaccctggtgaaccgcatcgagctgaagggcatcgact<br>tcaaggaggacggcaacatcctggggcacaagctggagtacaactacaa<br>cagccacaacgtctatatcatggccgacaagcagaagaacggcatcaag<br>gtgaacttcaagatccgccacaacatcgaggacggcagcgtgcagctcg<br>ccgaccactaccagcagaacacccccatcggcgacggccccgtgctgct<br>gcccgacaaccactacctgagcacccagtccgccctgagcaaagaccc<br>caacgagaagcgcgatcacatggtcctgctggagttcgtgaccgccgcc<br>gggatcactctcggcatggacgagctgtacaaggaattctaa<br>Actgaagggaagacactggcttcagtagtgaaggcttcagtttttcgcca<br>gctaggtaagtacttttatgccccaagtatatatagtgttgtgttctgccttatg<br>caggttaagaatctctctgaaacaatcaatacatgttatgctgagtga | |
| t-Δ*Vegfa*::*egfp*-9 (1444 bp)<br>(Exon1=**600** nt)<br>(Mus musculus,<br>chr XVII ~94.99 Mbp) | Acttttcgtccaacttctgggctcttctcgctccgtagtagccgtggtctgcg<br>ccgcaggagacaaaccgatcggagctgggagaagtgctagctcgggcc<br>tggagaagccggggcccgagaagagaggggaggaagagaaggaaga<br>ggagaggggccgcagtgggcgctcggctctcaggagccgagctcatg<br>gacgggtgaggcggccgtgtgcgcagacagtgctccagccgcgcgcg<br>cgccccaggccccggcccgggcctcggttccagaagggagaggagcc<br>cgccaaggcgcgcaagagagcgggctgcctcgcagtccgagccggag<br>agggagcgcgagccgcgccggccccggacgggcctccgaaaccatga<br>ac<br>Atggtgagcaagggcgaggagctgttcaccggggtggtgcccatcctg<br>gtcgagctggacggcgacgtaaacggccacaagttcagcgtgtccggc<br>gagggcgagggcgatgccacctacggcaagctgaccctgaagttcatct<br>gcaccaccggcaagctgcccgtgccctggcccaccctcgtgaccaccct<br>gacctacggcgtgcagtgcttcagccgctaccccgaccacatgaagcag<br>cacgacttcttcaagtccgccatgcccgaaggctacgtccaggagcgcac<br>catcttcttcaaggacgacggcaactacaagacccgcgccgaggtgaag<br>ttcgagggcgacaccctggtgaaccgcatcgagctgaagggcatcgact<br>tcaaggaggacggcaacatcctggggcacaagctggagtacaactacaa | UHA: 385 bp<br>*egfp*: 717+9 bp<br>DHA: 333 bp<br>Del (**600**-540+223): 283 bp |
| | cagccacaacgtctatatcatggccgacaagcagaagaacggcatcaag gtgaacttcaagatccgccacaacatcgaggacggcagcgtgcagctcg ccgaccactaccagcagaacacccccatcggcgacggccccgtgctgct gcccgacaaccactacctgagcacccagtccgccctgagcaaagaccc caacgagaagcgcgatcacatggtcctgctggagttcgtgaccgccgcc gggatcactctcggcatggacgagctgtacaaggaattctaa Ccacccaccagagtcaccgcacgtacgatctgggccgagcagcggag ggcgggagccagaggaggaggctgagggggctgggcttgtgccgag gctggcggcagaagtttgctccgggtcgcgggtccccggagaactggaa gtccgggcaaagggggcgggagtccggagcccagcgggcatgcctgg gggtgctcggaccttggacccggggagggcagagatcgtggaggggg cagggcgcgggcgaccgagggggctttgctgtcactgccgtttgggtctc tgaggcccttgcagtgagtttggggaaagttttaggatggattgctg | |

*Capital letters represent the site for restriction enzyme digestion; letters with dash-line represent complementary region with plasmid; letters with single underline represent promoter; letters with double underline represent terminator; boldface letters represent direct repeat; italic letters represent spacer; UHA: upstream homologous arm; DHA: downstream homologous arm; Fsc / Rsc: primers for the construction of Cas expression plasmids in genome editing of *S. cerevisiae* LYC4; UF / UR: primers used for amplification of UHA by PCR; DF / DR: primers used for amplification of DHA by PCR; vF / vR: primers used for verification of edited sequences (two fragment sequences in a genome normally locate at both sides of an edited sequence but not in t-DNA); -9: represents the Cas9-mediated genome editing; NCBI database: KY243071- KY243077.

**Table S3.** Genome editing efficiency in the template-based eukaryotic genome editing directed by the *Svi*Cas3*

| Cells \ Conditions | HCN | MCN | NCGEC | TE (%) | HRE (%) | GEE (%) |
|---|---|---|---|---|---|---|
| **RNA-guided genome editing** | | | | | | |
| *S. cerevisiae* LYC4 | | | | 6.8 | 94.0 | 6.4 |
| *Sc*$^c$ (×10$^8$/ml) | 3.77±0.61 | NT | NT | | | |
| *Sc*-M (×10$^7$/ml) | NT | 2.57±1.20 | NT | | | |
| *S. cerevisiae* LYC4-Δ*crtE* | NT | NT | 47 | | | |
| HEK293T | | | | 12.2 | 36.0 | 4.4 |
| Cell number in Bright filed | 584.2±67.5 | NT | NT | | | |
| Cell number with mCherry | NT | 71.3±33.6 | NT | | | |
| Cell number with EGFP | NT | NT | 25.7±7.3 | | | |
| NIH-3T3 | | | | 11.9 | 61.3 | 7.3 |
| Cell number in Bright filed | 337.3±98.3 | NT | NT | | | |
| Cell number with mCherry | NT | 40.3±24.1 | NT | | | |
| Cell number with EGFP | NT | NT | 24.7±12.5 | | | |
| **DNA-guided genome editing** | | | | | | |
| *S. cerevisiae* LYC4 | | | | 6.6 | 98.0 | 6.5 |
| *Sc*$^c$ (×10$^8$/ml) | 1.89±1.37 | NT | NT | | | |
| *Sc*-M (×10$^7$/ml) | NT | 1.25±0.28 | NT | | | |
| *S. cerevisiae* LYC4-Δ*crtE* | NT | NT | 49 | | | |
| HEK293T | | | | 11.8 | 65.8 | 7.8 |
| Cell number in Bright filed | 278.7±249.4 | NT | NT | | | |
| Cell number with mCherry | NT | 33.0±25.4 | NT | | | |
| Cell number with EGFP | NT | NT | 21.7±22.9 | | | |
| NIH-3T3 | | | | 14.8 | 37.3 | 5.5 |
| Cell number in Bright filed | 398.3±207.2 | NT | NT | | | |
| Cell number with mCherry | NT | 59.0±55.2 | NT | | | |
| Cell number with EGFP | NT | NT | 22.0±22.1 | | | |

*Concentration and number of cells are expressed as mean ± SEM of n ≥ 3; NT: not tested.

*Sc*$^c$: *S. cerevisiae* LYC4 competent cells growing on YPD plates (yeast extract peptone dextrose medium); *Sc*-M: mutants of *S. cerevisiae* LYC4 transformed by pRS415-*cas*3 and pYES2-NTA-t/g-Δ*crtE* growing on SD-Leu/Ura plates (synthetic dropout medium-leucine/uracil). HCN: host cell number; MCN: mutant cell number; NCGEC: number of correctly gene-edited cells identified from 50 potential gene-edited cells; TE: transformation efficiency; HRE: homologous recombination efficiency; GEE: genome editing efficiency.

**Table S4.** The main features of typical gene editing techniques [27,28,33,41]

| Item* | ZFN | TALEN | *Sp*Cas9 | *Svi*Cas3 |
|---|---|---|---|---|
| Reco-Mode | Protein | Protein | RNA | DNA or RNA |
| Size-TDS | ~24 bp | ~30 bp | ~20 bp | ≥40 bp? |
| Reco-element | ZFA | TALEA | sgRNA | D-/R-loop? |
| Endonuclease | Dimer of FokI | Dimer of FokI | *Sp*Cas9 | *Svi*Cas3 |
| MW (kDa) / AA | 65.4 / 587 | 65.4 / 587 | 158.4 / 1368 | 84.4 / 771 |
| Effector complex | ZFA::FokI | TALEA::FokI | sgRNA::Cas9 | D/R-loop::*Svi*Cas3 |
| Signature-TD | 3 bp / ZF array | T-5’ | PAM | unlimited |
| Resulted DNA | DSB | DSB | DSB | SSB ? |
| Cytotoxicity | high | mediate | mediate | Low ? |
| Off-target | low | low | high | not detected |
| NoTS | single | single | multiple | multiple |
| Efficiency | low | low | high | high |
| Operation | difficult | difficult | mediate | facile |
| Cast | high | high | mediate | low |

* Reco-Mode: Recognition mode of engineered-nucleases to recognize target DNA, Size-TDS: the size of target DNA sequence recognized by an endonuclease, Reco-element: recognition element, ZFA: zine finger array protein, TALEA: transcription activator like effector array protein, sgRNA: single guide RNA, MW: molecular weight, AA: the number of amino acid residues, Signature-TD: the signature of target DNA, NoTS: number of target site per round operation, DSB: double strand break; SSB: single strand break, ?: to be further determined.

## Supplementary References:

# Appendix

Prof. / Dr. Wang-Yu Tong
Integrated Biotechnology Laboratory
School of Life Sciences
Anhui University
111 Jiulong Road, Hefei 230601, China
Cell: (86) 150-56036299
E-mail: tongwy@ahu.edu.cn

**May 9, 2023**

Dear editors,

We are pleased to submit our original work, **"Template-based eukaryotic genome editing directed by *Svi*Cas3"**, written by *Wang-Yu Tong*, Yong Li, Shou-Dong Ye, An-Jing Wang, Yan-Yan Tang, Mei-Li Li, Zhong-Fan Yu, Ting-Ting Xia, Qing-Yang Liu and Si-Qi Zhu*, to you (arXiv) according to the "Author Guide" from Your website.

It is we all known that gene editing and base editing technologies are the most important tools for creating and transforming organisms and treating human genetic diseases. Here, we recommend this paper for publication as a Research Article in your "arXiv" journal for the following reasons:

(1) The discovery that the single *Svi*Cas3 can direct DNA-guided template-based gene editing and base editing further supports the conclusion that a full Cascade is not required in the RNA-guided genome editing conducted by the *Svi*Cas3 (see sister article: "Prokaryotic genome editing based on the subtype I-B-*Svi* CRISPR-Cas system").

(2) In the DNA-guided genome editing mediated by the single *Svi*Cas3, template-DNA (t-DNA) fragments are required, whereas guide-DNA (g-DNA) fragments are not, suggesting that CRISPR, the basis of crRNA design, is not meaningful for the *Svi*Cas3-based gene editing and base editing.

(3) Compared with its sister article (title: Prokaryotic genome editing based on the subtype I-B-*Svi* CRISPR-Cas system) focusing on RNA-guided genome editing in prokaryotic microorganisms, this article focusing on DNA-guided gene and base editing has the following advantages: (i)simple operation, (ii) low cost, and (iii) free selection of target sequence.

In conclusion, we believe that the *Svi*Cas3 from type I CRISPR-Cas system in *Streptomyces virginiae* IBL14 is highly likely to replace the *Sp*Cas9 from type II CRISPR-Cas system in *S. pyogenes* as the most important tool for creating and modifying organisms and treating human genetic diseases.

We hope that the authenticity of these results reported in "Template-based genome editing guided by *Svi*Cas3" should be tested by independent researchers as soon as possible. Most importantly, we sincerely hope that the *Svi*Cas3-based gene editing tools will soon be applied to the treatment of human genetic diseases and basic research in life sciences (see patent: Type I-B CRISPR-Cas system gene *cas*3-based gene editing method, US 11,286,506 B2).

Reviewers are welcome to point out academic mistakes in this research paper objectively, rather than subjectively. In particular, relevant researchers are welcome to verify the functional authenticity of the *SviCas3* protein.

All data for this study are included in this Manuscript or Supplementary Information, or on the websites marked in the paper.

A patent application (CN107557373A / WO2019056848A1 / EP3556860A1 / US11286506 B2) has been filed for the content disclosed in this study.

The work was not funded by any agency.

We obey the copyright policy of your editorial office as to the copyright of the paper.

We look forward to hearing from you.

Yours sincerely,

Wang-Yu Tong